\documentclass[3p,times]{elsarticle}
\graphicspath{{Figures/}}
\DeclareGraphicsExtensions{.pdf,.png,.jpg}
\usepackage{graphicx}
\usepackage{amsmath,amsthm,amsbsy,amssymb}
\usepackage{calligra,calrsfs,mathrsfs}
\usepackage{multirow}
\usepackage{ulem,cancel}
\usepackage{cases}
\usepackage{pgfplots}
\pgfplotsset{compat=1.18}
\usepackage{subcaption}
\usepackage{tikz}
\usepackage{tkz-euclide}
\usetikzlibrary[patterns]

\usepackage{array}
\usepackage{enumerate}
\usepackage[inline]{enumitem}

\usepackage[U]{fontenc}
\usepackage[latin9]{inputenc}
\usepgfplotslibrary{fillbetween}
\usetikzlibrary{patterns}

\usepackage{tikz-dimline}
\usepackage{color}
\usepackage{textcomp}
\usepackage{soul,xcolor}
\biboptions{sort&compress}
\usepackage[font=normalsize,labelfont=bf]{caption}

\usepackage{csvsimple}

\pgfplotsset{
/pgfplots/bar cycle list/.style={/pgfplots/cycle list={%
    {blue,fill=blue!30!white,mark=none},%
    {red,fill=red!30!white,mark=none},%
    {brown!60!black,fill=brown!30!white,mark=none},%
    {black,fill=gray,mark=none},%
    {violet!80!black,fill=violet,mark=none},%
    {orange,fill=orange!50!white,mark=none}%
    }
},
}
\usepackage[noend]{algpseudocode}
\usepackage{algorithm}
\usepackage{pgfplotstable}
\usepackage{booktabs}
\usepackage{csvsimple}

\usepackage{hyperref}
\hypersetup{
    colorlinks,
    linkcolor={magenta},
    citecolor={blue},
    urlcolor={blue!80!black},
    breaklinks=true,
	plainpages=true
}

\usepackage{listings}
\definecolor{codegreen}{rgb}{0,0.6,0}
\definecolor{codegray}{rgb}{0.5,0.5,0.5}
\definecolor{codepurple}{rgb}{0.58,0,0.82}
\definecolor{backcolour}{rgb}{0.95,0.95,0.92}

\lstdefinestyle{mystyle}{
	backgroundcolor=\color{backcolour},   
	commentstyle=\color{codegreen},
	keywordstyle=\color{magenta},
	stringstyle=\color{codepurple},
	basicstyle=\ttfamily\footnotesize,
	breakatwhitespace=false,         
	breaklines=true,                 
	captionpos=b,                    
	keepspaces=true,                  
	showspaces=false,                
	showstringspaces=false,
	showtabs=false,                  
	tabsize=2
}

\usetikzlibrary{positioning} 

\newcommand{\ifcomment}{\iffalse}
\newcommand{\bs}[1]{\boldsymbol{#1}}

\newcommand{\avg  }[1]{\langle #1 \rangle}

\newcommand{\ti}[1]{\tilde{#1}}
\newcommand{\G}{\Gamma}

\newcommand{\Om}{\Omega}

\newcommand{\cT}{{\mathcal T}}
\newcommand{\tO}{\tilde{\Omega}_h}

\newcommand{\tG}{\tilde{\Gamma}_h}

\newcommand{\tGD}{\tilde{\Gamma}_{D,h}}

\newcommand{\tx}{{\tilde{\bs{x}}}}

\newcommand{\Mw}{\boldsymbol{w}^h}
\newcommand{\Mu}{\boldsymbol{u}^h}

\newdefinition{rem}{Remark}

\newtheorem*{remark}{Remark}

\newcommand{\cref}[2]{\hyperref[#2]{#1~\ref*{#2}}}
\newcommand{\colref}[2]{\hyperref[#2]{#1~\ref*{#2}}}
\newcommand{\eqnref}[1]{\colref{Eq.}{#1}}
\newcommand{\figref}[1]{\colref{Figure}{#1}}

\newcommand{\secref}[1]{\colref{Section}{#1}}
\newcommand{\tabref}[1]{\colref{Table}{#1}}

\newcommand{\Algref}[1]{\hyperref[#1]{Algorithm~\ref*{#1}}}

\newcolumntype{M}[1]{>{\centering\arraybackslash}m{#1}}

\newcommand{\Frontera}{\href{https://www.tacc.utexas.edu/systems/frontera}{Frontera}}

\newcommand{\petsc}{\href{https://petsc.org/release/}{\textsc{PETSc}}}
\newcommand{\dendroKT}{\textsc{Dendro-KT}}

\newcommand{\Intercepted}{\textsc{Intercepted}}

\newcommand{\Exterior}{\textsc{Exterior}}
\newcommand{\Interior}{\textsc{Interior}}
\newcommand{\FalseIntercepted}{\textsc{FalseIntercepted}}
\newcommand{\TrueIntercepted}{\textsc{TrueIntercepted}}

\definecolor{ActiveElement}{RGB}{154,203,78}
\definecolor{InterceptedElement}{RGB}{255,212,25}
\definecolor{FalseInterceptedElement}{RGB}{96,95,255}
\definecolor{lightprofgreen}{RGB}{60,179,113} 
\definecolor{NodalStrongBC}{RGB}{74,77,248}

\definecolor{cpu1}{HTML}{4CAF50}
\definecolor{cpu2}{HTML}{FFC107}
\definecolor{cpu3}{HTML}{F44336}
\definecolor{cpu4}{HTML}{2196F3}
\definecolor{cpu5}{HTML}{9932CC}

\definecolor{gpu1}{HTML}{A5D6A7}
\definecolor{gpu2}{HTML}{FFE082}
\definecolor{gpu3}{HTML}{EF9A9A}
\definecolor{gpu4}{HTML}{90CAF9}

\definecolor{TruePt}{RGB}{255,0,0}
\definecolor{Exp1}{RGB}{85,170,255}
\definecolor{Exp2}{RGB}{126,0,0}
\definecolor{SurrogateGP}{RGB}{159,159,159}

\definecolor{sq_b1}{RGB}{37,52,148}
\definecolor{sq_b2}{RGB}{44,127,184}
\definecolor{sq_b3}{RGB}{65,182,196}
\definecolor{sq_b4}{RGB}{127,205,187}
\definecolor{sq_b5}{RGB}{199,233,180}
\definecolor{sq_b6}{RGB}{255,255,204}

\definecolor{sq_r1}{RGB}{189,0,38}
\definecolor{sq_r2}{RGB}{240,59,32}
\definecolor{sq_r3}{RGB}{253,141,60}
\definecolor{sq_r4}{RGB}{254,178,76}
\definecolor{sq_r5}{RGB}{254,217,118}
\definecolor{sq_r6}{RGB}{255,255,178}

\definecolor{sq_g1}{RGB}{0,104,55}
\definecolor{sq_g2}{RGB}{49,163,84}
\definecolor{sq_g3}{RGB}{120,198,121}
\definecolor{sq_g4}{RGB}{173,221,142}
\definecolor{sq_g5}{RGB}{217,240,163}
\definecolor{sq_g6}{RGB}{255,255,204}

\definecolor{div_c1}{RGB}{230,171,2}
\definecolor{div_c2}{RGB}{102,166,30}
\definecolor{div_c3}{RGB}{231,41,138}
\definecolor{div_c4}{RGB}{117,112,179}
\definecolor{div_c5}{RGB}{217,95,2}
\definecolor{div_c6}{RGB}{27,158,119}
\definecolor{div_c7}{RGB}{215,48,39}

\definecolor{div_d1}{RGB}{215,25,28}
\definecolor{div_d2}{RGB}{253,174,97}
\definecolor{div_d3}{RGB}{255,255,191}
\definecolor{div_d4}{RGB}{171,217,233}
\definecolor{div_d5}{RGB}{44,123,182}

\definecolor{lineclr}{RGB}{0,0,0}
\definecolor{utorange}{RGB}{0,0,255}
\definecolor{utsecblue}{RGB}{255,255,0}
\definecolor{utsecgreen}{RGB}{255,0,0}
\definecolor{red!15}{RGB}{0,255,255}
\definecolor{fillclr5}{RGB}{0,255,0}
\definecolor{fillclr6}{RGB}{255,0,255}
\definecolor{fillclr7}{RGB}{255,255,255}
\definecolor{fillclr8}{RGB}{0,0,0}

\definecolor{armygreen}{rgb}{0.29, 0.33, 0.13}
\definecolor{aurometalsaurus}{rgb}{0.43, 0.5, 0.5}
\definecolor{applegreen}{rgb}{0.55, 0.71, 0.0}
\definecolor{darkgreen}{rgb}{0.0, 0.4, 0.25}

\def\drawcubeI(#1,#2,#3,#4,#5){ 
\coordinate (O) at (#1,#2,#3);
\coordinate (A) at (#1,#2+#4,#3);
\coordinate (B) at (#1,#2+#4,#3+#4);
\coordinate (C) at (#1,#2,#3+#4);
\coordinate (D) at (#1+#4,#2,#3);
\coordinate (E) at (#1+#4,#2+#4,#3);
\coordinate (F) at (#1+#4,#2+#4,#3+#4);
\coordinate (G) at (#1+#4,#2,#3+#4);
\draw[#5] (O) -- (C) -- (G) -- (D) -- cycle;
\draw[#5] (O) -- (A) -- (E) -- (D) -- cycle;
\draw[#5] (O) -- (A) -- (B) -- (C) -- cycle;
\draw[#5] (D) -- (E) -- (F) -- (G) -- cycle;
\draw[#5] (C) -- (B) -- (F) -- (G) -- cycle;
\draw[#5] (A) -- (B) -- (F) -- (E) -- cycle;
}

\def\drawcubeII(#1,#2,#3,#4,#5,#6,#7){ 
\coordinate (O) at (#1,#2,#3);
\coordinate (A) at (#1,#2+#4,#3);
\coordinate (B) at (#1,#2+#4,#3+#4);
\coordinate (C) at (#1,#2,#3+#4);
\coordinate (D) at (#1+#4,#2,#3);
\coordinate (E) at (#1+#4,#2+#4,#3);
\coordinate (F) at (#1+#4,#2+#4,#3+#4);
\coordinate (G) at (#1+#4,#2,#3+#4);
\draw[#5,fill=#6,opacity=#7] (O) -- (C) -- (G) -- (D) -- cycle;
\draw[#5,fill=#6,opacity=#7] (O) -- (A) -- (E) -- (D) -- cycle;
\draw[#5,fill=#6,opacity=#7] (O) -- (A) -- (B) -- (C) -- cycle;
\draw[#5,fill=#6,opacity=#7] (D) -- (E) -- (F) -- (G) -- cycle;
\draw[#5,fill=#6,opacity=#7] (C) -- (B) -- (F) -- (G) -- cycle;
\draw[#5,fill=#6,opacity=#7] (A) -- (B) -- (F) -- (E) -- cycle;
}

\def\drawNodes(#1,#2,#3,#4,#5,#6,#7){ 
\foreach \x in {#1,#7,...,#2}{
	\foreach \y in {#3,#7,...,#4}{
		\foreach \z in {#5,#7,...,#6}{
				\draw[fill=red!60] (\x,\y,\z) circle (0.15);
				}
			}
	}				
		
}

\pgfplotsset{
  log x ticks with fixed point/.style={
      xticklabel={
        \pgfkeys{/pgf/fpu=true}
        \pgfmathparse{exp(\tick)}%
        \pgfmathprintnumber[fixed relative, precision=3]{\pgfmathresult}
        \pgfkeys{/pgf/fpu=false}
      }
  },
  log y ticks with fixed point/.style={
      yticklabel={
        \pgfkeys{/pgf/fpu=true}
        \pgfmathparse{exp(\tick)}%
        \pgfmathprintnumber[fixed relative, precision=3]{\pgfmathresult}
        \pgfkeys{/pgf/fpu=false}
      }
  }
}

\makeatletter
\newcommand\resetstackedplots{
\makeatletter
\pgfplots@stacked@isfirstplottrue
\makeatother
\addplot [forget plot,draw=none] coordinates{(48,0) (96,0) (192,0) (384,0) (768,0) (1536,0) (3072,0) (6144,0)};
}
\makeatother

\makeatletter
\newcommand\resetstackedplotsOne{
\makeatletter
\pgfplots@stacked@isfirstplottrue
\makeatother
\addplot [forget plot,draw=none] coordinates{(384,0) (768,0) (1536,0) (3072,0) (6144,0)};
}
\makeatother

\makeatletter
\newcommand\resetstackedplotsTwo{
\makeatletter
\pgfplots@stacked@isfirstplottrue
\makeatother
\addplot [forget plot,draw=none] coordinates{(16,0) (32,0) (64,0) (128,0) (256,0) (512,0) (1024,0) (2048,0) (4096,0) (8192,0) (16384,0) (32768,0)};
}
\makeatother

\makeatletter
\newcommand\resetstackedplotsThree{
\makeatletter
\pgfplots@stacked@isfirstplottrue
\makeatother
\addplot [forget plot,draw=none] coordinates{(2,0) (4,0) (8,0) (16,0) (32,0) (64,0)};
}
\makeatother

\makeatletter
\newcommand\resetstackedplotsFour{
\makeatletter
\pgfplots@stacked@isfirstplottrue
\makeatother
\addplot [forget plot,draw=none] coordinates{(4,0) (8,0) (16,0) (32,0) (64,0)};
}
\makeatother

\makeatletter
\newcommand\resetstackedplotsFive{
\makeatletter
\pgfplots@stacked@isfirstplottrue
\makeatother
\addplot [forget plot,draw=none] coordinates{(1,0) (2,0) (4,0) (8,0) (16,0) (32,0) (64,0) (128,0)};
}
\makeatother

\makeatletter
\newcommand\resetstackedplotsSix{
\makeatletter
\pgfplots@stacked@isfirstplottrue
\makeatother
\addplot [forget plot,draw=none] coordinates{(2,0) (4,0) (8,0) (16,0) (32,0) (64,0) (128,0)};
}
\makeatother

\newcolumntype{P}[1]{>{\centering\arraybackslash}p{#1}}

\allowdisplaybreaks

\begin{document}

\begin{frontmatter}


\title{Simulating incompressible flows over complex geometries using the shifted boundary method with incomplete adaptive octree meshes}

\author[ISU]{Cheng-Hau Yang}
\ead{chenghau@iastate.edu}
\author[Duke]{Guglielmo Scovazzi}
\ead{guglielmo.scovazzi@duke.edu}
\author[ISU]{Adarsh Krishnamurthy}
\ead{adarsh@iastate.edu}
\author[ISU]{Baskar Ganapathysubramanian\corref{cor}}
\ead{baskarg@iastate.edu}
\cortext[cor]{Corresponding author}
\address[ISU]{Iowa State University, Ames, IA}
\address[Duke]{Department of Civil and Environmental Engineering, Duke University, Durham, North Carolina 27708, USA}

\begin{abstract}
We extend the shifted boundary method (SBM) to the simulation of incompressible fluid flow using immersed octree meshes. Previous work on SBM for fluid flow primarily utilized two- or three-dimensional unstructured tetrahedral grids. Recently, octree grids have become an essential component of immersed CFD solvers, and this work addresses this gap and the associated computational challenges. We leverage an optimal (approximate) surrogate boundary constructed efficiently on incomplete and adaptive octree meshes. The resulting framework enables the simulation of the incompressible Navier-Stokes equations in complex geometries without requiring boundary-fitted grids. Simulations of benchmark tests in two and three dimensions demonstrate that the Octree-SBM framework is a robust, accurate, and efficient approach to simulating fluid dynamics problems with complex geometries.

\end{abstract}
\end{frontmatter}

\section{Introduction}
\label{Sec:Intro}

Simulating incompressible flow poses significant challenges, especially when geometries are complex~\cite{verzicco2023immersed, mittal2023origin}. Traditional methods that rely on body-fitted meshes are ultimately constrained by the complexity of generating meshes that accurately conform to these geometries. Creating suitable body-fitted meshes, particularly for intricate or irregular shapes, is a labor-intensive process that often requires significant manual effort and expertise~\cite{george1991automatic, owen1998survey, persson2005mesh}. Engineers and researchers must iteratively adjust and refine the mesh to capture the geometric details accurately while ensuring numerical stability and convergence~\cite{boelens2009f16, kurtulus2015unsteady}. This process can be incredibly cumbersome when dealing with large-scale simulations, where the sheer size and complexity of the computational domain make manual mesh generation infeasible or prohibitively expensive. The overall cost and time associated with manual mesh generation not only slows down the simulation workflow but also limits the ability to perform extensive parametric studies, optimize designs efficiently, or train reduced-order models and machine-learning algorithms, for example. This poses a bottleneck when conducting large-scale simulations with traditional body-fitted meshes. Despite advances in automated mesh generation algorithms~\cite{george1991automatic, simonovski2011automatic, beaufort2020automatic}, achieving the desired mesh quality for complex geometries without human intervention remains a challenge.

One promising approach to address these difficulties is through the family of methods that started with the Immersed Boundary Method (IBM). First developed by Peskin~\cite{peskin1972flow}, the Immersed Boundary Method revolutionized how flow simulations across complex geometries are approached by embedding the geometry within the fluid domain and using a fixed, non-body-fitted mesh. This method alleviates the need for complex mesh generation by allowing the use of simple, structured grids, such as Cartesian grids, irrespective of the geometry's complexity. By simplifying the time-consuming mesh generation process, IBM significantly reduces manual effort and computational costs, making large-scale simulations more practical and efficient. The IBM and its variants have been extensively applied to simulate flows around complex geometries and moving boundaries. For a comprehensive review of the IBM, we refer the interested reader to \citet{mittal2005immersed}, who categorize implementations of the IBM into two categories based on how the forcing function between the geometry and the fluid is imposed:
\begin{itemize}[topsep=3pt,itemsep=0pt,left=0pt]
\item \textbf{Continuous forcing}: The force term is incorporated directly into the strong form of the Navier-Stokes equations, with the force represented by a smoothed Dirac delta function distributed over a region of the fluid domain near the boundary. Techniques such as the Immersed Boundary Method (IB)~\citep{peskin1972flow, peskin2002immersed,lai2000immersed}, the Immersed Boundary Projection Method (IBPM)~\citep{taira2007immersed,colonius2008fast,taira2008immersed,ong2020immersed}, the Extended Immersed Boundary Method (EIBM)~\citep{wang2004extended}, and the Immersed Finite Element Method (IFEM)~\citep{zhang2004immersed,liu2006immersed,wang2013modified,cheng2019openifem} are notable examples that utilize this strategy.

\item \textbf{Discrete forcing}: This approach involves taking the original strong form of the Navier-Stokes equations without adding any new force term and discretizing it. After discretization, boundary terms are added at discrete points on the boundaries within in the cut cells (often called the \Intercepted{} cells) to apply the boundary conditions. Methods like the Sharp-Interface Immersed Boundary Method (SIIB)~\citep{mittal2008versatile,seo2011sharp,angelidis2016unstructured,turner2024high}, the Curvilinear Immersed Boundary Method (CURVIB)~\citep{ge2007numerical,borazjani2008curvilinear,khosronejad2011curvilinear}, the Finite Cell Method (FCM)~\citep{Parvizian:07.1,Duester:08.1,schillinger2013review,stavrev2016geometrically,de2017condition, jomo2021hierarchical}, Immersogeometric Analysis (IMGA)~\citep{kamensky2015immersogeometric,xu2016tetrahedral, wang2017rapid, HOANG2019421, DEPRENTER2019604, ZhuQiming201911, XU2021103604, kamensky2021open, balu2023direct}, and the Shifted Boundary Method (SBM)~\citep{main2018shifted,Main2018TheSB,KARATZAS2020113273,atallah2020second,atallah2021shifted,atallah2021analysis,colomes2021weighted,saurabh2021scalable,atallah2022high,ZENG2022115143,heisler2023generating,yang2024optimal}  exemplify this category.
\end{itemize}


Building on the foundation laid by IBM, the Shifted Boundary Method (SBM) is an innovative numerical approach designed to address complex geometries and boundary conditions using non-boundary fitted meshes. Within the category of discrete forcing, SBM is a Finite Element Method (FEM) that employs Nitsche's method~\citep{nitsche1971variationsprinzip} to enforce boundary conditions weakly. Unlike methods such as FCM and IMGA, which directly apply boundary conditions to the actual boundary, SBM applies them at a surrogate boundary\footnote{The central concept of the SBM approach -- to "shift" the application of the boundary condition from the true surface (which requires a body fitted mesh) to a surrogate surface (which could be a non-body fitted envelope around the geometry) -- has been successfully applied to FEM~\citep{main2018shifted,Main2018TheSB}, VEM~\citep{hou2024high,bertoluzza2024virtualelementmethodpolygonal}, and IGA~\citep{antonelli2024shifted} formulations}. This technique offers significant advantages in fluid dynamics simulations, particularly its implementation and ability to address these computational challenges:
\begin{itemize}[topsep=3pt,itemsep=0pt,left=0pt]

    \item \textbf{Ease of Implementation}: SBM is notably more straightforward to implement than other immersed methods, such as FCM and IMGA. These methods involve surface quadrature (usually Gauss points) over the actual boundary, requiring additional data structures into the existing code framework. Conversely, SBM simplifies the process by utilizing Gauss points along the surrogate boundary, which naturally aligns with the existing element edges (faces) in the original framework. Additionally, other discrete forcing approaches (e.g., IMGA and FCM) require a higher number of quadrature points in the cut elements for accurate volume integration, which increases their complexity of implementation and often creates a computational load imbalance across processors during parallel (distributed memory) simulations~\cite{saurabh2021industrial}.

    \item \textbf{Resolving Sliver Cut-cell Issue}: A notable advantage of SBM is its capability to eliminate the poor conditioning of the stiffness matrix caused by sliver cut cells~\citep{antonelli2024shifted}. SBM addresses this by integrating over all volume Gauss points, thus avoiding the issues arising from these tiny cut cells.

\end{itemize}

This work extends the SBM approach to performing incompressible flow simulations utilizing an octree-based adaptive mesh\footnote{In addition to IBM in octree-based meshes, several researchers apply sharp forcing to enforce boundary conditions on octree meshes using the level-set method~\cite{osher1988fronts}. This method, which implicitly represents irregular free boundaries, is particularly well-suited for solving a range of problems, including Poisson equations~\cite{losasso2004simulating,kim2023super}, linear elasticity~\cite{theillard2013second}, Euler~\cite{popinet2003gerris}, and Navier-Stokes~\cite{guittet2015stable,egan2021direct} equations, as well as the Stefan problem~\cite{chen2009numerical,papac2013level,bayat2022sharp}, especially in the context of complex geometries and topologies.}. Octree meshes are ideally suited for flow simulations, due to their inherent flexibility and scalability when handling complex geometries~\citep{popinet2003gerris,guittet2015stable,sousa2019finite,egan2021direct,saurabh2021industrial,egan2021direct,van2022fourth,yu2022multi,blomquist2024stable}. Octree meshes allow for adaptive refinement, enabling efficient resolution of detailed features in the immersed geometry without requiring body-fitted meshes. This makes octree meshes particularly advantageous in applications where high levels of local refinement are necessary around complex or evolving boundaries. Moreover, octree meshes maintain a Cartesian structure, simplifying the implementation of fast solvers and parallel algorithms, which is critical in large-scale simulations. In recent work, researchers have extended the octree data structure to account for incomplete octrees~\cite{saurabh2021scalable}. Incomplete octree meshes are particularly useful because they efficiently manage regions that are irrelevant to the solution, thus reducing unnecessary computations. These octrees, by design, exclude elements outside the solution domain, enabling adaptive refinement only where necessary. This selective refinement makes incomplete octree meshes particularly suitable for immersed boundary problems where complex geometries are present. By focusing computational effort on regions near boundaries or high-gradient areas, incomplete octrees ensure computational efficiency without sacrificing accuracy.

\textbf{Technical contributions}: To the best of our knowledge, this is the first instance of incompressible flow simulations utilizing the SBM on an octree-based adaptive mesh. This work extends the application of the SBM using optimal surrogate boundaries, which has previously demonstrated enhanced solution accuracy for Poisson and Linear Elasticity equations within FEM~\citep{yang2024optimal} and for the Poisson equation within Isogeometric Analysis (IGA)~\citep{antonelli2024shifted}, to the realm of Navier-Stokes equations. This extension is novel, as it incorporates the use of an (incomplete) octree mesh in conjunction with an optimal surrogate boundary, a methodology not yet explored in this context. We illustrate the capability of the octree-SBM approach by solving various benchmark problems in fluid dynamics. These cases, including simulations of flow around circular geometries~\citep{zhang1995transition, liu1998preconditioned, uhlmann2005immersed, posdziech2007systematic, wang2009immersed, wu2009implicit, yang2009smoothing, rajani2009numerical, kanaris2011three, Kamensky:2015ch, main2018shifted_2, kang2021variational}, lid-driven cavity flow with a circular obstacle~\citep{huang2020simulation}, aerodynamic problem~\citep{kurtulus2015unsteady,di2018fluid}, flow past a sphere at Reynolds numbers of 300 and 3700~\citep{le1970numerical, roos1971some, yun2006vortical, park2006dynamic, Johnson1999FlowPA, Marella2005, vanella2010direct, wang2011immersed, rodriguez2011direct, angelidis2016unstructured, schlichting2016boundary, dorschner2016grid, kang2021variational}, flow past a cylinder at Reynolds numbers of 3900~\citep{lourenco1994characteristics,norberg1994experimental,beaudan1995numerical,franke2002large,ma2000dynamics,parnaudeau2008experimental,meyer2010conservative,Main2018TheSB,lysenko2012large}, and internal Flow through a nozzle. We finally illustrate the capability of the octree-SBM approach in managing complex boundaries and geometries by simulating flow through a complex gyroid topology. Creating a body-fitted mesh for such a geometry is non-trivial and requires significant human intervention. Additionally, we also show the parallel scaling performance of the approach on up to 500 processors. 

The rest of the paper is structured as follows: \secref{sec:math} outlines the specific variational problem in the context of body-fitted FEMs and discusses the SBM variational formulation. \secref{sec:implementation} covers the implementation of the SBM, and the data structures required in this framework. \secref{sec:results} evaluates the proposed Octree-SBM framework through several canonical benchmarks, as well as the problem of flow past a gyroid geometry. This last test demonstrates in particular the robustness of Octree-SBM in handling complex geometries. Finally, in \secref{sec:conclusion}, we conclude and suggest directions for future research.

\section{Mathematical Formulation}\label{sec:math}

\begin{figure}[t!]
    \centering
        \centering
        \includegraphics[width=0.35\linewidth,trim=0 0 0 0,clip]{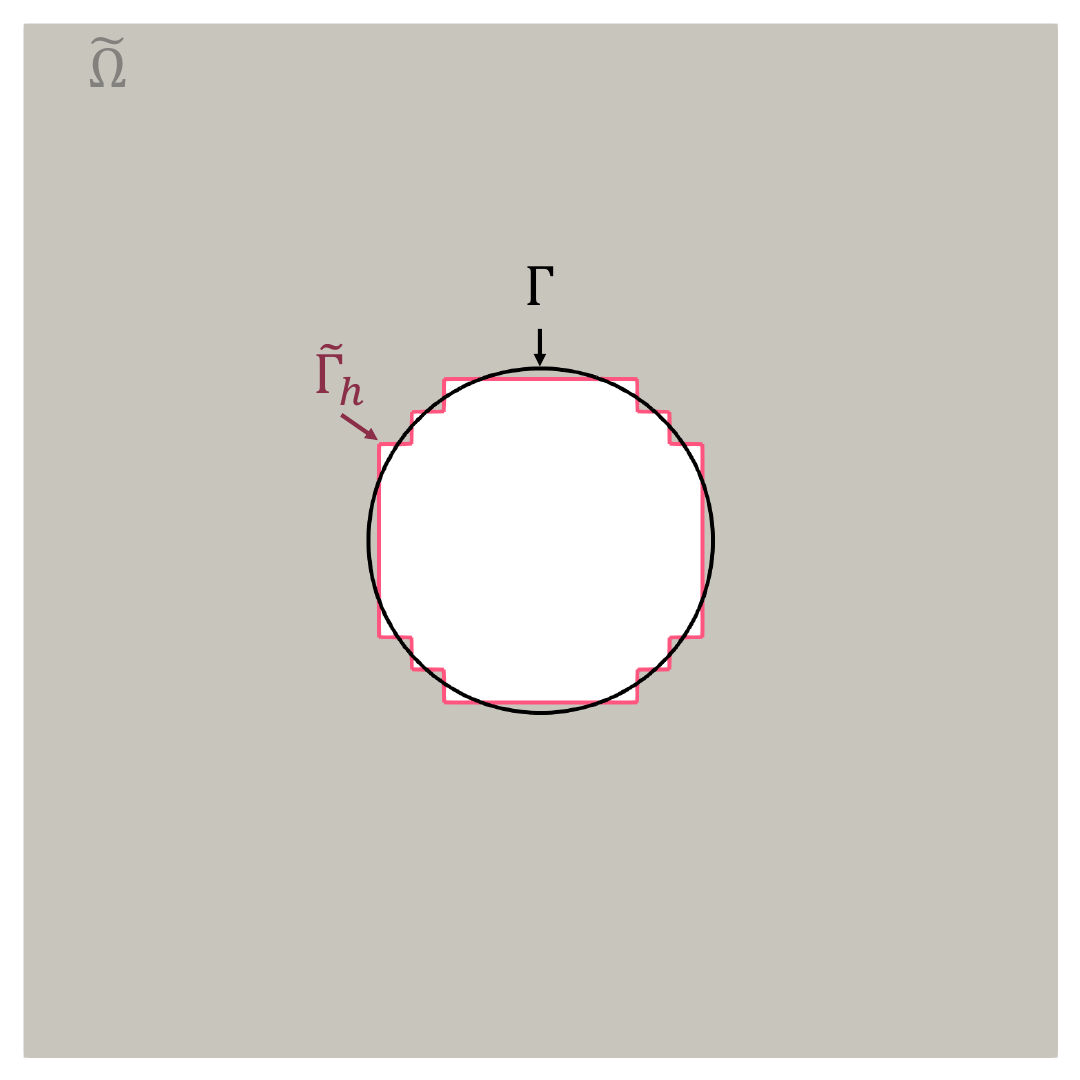}
        \caption{Surrogate domain ($\tO$), surrogate boundary ($\tG = \partial \tO$), and true boundary ($\G$)}
    \label{fig:TrueAndSurrogate}
\end{figure}

\subsection{Governing equations}\label{sec:GoverningEqs}

Consider a fluid domain $\Omega \subset \mathbb{R}^d$, with Lipschitz boundary $\G = \partial \Omega$ (here $d$ indicates the number of space dimensions, equal to either two or three).
The non-dimensional form of the incompressible Navier-Stokes equations can be stated as:
\begin{subequations}
\label{eq:NSEq}
\begin{align}
    \text{Momentum conservation:} \quad & \partial_t \bs{u} + (\bs{u} \cdot \nabla) \bs{u} + \nabla p - \frac{1}{Re} \nabla^2 \bs{u} - \bs{f} = 0 \; , 
    \label{eq:MomentumEq} \\
    \text{Mass conservation (incompressibility):} \quad & \nabla \cdot \bs{u} = 0
    \; ,
    \label{eq:Solenoidality}
\end{align}
\end{subequations}
where $\bs{u}$ represents the non-dimensional (normalized) velocity vector, $p$ the non-dimensional pressure, and $\bs{f}$ the non-dimensional forcing vector. $Re = \frac{\rho_r u_r L_r}{\mu_r}$ stands for the Reynolds number, where $\rho_r$ is the reference density, $u_r$ is the reference velocity, $L_r$ is the reference length, and $\mu_r$ is the reference dynamic viscosity.

\subsection{Boundary conditions}\label{sec:BCs}

The governing equations~\eqnref{eq:NSEq} are complemented with boundary conditions, which, for the purpose of this work, can be categorized as inflow, outflow, and no-slip boundary conditions.
We assume a non-overlapping partition of the boundary $\G$ into an inflow boundary $\G_{in}$, an outflow boundary $\G_{out}$, and a no-slip boundary $\G_{ns}$.
For simplicity, we will impose inflow boundary conditions prescribing all components of the velocity $\bs{u}$ on the inflow boundary $\G_{in}$, hence reducing the inflow condition to a Dirichlet condition on the velocity. The no-slip boundary condition also involves prescribing all the components of the velocity (to zero) and is also of Dirichlet type. Hence we will denote by $\G_D = \G_{in} \cup \G_{ns}$ the Dirichlet boundary, which incorporates both the inflow and no-slip portions of the boundary $\G$, and impose:
\begin{align}
    \bs{u} &=\; \bs{u}_D \; , \quad \mbox{on } \G_D \; .
\end{align}
An outflow boundary condition allows the fluid to freely exit the flow domain. This result is achieved with two different strategies in the present work.
The first strategy involves imposing the pressure at the outlet, as 
\begin{align}
   p &=\; p_{out} \; , \quad \mbox{on } \G_{out} \; ,
\end{align}
where $p_{out}$ is typically set to zero, strongly in the variational forms.
This outflow condition is effective and robust when strong vortical structures are absent from the fluid, so that reverse flow at the outlet is unlikely.  
Examples of application of this boundary condition can be seen in~\secref{subsec:2D_cylinder} and \secref{subsubsec:Gyroid},

Instead, when strong recirculating, vortical flow is expected at the outlet, an alternative stabilized outflow boundary condition is preferred, as discussed in \secref{sec:NS_backflow}: 
Stabilized backflow outlet boundary conditions are applied in \secref{subsub:airfoil}, \secref{subsubsec:Re300}, \secref{subsubsec:Re3700}, and \secref{subsubsec:Cylinder}.

\subsection{Variational Multiscale formulation for body-fitted grids}\label{sec:VMS}

{\it Notation.} Let us denote by $L^2(\Om)$ the space of square integrable functions over $\Om$. 
Here and in the following, $(v,w)_\omega = \int_\omega v \,w$ denotes the $L^2$-inner product on a subset $\omega \subseteq \Om$, and $\avg{v,w}_{\gamma} = \int_{\gamma} v \, w$ denotes the $L^2$-inner product on a subset $\gamma \subseteq \G$. 
Let $H^m(\Om)=W^{m,p}(\Om)$ indicate the Sobolev spaces of index of regularity $m \geq 0$ and index of summability $p = 2$, equipped with the (scaled) norm
\begin{equation}
\|v \|_{H^{m}(\Om)} 
= \left( \| v \|^2_{L^2(\Om)} + \sum_{k = 1}^{m} \| l(\Om)^k  \bs{D}^k v \|^2_{L^2(\Om)} \right)^{1/2} \; ,
\end{equation}
where $\bs{D}^{k}$ is the $k$th-order spatial derivative operator and $l({\Om})= (\mathrm{meas}_{n_d}({\Om}))^{1/{n_d}}$ is a characteristic length of the domain ${\Om}$ (introduced here to maintain dimensional consistency in the definitions of the Sobolev norms). Note that $H^0(\Om)=L^{2}(\Om)$.  As usual, we use a simplified notation for norms and semi-norms, i.e., we set $\| v \|_{m;\Om}=\| v \|_{H^m(\Om)}$ and $| v |_{k;\Om}= 
\| \bs{D}^k v \|_{0;\Om}= \| \bs{D}^k v \|_{L^2(\Om)}$.
Note that the above definitions are general, and the sets $\Om$ and $\G$ can be replaced by other sets, depending on the situation.

To start the discussion, we introduce here a body-fitted FEM formulation of the incompressible Navier-Stokes equations, which will serve as a benchmark for the the Shifted Boundary Method, soon to be developed.
We will consider in particular a Variational Multiscale (VMS) formulation of the incompressible Navier-Stokes equations~\cite{TJRHughes:1998a,HugScoFr04,Bazilevs07b}, which can be interpreted as both a form of numerical stabilization (i.e., the SUPG and PSPG stabilization) or as a turbulence LES model.

Assuming that the domain $\Om$ is discretized using finite elements, we can introduce the family $\cT_h(\Om)$ of shape-regular discrete decompositions (meshes/grids) of $\Om$.
Here, with some abuse of notation, we have assumed that $\cT_h(\Om)$ exactly decomposes $\Om$, which is true only if the boundary $\G$ of $\Om$ is polygonal/polyhedral.
It would be more appropriate to write $\cT_h(\bar{\Om})$, where $\bar{\Om}$ is a domain that approximates $\Om$ and has a polygonal/polyhedral boundary $\bar{\G}$ whose vertices lie on $\G$.
In what follows, for the sake of simplicity, we will omit this more precise, although complex, notation.

We introduce the discrete space of continuous, piecewise-linear, vector-valued test functions over the decompositions $\cT_h(\Om)$:
\begin{equation}
\bs{V}^h(\Om) = \left\{ \Mw \mid \Mw \in (C^0(\Om))^d \cap (\mathcal{P}^1(T))^d \,, \mbox{ with } T \in \cT_h(\Om) \mbox{ and } \Mw_{| \G_D} = \bs{0} \right\} 
\end{equation}
and the corresponding space of trial functions 
\begin{equation}
\bs{S}^h(\Om) = \left\{ \bs{v}^h \mid \bs{v}^h \in (C^0(\Om))^d \cap (\mathcal{P}^1(T))^d \,, \mbox{ with } T \in \cT_h(\Om) \mbox{ and } \bs{v}^h_{| \G_D} = \bs{u}_D \right\} 
\; .
\end{equation}
More precisely, imposing the value $\bs{u}_D$ on $\G_D$ for a $ \bs{v}^h \in \bs{S}^h(\Om)$ holds only if the boundary conditions $\bs{u}_D$ is piece-wise linear along the edges/faces lying on $\G_D$.
Also in this case, we simplify our notation, which is slightly abused.
We also introduce the discrete space of continuous, piecewise-linear, scalar-valued functions over $\cT_h(\Om)$:
 \begin{equation}
V^h(\Om) = \left\{ w^h \mid w^h \in C^0(\Om) \cap \mathcal{P}^1(T) \,, \, T \in \cT_h(\Om) \right\} 
\; ,
\end{equation}
in which no boundary conditions are incorporated.
The VMS formulation of the incompressible Navier-Stokes equations on body-fitted grids reads: \\[.2cm]

Find $\Mu \in \bs{V}^h(\Om)$ and $p^h \in S^h(\Om)$ such that, for any $\Mw \in \bs{V}^h(\Om)$ and $q^h \in V^h(\Om)$,
\begin{equation}
\label{eq:VMS_formulation}
{NS}[\Om \,; \cT_h(\Om)](\Mw,q^h;\Mu,p^h) \; = \; 0
\end{equation}
where
 \begin{align}
\label{eq:VMS_formulation_2}
{NS}[\Om \, ; \cT_h(\Om)](\Mw,q^h;\Mu,p^h)  
& = \; 
(\Mw , \partial_t \Mu + \Mu \cdot \nabla \Mu )_{\Om} 
+ \frac{1}{Re} (\nabla^s \Mw , \nabla^s \Mu )_{\Om} 
- ( \nabla \cdot \Mw, p^h )_{\Om} 
+ ( q^h, \nabla \cdot \Mu )_{\Om}  
- (\Mw, \bs{f}^h )_{\Om}  
\nonumber 
\\ 
& \phantom{=} \; 
-  \sum_{T \in \cT_h(\Om)}( \Mu \cdot \nabla \Mw, \bs{u'} )_{T}
+ \sum_{T \in \cT_h(\Om)}( \Mw, \bs{u'} \cdot \nabla \Mu )_{T} 
- \sum_{T \in \cT_h(\Om)}( \nabla \Mw, \bs{u'} \otimes \bs{u'} )_{T} 
\nonumber 
\\ 
& \phantom{=} \; 
- \sum_{T \in \cT_h(\Om)}( \nabla \cdot \Mw, p' )_{T} 
- \sum_{T \in \cT_h(\Om)}(\nabla q^h, \bs{u'} )_{T} \; ,
\end{align}
where $\nabla^s \Mu =1/2( \nabla \Mu + (\nabla \Mu)^t )$ is the symmetric part of the gradient of $\Mu$ and $(\nabla \Mu)^t$ is the transpose of the gradient of $\Mu$.
The terms in the first row of \eqnref{eq:VMS_formulation_2} represent the Galerkin formulation of the Navier-Stokes equations, while the other terms are associated with the VMS stabilization, as elaborated in \cite{bazilevs2007variational}. From the VMS literature, the primed terms, also known as fine-scale variables, are defined as follows:
\begin{subequations}
\begin{align}
\bs{u'} & = -\tau_M \bs{r}_M(\Mu,p^h) \; , \\
p' & = -\tau_C r_C(\Mu) \; , 
\end{align}
where
\begin{align}
\bs{r}_M & = \;\partial_t \Mu + \Mu \cdot \nabla \Mu + \nabla p^h - \frac{1}{Re} \nabla^2 \Mu  - \bs{f}\; , \\
r_C& = \;\nabla \cdot \Mu\; , \\
\tau_M& = \;\Big(\frac{4}{\Delta t^2} + \Mu \cdot \bs{G} \Mu + \frac{C_M}{Re^2} \bs{G} : \bs{G}\Big)^{-\frac{1}{2}}\; , \\
\tau_C& = \;(\tau_M \bs{g} \cdot \bs{g})^{-1}\; , \\
G_{ij} & = \; \sum_{k=1}^{d}\frac{\partial \xi_k}{\partial x_i} \frac{\partial \xi_k}{\partial x_j}\; , \\
g_i & = \; \sum_{j=1}^{d}\frac{\partial \xi_j}{\partial x_i}\; .
\end{align}
\end{subequations}

In this context, $\bs{r}_M$ and $r_C$ are residuals of the momentum and mass conservation equations calculated using the coarse-scale (i.e., numerical) solutions $\Mu$ and $p^h$. The constant $C_M$ is set to 36 and the quantities $G_{ij}$ and $g_i$ involve isoparametric maps from the reference configuration of an element to its physical counterpart.
Time integration is performed using the implicit Euler time integrator.

\subsection{Backflow stabilization for Navier-Stokes}\label{sec:NS_backflow}
We discuss now the stable treatment of outflow boundary conditions.
Backflow stabilization~\cite{Moghadam2011, lanzendorfer2011pressure, feistauer2013existence, ismail2014stable, Braack2014, bertoglio2014tangential, dong2015convective, liu2020simple} has been proposed to obviate numerical instabilities at outflow or open boundaries in fluid dynamics simulations, especially at moderate to high Reynolds numbers. In such cases, flow recirculation or strong vortices can cause reverse flow at the outflow boundary, also known as backflow. This backflow destabilizes the simulation, leading to inaccurate results or computational divergence. Backflow stabilization techniques involve adding a boundary term at the outlet (or open boundary) that activates when backflow occurs and introduces a dissipative mechanism to stabilize the simulation. In our work, we implement the backflow stabilization techniques proposed in~\citep{Moghadam2011} and~\citep{Braack2014}. While both methods add a boundary term to handle outflow instabilities, they differ in the treatment of the tunable parameter $\beta_o$. In the backflow stabilization proposed in~\citep{Moghadam2011}, $\beta_o$ can be adjusted between 0 and 1, providing flexibility, whereas in the Directional Do-Nothing (DDN) method~\citep{Braack2014}, it is fixed at 0.5. Both methods involve adding to the right side of \eqnref{eq:VMS_formulation} the term:
\begin{align}
& - \avg{\Mw, \beta_o \min(0, \Mu \cdot \bs{n}) \Mu }_{\G_o} .
\end{align}

For the simulations presented in \secref{subsub:airfoil}, \secref{subsubsec:Re300}, \secref{subsubsec:Re3700}, and \secref{subsubsec:Cylinder}, we applied backflow stabilization at the outlet boundary, with $\beta_o$ set to 0.5.

\subsection{Preliminaries on the Shifted Boundary Method: The true domain, the surrogate domain and maps}
\label{sec:sbmDef}

\begin{figure}[b!]
	\centering
     \begin{subfigure}[b]{.5\linewidth}\centering
    \begin{tikzpicture}[scale=0.6]
    
    \draw[line width = 0.25mm,densely dashed,gray] (-2,-4) -- (8,-4); 
    \draw[line width = 0.25mm,densely dashed,gray] (-2,-2) -- (8,-2); 
    \draw[line width = 0.25mm,densely dashed,gray] (-2,0) -- (8,0); 
    \draw[line width = 0.25mm,densely dashed,gray] (-2,2) -- (8,2); 
    \draw[line width = 0.25mm,densely dashed,gray] (-2,4) -- (8,4); 
    \draw[line width = 0.25mm,densely dashed,gray] (-2,-4) -- (-2,4); 
    \draw[line width = 0.25mm,densely dashed,gray] (0,-4) -- (0,4); 
    \draw[line width = 0.25mm,densely dashed,gray] (2,-4) -- (2,4); 
    \draw[line width = 0.25mm,densely dashed,gray] (4,-4) -- (4,4); 
    \draw[line width = 0.25mm,densely dashed,gray] (6,-4) -- (6,4); 
    \draw[line width = 0.25mm,densely dashed,gray] (8,-4) -- (8,4); 
    
    \draw[line width = 0.5mm,red] (8,2) -- (2,2);  
    \draw[line width = 0.5mm,red] (2,0) -- (2,2);  
    \draw[line width = 0.5mm,red] (0,0) -- (2,0);  
    \draw[line width = 0.5mm,red] (0,0) -- (0,-2);  
    \draw[line width = 0.5mm,red] (0,-2) -- (0,-4);  
    
    \draw [line width = 0.5mm,blue] plot[smooth] coordinates {(-0.05,-4.5) (2.65,0.35) (9.25,1.55)};
    
    \node[text width=0.5cm] at (8.5,2.2) {\large${\color{red}\tG}$};
    \node[text width=3cm] at (2.05,1.25) {\large${\color{red}\tO}$};
    \node[text width=3cm] at (1,-2.5) {\large${\color{blue}\Om}$};
    \node[text width=1cm] at (7.7,-3.5) {\large${\color{black}{\cal D}}$};
    \node[text width=0.5cm] at (9.75,1.65) {\large${\color{blue}\G}$};
    \node[text width=3cm] at (4.325,1.35) {\large$\Om \setminus \tO $};
    \end{tikzpicture}
    \caption{The true domain $\Om$, the surrogate domain $\tO \subset \Om$, the true boundary $\G$, and the surrogate boundary $\tG$ (in the case $\lambda=0$).}
    \label{fig:SBM}
\end{subfigure}
\qquad
\begin{subfigure}[b]{.3\linewidth}\centering
	\begin{tikzpicture}[scale=0.6]
	\draw[line width = 0.25mm,densely dashed,gray] (0,0.5) rectangle (-4.5,5); 
	\draw[line width = 0.25mm,densely dashed,gray] (0,5) rectangle (4.5,0.5); 

	\draw [line width = 0.5mm,blue, name path=true] plot[smooth] coordinates {(0.4-0.5,-0.5) (2.16-0.5,2.55) (1.0-0.5,6)};
	
	\draw[line width = 0.5mm,red] (0,0.5) -- (0,5);
	
	\node[text width=0.5cm] at (-0.125,4.0) {\large${\color{red}\ti{\G}}$};
	\node[text width=0.5cm] at (1.8,4.5) {\large${\color{blue}\G}$};
	\node[text width=0.5cm] at (0.75,3.25) {\large$\bs{d}$};
	\node[text width=0.5cm,blue] at (2.5,3.5) {\large$\bs{n}$};
	\node[text width=0.5cm,blue] at (2.2,2.25) {\large$\bs{\tau}$};
	\node[text width=0.5cm,red] at (0.6,2.45) {\large$\ti{\bs{n}}$}; 
	
	\draw[->,line width = 0.25mm,-latex,red] (0,2.75) -- (1,2.75); 
	\draw[->,line width = 0.25mm,-latex] (0,2.75) -- (1.62,3.1);
	\draw[->,line width = 0.25mm,-latex,blue] (1.62,3.1) -- (1.78,2.29);
	\draw[->,line width = 0.25mm,-latex,blue] (1.62,3.1) -- (2.45,3.25);
	
	\end{tikzpicture}
    \caption{The distance vector $\bs{d}$, the true normal $\bs{n}$, the true tangent $\bs{\tau}$, and the surrogate normal $\ti{\bs{n}}$ (horizontal).}
    \label{fig:ntd}
\end{subfigure}
    \caption{The surrogate domain, its boundary, and the distance vector $\bs{d}$.}
    \label{fig:surrogates}
\end{figure}
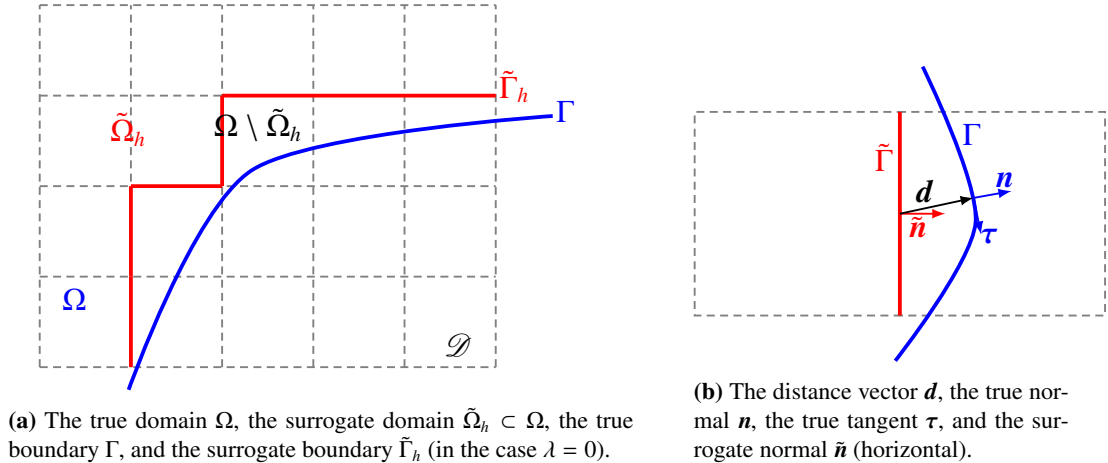

Consider a closed region ${\cal D}$ such that $\text{clos}(\Om) \subseteq {\cal D}$ (here $\text{clos}(\Om)$ indicates the {\it closure} of $\Om$) and the family $\cT_h({\cal D})$ of admissible and shape-regular discrete decompositions (meshes/grids) of ${\cal D}$. 
In this work, we specifically focus on octree grids aligned along the axis of the Cartesian space (see \figref{fig:SBM}). 
Then, we restrict each $\cT_h({\cal D})$  by selecting those elements $T \in \cT_h({\cal D})$ such that 
\begin{align}
\label{eq:lambda_def}
\mathrm{meas}(T \cap \Om) > (1-\lambda) \, \mathrm{meas}(T) \; , \qquad \mbox{for some } \lambda \in [0,1] \; .
\end{align}
In other words, these are elements that have an intersection with the domain of interest $\Om$ with an area/volume larger than $1-\lambda$ with respect to their total area/volume, respectively in two/three dimensions.
For example, choosing $\lambda=0$ selects the elements that are strictly contained in the computational domain $\Om$ (see, e.g.,~\figref{fig:SBM}), choosing $\lambda =1$ selects the elements that have a non-empty intersection with $\Om$, and choosing $\lambda=0.5$ selects elements whose intersection with $\Om$ includes at least $50\%$ of their area/volume (see an example of this last case in~\figref{fig:TrueAndSurrogate}).
We define the family of grids that satisfies~\eqnref{eq:lambda_def} as  
$$
\ti{\cT}_h(\lambda) := \{ T \in \cT_h({\cal D}) : \mathrm{meas}(T \cap \Om) > (1-\lambda) \, \mathrm{meas}(T) \}\,.
$$ 
This identifies the {\sl surrogate domain}
$$
\tO(\lambda) := \text{int} \, \Bigl( \bigcup_{T \in \ti{\cT}_h(\lambda)}  T \Bigr)  \,,
$$
or, more simply, $\tO$, with {\sl surrogate boundary} $\tG:=\partial \tO$ and outward-oriented unit normal vector $\ti{\bs{n}}$ to $\tG$. Obviously, $\ti{\cT}_h(\lambda)$ is an admissible and shape-regular family of decompositions of $\tO$ (see again~\figref{fig:SBM}). 
Here, we make the optimal choice $\lambda=0.5$, which minimizes the average distance between the surrogate and true boundaries, and consequently the numerical error in computations. The specific details of the optimal construction of the surrogate boundary will be discussed in Section~\ref{sec:implementation}, and the interested reader is also pointed to~\cite{yang2024optimal}, where a detailed analysis of the optimality of the choice $\lambda=0.5$ is carried out.

Consider now the mapping, sketched in \figref{fig:ntd},
\begin{subequations}\label{eq:defMmap}
\begin{align}
\bs{M}_{h}:&\; \tG \to \G \; ,  \\
&\; \ti{\bs{x}} \mapsto \bs{x}   \; ,
\end{align}
\end{subequations}
which associates to any point $\ti{\bs{x}} \in \tG$ on the surrogate boundary a point $\bs{x} = \bs{M}_{h}(\ti{\bs{x}})$ on the physical boundary $\G$. 
In this work, $\bs{M}_{h}$ is defined as the closest-point projection of $\ti{\bs{x}}$ on $\G$, as shown in~\figref{fig:ntd}.
Through $\bs{M}_{h}$, a distance vector function $\bs{d}_{\bs{M}_{h}}$ can be defined as
\begin{align}
\label{eq:Mmap}
\bs{d}_{\bs{M}_{h}} (\ti{\bs{x}})
\, = \, 
\bs{x}-\ti{\bs{x}}
\, = \, 
[ \, \bs{M}-\bs{I} \, ] (\ti{\bs{x}})
\; .
\end{align}
For the sake of simplicity, we set $\bs{d} = \bs{d}_{\bs{M}_{h}} $ where $\bs{d} = \|\bs{d}\| \bs{\nu}$  and $\bs{\nu}$ is a unit vector. 
\begin{remark}
	There are many strategies to define the map $\bs{M}_{h}$ and, correspondingly, the distance vector $\bs{d}$. Whenever uniquely defined, the closest-point point projection of $\ti{\bs{x}}$ upon $\G$ is a natural choice for $\bs{x}$. But other and more sophisticated choices may be preferable, depending on which geometric data structures are available. Among them is a level set description of the true boundary, in which $\bs{d}$ is defined by means of a distance function.
See also the discussion in~\cite{atallah2021analysis,atallah2021shifted} for the case of a domain $\Om$ with corners and for the case of a surrogate boundary where multiple types of boundary conditions are applied.
\end{remark}

\subsection{A shifted boundary formulation of the incompressible Navier-Stokes equations}\label{sec:SBM_NS}

We will develop next a shifted boundary formulation of the incompressible Navier-Stokes equations, in the specific case of Dirichlet boundary conditions.
Other types of boundary conditions can be treated with the SBM, but these are somewhat less interesting in the context of the incompressible Navier-Stokes equations.
This is because the shapes that are immersed into a fluid domain typically require the no-slip condition, which is a Dirichlet-type condition.

Consider then a surrogate boundary $\tGD$ in proximity of a true Dirichlet boundary $\G_D$.
Using the construction of the distance between the two boundaries, it is possible to introduce the Taylor expansion of the velocity vector:
\begin{align}
	\label{eq:SBM_Dirichlet1}
	\bs{u}(\tx)
	 + (\nabla \bs{u} \cdot \bs{d}) (\tx)  
	 + (\bs{R}_{D}(\bs{u}, \bs{d}))(\tx) 
	&=\;  
	\bs{u}_{D}(\bs{M}_h (\tx))
	\; , \quad  \mbox{ on } \tGD \; , 
\end{align}
where the remainder $\bs{R}_{D}(\bs{u}, \bs{d})$ satisfies $\| \bs{R}_{D}(\bs{u}, \bs{d}) \| = o(\| \bs{d}  \|^2)$ as $\| \bs{d} \, \| \to 0$.
We can then define, on $\tGD$, the {\it extension} operator 
\begin{align}
\label{eq:def-extS}
\mathbb{E}\bs{u}_{D}(\tx) 	 
&:= \; 
\bs{u}_{D}(\bs{M}_h (\tx)) 
\end{align}
and with {\it shift} operator
\begin{align}
\label{eq:def-bndS}
\bs{S}_{D,h} \, \bs{u} (\tx)	 
&:= \; 
\bs{u} (\tx)
+ \nabla \bs{u} (\tx) \, \bs{d} (\tx)
\; .
\end{align}
Neglecting the higher-order residual term in~\eqnref{eq:SBM_Dirichlet1}, we obtain the final expression of the {\it shifted} boundary conditions
\begin{align}
\label{eq:SBM_Dirichlet}
\bs{S}_{D,h} \, \bs{u} 	
&=\;  
\mathbb{E}\bs{u}_{D}
\; , \qquad \mbox{ on } \tGD \; .
\end{align}
In what follows - for the sake of simplicity and whenever it does not cause confusion - we will omit the symbol $\mathbb{E}$ from $\mathbb{E}\bs{u}_{D}(\tx)$ and simply write $\bs{u}_{D}(\tx)$.
Introducing now the scalar and vector discrete function spaces
\begin{align}
\ti{V}^h(\tO(\lambda)) &=\; \left\{ q^h \mid q^h \in C^0(\Om) \cap \mathcal{P}^1(T) \,, \mbox{ with } T \in \ti{\cT}_h(\lambda) \right\}  \; ,
\\
\ti{\bs{V}}^h(\tO(\lambda)) &=\; \left\{ \Mw \mid \Mw \in (C^0(\Om))^d \cap (\mathcal{P}^1(T))^d \,, \mbox{ with } T \in \ti{\cT}_h(\lambda) \right\}  \; ,
\end{align}
which does not incorporate any boundary conditions, we can present the shifted boundary variational formulation of the Navier-Stokes equations: \\[.2cm]

Find $\Mu \in \ti{\bs{V}}^h(\tO(\lambda))$ and $p^h \in V^h(\tO(\lambda))$ such that, for any $\Mw \in \ti{\bs{V}}^h(\tO(\lambda))$ and $q^h \in V^h(\tO(\lambda))$,
\begin{align}
\label{eq:SBMNS}
0 
\;  = & \; 
{NS}[\tO(\lambda) \,; \ti{\cT}_h(\lambda)](\Mw,q^h;\Mu,p^h) 
\nonumber \\
&
- \underbrace{\avg{\Mw,(2 \,  Re^{-1} \nabla^s \Mu - p^h \bs{I}) \ti{\bs{n}}}_{\tGD}}_{Consistency \; term} 
-\underbrace{\avg{( 2 \,  Re^{-1} \nabla^s \Mw + q^h \bs{I}) \ti{\bs{n}},\Mu + \nabla \Mu \, \bs{d} -\bs{u_D}}_{\tGD}}_{Adjoint\; consistency \; term}
\nonumber \\
&
+ \underbrace{ \beta \, Re^{-1} \avg{ h^{-1} (\Mw + \nabla \Mw \, \bs{d}) , \Mu + \nabla \Mu \, \bs{d} -\bs{u_D}}_{\tGD}}_{Penalty \; term}
\end{align}
where the term ${NS}[\tO(\lambda) \,; \ti{\cT}_h(\lambda)](\Mw,q^h;\Mu,p^h)$ is defined in analogy to \eqnref{eq:VMS_formulation_2}.
This formulation is inspired by Nitsche's method for the weak imposition of Dirichlet-type boundary conditions.
In particular, the term in the second row of~\eqnref{eq:SBMNS} is due to integration by parts and the fact that now $\Mu, \, \Mw \in \ti{\bs{V}}^h(\tO(\lambda))$.
The parameter $\beta$ in the last term of~\eqnref{eq:SBMNS} is a penalty that, if sufficiently large, ensures the numerical stability of the overall formulation.

The proposed SBM formulation satisfies a principle of global conservation of mass and momentum, as already described in~\cite{main2018shifted_2}. We briefly summarize the principle of conservation of mass (the momentum conservation can be proved in a similar way): taking $\Mw=0$ and $q^h=1$ in~\eqnref{eq:SBMNS}, we obtain, using the Gauss divergence theorem,
\begin{align}
\label{eq:SBMNS_masscons}
0 
\;  = & \; 
( 1 , \nabla \cdot \Mu )_{\tO}  
-\avg{ 1 , ( \Mu + \nabla \Mu \, \bs{d} -\bs{u}_D) \cdot \ti{\bs{n}} }_{\tGD}
\nonumber \\
\;  = & \; 
\avg{ 1 , \Mu \cdot \ti{\bs{n}} }_{\tG}  
-\avg{ 1 , ( \Mu + \nabla \Mu \, \bs{d} -\bs{u}_D) \cdot \ti{\bs{n}} }_{\tGD}
\nonumber \\
\;  = & \; 
 \avg{ 1 , \Mu \cdot \ti{\bs{n}} }_{\tG \setminus \tGD }  
+\avg{ 1 , ( \bs{u}_D - \nabla \Mu \, \bs{d} ) \cdot \ti{\bs{n}} }_{\tGD}
\; ,
\end{align}
which corresponds to a statement of conservation of mass fluxes entering and exiting the surrogate domain. 
Note that at surrogate Dirichlet boundaries the mass flux is given by $( \bs{u}_D - \nabla \Mu \, \bs{d} ) \cdot \ti{\bs{n}}$, which includes the Taylor expansion correction.
This is expected, since the surrogate boundary is at a distance from the true boundary.

\section{Implementation}
\label{sec:implementation}

\subsection{SBM data structures, algorithms, and optimal surrogate boundary}
\label{sec:opt}

In this section we elaborate on some important terminology, referencing concepts detailed in~\citet{yang2024optimal}. \figref{fig:ElementMarkers} visually delineates the varied domains (namely, the true and surrogate domains) and their linkage with distinct mesh elements.
Specifically, this figure deptics:
\begin{itemize}[left=0pt,itemsep=0pt]
\item \Interior{} elements (\textcolor{ActiveElement}{$\blacksquare$}): These mesh elements are entirely encompassed within the physical boundary over which we simulate the Navier-Stokes equations ($\Omega$).
\item \Intercepted{} elements (\textcolor{InterceptedElement}{$\blacksquare$} $\cup$ \textcolor{FalseInterceptedElement}{$\blacksquare$}): These elements intersect the (true) physical boundary $\G$, with some of their nodes located inside and others outside. Further classification of \Intercepted{} elements into two distinct groups is based on a precise strategy, based on the choice of $\lambda$, as defined in \eqnref{eq:lambda_def}:
\begin{itemize}
\setlength\itemsep{0em}
\item \TrueIntercepted{} elements (\textcolor{InterceptedElement}{$\blacksquare$}): \Intercepted{} elements within the surrogate domain $\tO(\lambda)$, incorporated into the SBM analysis.
\item \FalseIntercepted{} elements (\textcolor{FalseInterceptedElement}{$\blacksquare$}): \Intercepted{} elements outside the surrogate domain $\tO(\lambda)$, omitted from the SBM analysis. As already indicated in Section~\ref{sec:sbmDef}, an element is classified as \FalseIntercepted{} if the proportion of its volume that lies outside $\Om$ relative to its total volume exceeds a certain threshold $\lambda$. Setting $\lambda = 0$ categorizes all \Intercepted{} elements as \FalseIntercepted{}, resulting in a surrogate domain that is entirely encompassed within the true domain. Conversely, a $\lambda = 1$ setting includes every \Intercepted{} element, thus defining a surrogate domain that completely envelops the true domain. A threshold of $\lambda=0.5$ is considered ideal for establishing an optimal surrogate boundary. Additionally, $\lambda = 0.5$ has been proved to guarantee the minimal RMS distance between the surrogate boundary ($\tG$) and true boundary ($\G$)~\citep{yang2024optimal}. In this paper, unless otherwise specified, the simulations are conducted with a $\lambda$ value of 0.5.

\begin{figure}[t!]
    \centering
    \includegraphics[width=0.3\linewidth,trim=0 0 0 0,clip]{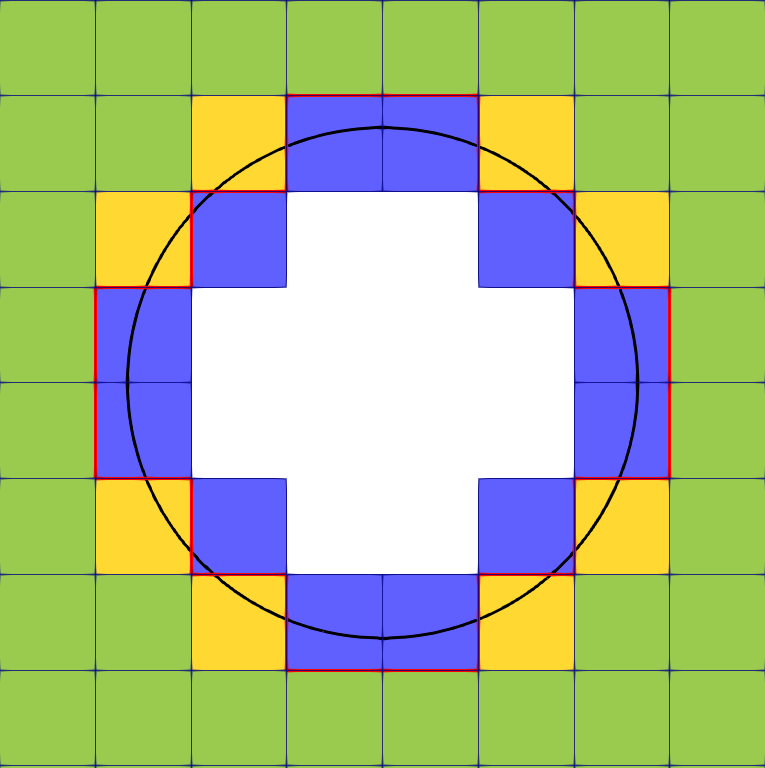}
    \caption{Color illustration for three distinct element types within the SBM algorithms: \Interior{} elements represented by (\textcolor{ActiveElement}{$\blacksquare$}), \TrueIntercepted{} elements denoted with (\textcolor{InterceptedElement}{$\blacksquare$}), and \FalseIntercepted{} elements indicated by (\textcolor{FalseInterceptedElement}{$\blacksquare$}).}
    \label{fig:ElementMarkers}
\end{figure}

\end{itemize}
\end{itemize}
Diverging from~\citet{yang2024optimal}, we exclude \Exterior{} elements, as our focus narrows to exclude those elements represented inside \figref{fig:ElementMarkers}, effectively removing the elements within the circular region through a process we term as "carving out."


Moreover, the illustration in~\figref{fig:ElementMarkers} emphasizes two primary domains of interest:
\begin{itemize}
\setlength\itemsep{0em}
\item The physical (or true) domain $\Omega$, enclosed by the physical (or true) boundary $\G$ (\textcolor{black}{---}).
\item The surrogate domain $\tO(\lambda) = $ \textcolor{ActiveElement}{$\blacksquare$} $\cup$ \textcolor{InterceptedElement}{$\blacksquare$}, which is enclosed by the surrogate boundary $\tG$ (\textcolor{red}{---}), comprising both \Interior{} elements (\textcolor{ActiveElement}{$\blacksquare$}) and \TrueIntercepted{} elements (\textcolor{InterceptedElement}{$\blacksquare$}). This is the primary domain for SBM computations.
\end{itemize}
It is noteworthy that our selection of plots differs from those in~\citet{yang2024optimal} because, in the context of the Navier-Stokes equation, interest typically lies in the region outside the enclosed geometry. This preference stems from the common objective of studying the flow around an obstacle and measuring various quantities. Consequently, the area outside the circle in~\figref{fig:ElementMarkers} represents elements that are \Interior{} to the computational domain $\tO(\lambda)$, as the solution is defined over this domain.
%

\subsection{Evaluating quantities on true boundary}\label{sec:Eval_quantity}

\begin{figure}[t!]
  \centering
  \begin{subfigure}{0.68\linewidth}
    \centering
    \includegraphics[width=0.99\linewidth]{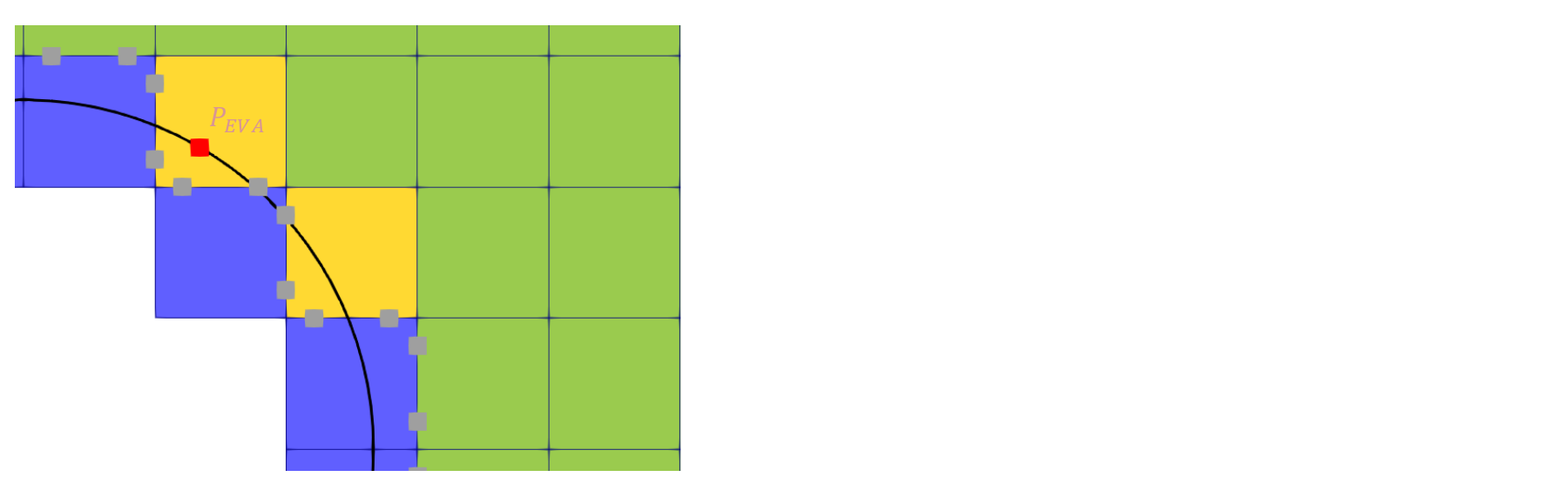}
    \caption{Case 1: The evaluated point (\textcolor{TruePt}{\scalebox{0.5}{$\blacksquare$}}) is in the  \TrueIntercepted{} element(\textcolor{InterceptedElement}{$\blacksquare$}).}
    \label{fig:EvaluateTrueBoundaryPt_TrueIntercept}
  \end{subfigure}
  \begin{subfigure}{0.68\linewidth}
    \centering
    \includegraphics[width=0.99\linewidth]{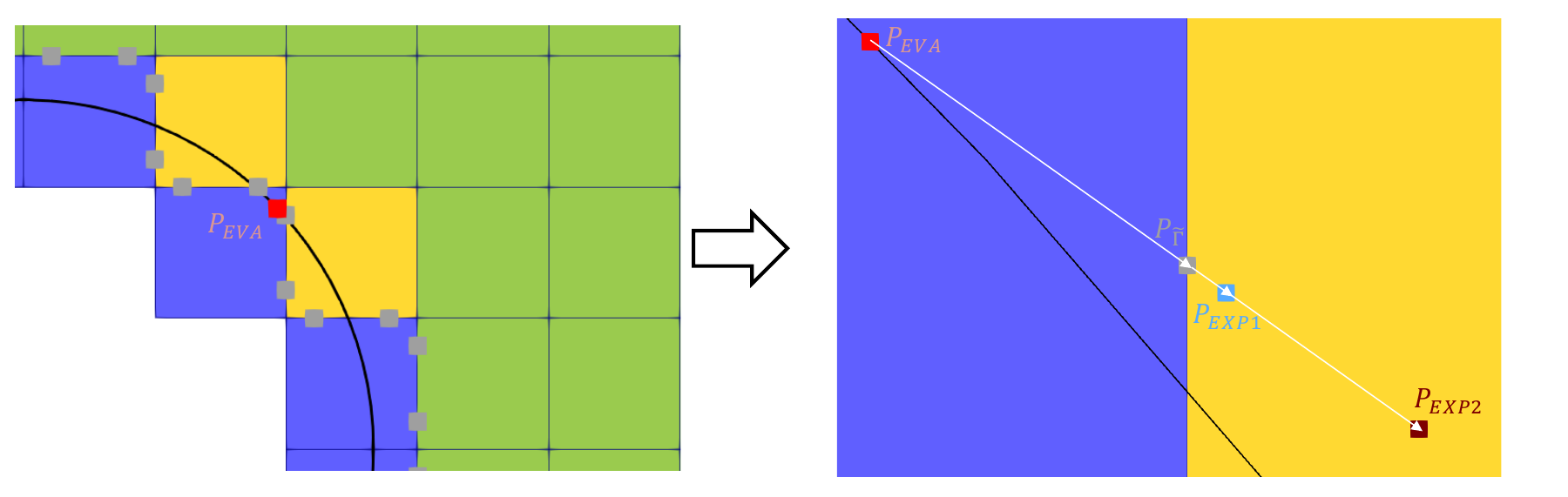}
    \caption{Case 2: The evaluated point (\textcolor{TruePt}{\scalebox{0.5}{$\blacksquare$}}) is in the  \FalseIntercepted{} element(\textcolor{FalseInterceptedElement}{$\blacksquare$}). The extrapolation point 1 (\textcolor{Exp1}{\scalebox{0.5}{$\blacksquare$}}) and extrapolation point 2 (\textcolor{Exp2}{\scalebox{0.5}{$\blacksquare$}}) fall on the \TrueIntercepted{}{} element (\textcolor{InterceptedElement}{$\blacksquare$}).}
    \label{fig:EvaluateTrueBoundaryPt_FalseIntercept_TrueIntercepted}
  \end{subfigure}
  \begin{subfigure}{0.68\linewidth}
    \centering
    \includegraphics[width=0.99\linewidth]{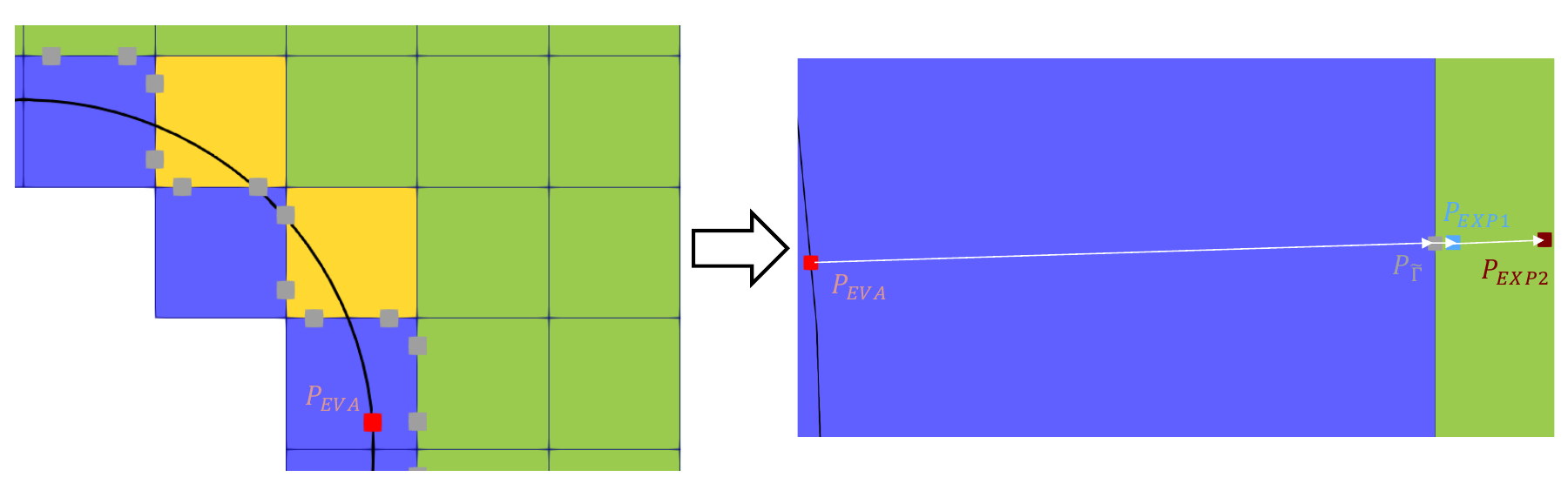}
    \caption{Case 3: The evaluated point (\textcolor{TruePt}{\scalebox{0.5}{$\blacksquare$}}) is in the  \FalseIntercepted{} element(\textcolor{FalseInterceptedElement}{$\blacksquare$}). The extrapolation point 1 (\textcolor{Exp1}{\scalebox{0.5}{$\blacksquare$}}) and extrapolation point 2 (\textcolor{Exp2}{\scalebox{0.5}{$\blacksquare$}}) fall on the \Interior{} element (\textcolor{ActiveElement}{$\blacksquare$}).}  \label{fig:EvaluateTrueBoundaryPt_FalseIntercept_Interior}
    \end{subfigure}
    \caption{The approaches to determining the value at a point of interest (\textcolor{TruePt}{\scalebox{0.5}{$\blacksquare$}}) within the SBM framework.}
    \label{fig:EvaluateTrueBoundaryPt}
\end{figure}

In the SBM calculations, \FalseIntercepted{} elements outside the surrogate domain $\tO(\lambda)$ are excluded, meaning no volume Gauss points are infilled within these elements. However, when performing post-processing tasks that involve computing physical quantities like the drag coefficient ($C_d$), pressure around the geometry, and boundary error, which are relevant at the true boundary ($\G$), we encounter situations where the points we are interested at the true boundary ($\G$) may land within \FalseIntercepted{} elements if we choose $\lambda$ to be less than 1. Consequently, this subsection explains the methodology for accurately computing physical values at the true boundary when $\lambda$ is set to a value below 1. To address these situations, the methodology depicted in \figref{fig:EvaluateTrueBoundaryPt} outlines three distinct cases for clarifying the evaluation process:

\vspace{0.5em}
\noindent\textbf{Case~1} (\figref{fig:EvaluateTrueBoundaryPt_TrueIntercept}): the evaluation point ($P_{EVA}$) (\textcolor{TruePt}{\scalebox{0.5}{$\blacksquare$}}) falls on the \TrueIntercepted{} element (\textcolor{InterceptedElement}{$\blacksquare$}). We interpolate the solution at the evaluation point using the FEM shape functions and the nodal values of the solution.

\vspace{0.5em}
\noindent\textbf{Case~2} (\figref{fig:EvaluateTrueBoundaryPt_FalseIntercept_TrueIntercepted}) and \textbf{Case~3} (\figref{fig:EvaluateTrueBoundaryPt_FalseIntercept_Interior}): the evaluation point ($P_{EVA}$) (\textcolor{TruePt}{\scalebox{0.5}{$\blacksquare$}}) falls on a \FalseIntercepted{} element (\textcolor{FalseInterceptedElement}{$\blacksquare$}). In such instances, we identify the Gauss point ($P_{\tG}$) (\textcolor{SurrogateGP}{\scalebox{0.5}{$\blacksquare$}}) on the nearby surrogate boundary that is closest to $P_{EVA}$. Upon locating the nearest surrogate boundary Gauss point $P_{\tG}$, we slightly extend the vector from the point of interest towards this Gauss point inward to identify extrapolation point 1 ($P_{EXP1}$) (\textcolor{Exp1}{\scalebox{0.5}{$\blacksquare$}}), ensuring it falls within a \TrueIntercepted{} (\textcolor{InterceptedElement}{$\blacksquare$}) or \Interior{} (\textcolor{ActiveElement}{$\blacksquare$}) element. Furthermore, we extend the vector from the point of interest towards the surrogate Gauss point further inward to locate extrapolation point 2 ($P_{EXP2}$) (\textcolor{Exp2}{\scalebox{0.5}{$\blacksquare$}}). The distinction between Case~2 and Case~3 lies in the type of element found for extrapolation: a \TrueIntercepted{} element (\textcolor{InterceptedElement}{$\blacksquare$}) in Case~2 and an \Interior{} element (\textcolor{ActiveElement}{$\blacksquare$}) in Case~3.~After identifying extrapolation points 1 and 2, denoted as $P_{EXP1}$ and $P_{EXP2}$ respectively, we define two characteristic lengths: the distance from the point under evaluation to $P_{EXP1}$, labeled as $d_{EXP1}$, and the distance to $P_{EXP2}$, labeled as $d_{EXP2}$, where $d_{EXP1} = |P_{EVA}P_{EXP1}|$ and $d_{EXP2} = |P_{EVA}P_{EXP2}|$. Utilizing the FEM shape functions, we interpolate to obtain the values at $P_{EXP1}$ and $P_{EXP2}$, which are $u_{EXP1}$ and $u_{EXP2}$, respectively. With these interpolated values, we employ a straightforward extrapolation formula to calculate the evaluated value at the point of interest as follows:
\begin{equation}
u_{EVA} = \frac{u_{EXP1} \cdot d_{EXP2} - u_{EXP2} \cdot d_{EXP1}}{d_{EXP2} - d_{EXP1}}.
\end{equation}
This equation effectively utilizes the distances and interpolated values from the extrapolation points to determine the value at the evaluation point.

We do not use a normal-based method to determine the extrapolation points, i.e., using the normal of the true boundary to locate the extrapolation points in the normal direction. The reason for not adopting this approach is that it requires varying distances to identify the extrapolation points within the \TrueIntercepted{} and \Interior{} regions. Selecting an appropriate distance for this purpose is challenging. There is a risk of identifying an element that is either too distant from the true boundary or selecting a distance so short that the extrapolation points remain within the \FalseIntercepted{} element. Consider the case where $\lambda = 0$, signifying that all \Intercepted{} elements are categorized as \FalseIntercepted{} elements. Likely, the normal distance will not suffice to locate suitable extrapolation points accurately. While the geometry used here for demonstration purposes is merely a circle, in practical scenarios, the geometry could be far more complex and irregularly shaped, making the implementation of normal-based extrapolation point finding challenging. The alternative approach we adopt is simpler to implement and can be applied robustly across various cases and values of the parameter $\lambda$.

\subsection{Numerical implementation}

The numerical framework used in this study is built upon \dendroKT~\cite{ishii2019solving,saurabh2021scalable}, a highly scalable, parallel octree-based library, and \petsc, a widely used toolkit for solving large-scale scientific problems.

\dendroKT\ is a massively parallel, adaptive mesh refinement (AMR) codebase that uses octree-based domain decomposition. We have previously used it to model a variety of multiphysics simulations, including two-phase flow~\cite{khanwale2023projection}, electrokinetic transport phenomena~\cite{kim2024direct}, and transmission risk assessments~\cite{tan2023computational}. Our framework employs in-out tests~\cite{haines1994point,borazjani2008curvilinear,saurabh2021industrial} to quickly determine whether a point is inside or outside a given geometry. This approach enables the creation of an incomplete octree~\cite{saurabh2021scalable}, where regions that are not part of the solution domain are carved out. By avoiding unnecessary computations in these void areas, the incomplete octree significantly reduces computational overhead. Additionally, \dendroKT\ supports local mesh refinement, allowing us to focus computational resources on regions of interest, such as wake regions and boundary layers. By using space-filling curves (SFC) to partition octants across processors, \dendroKT\ ensures efficient load balancing in distributed memory environments. The library's top-down and bottom-up traversal algorithms further streamline operations like matrix assembly, eliminating the need for traditional element mappings. Furthermore, \dendroKT\ constructs 2:1-balanced~\cite{sundar2008bottom,fernando2017machine,ishii2019solving} octrees, ensuring that neighboring octants differ in refinement level by at most one, which improves the stability and accuracy of the numerical solution. For more details on the \dendroKT\ framework, readers are referred to~\cite{saurabh2021industrial,saurabh2021scalable}.

\petsc~(Portable, Extensible Toolkit for Scientific Computation) is the primary (non)linear solver (SNES and KSP constructs) used in this framework. It efficiently handles the parallel solution of the discretized systems resulting from the Navier-Stokes equations. \petsc's flexible interface and optimized data structures allow for efficient management of the sparse matrices generated by \dendroKT. Additionally, \petsc's support for distributed memory systems ensures scalability, making it ideal for large-scale simulations. The integration with \dendroKT\ enables seamless communication of nodal data between processes, facilitating the efficient assembly and solution of the linear systems arising from finite element discretization.

Beyond \dendroKT's role in finite element discretization and \petsc's role in parallel system solving, the implementation of the Shifted Boundary Method (SBM) requires two key components: (a) optimal surrogate boundary generation, and (b) fast distance function calculation based on k-d tree using the \texttt{nanoflann} library~\cite{blanco2014nanoflann}. For more details on the implementation of these components, readers are referred to~\citet{yang2024optimal}.

\section{Results}
\label{sec:results}
To evaluate the performance of the Octree-SBM approach, we simulate a series of benchmark simulations, as outlined in \figref{fig:NS}. The corresponding command-line arguments and solver tolerances for \petsc~are described in \ref{appendix:solver}. We compare simulation results with literature results, and when such results are unavailable, we perform simulations using body-fitted meshes to produce comparative baselines.

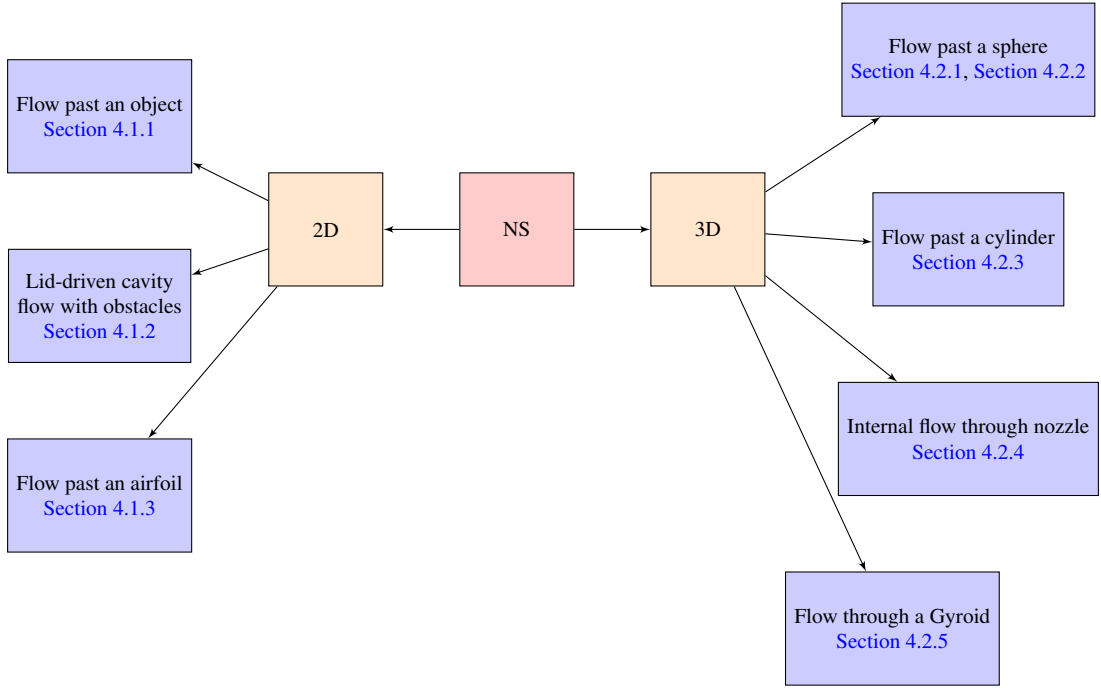
\begin{figure}
    \centering
    \tikzstyle{box} = [rectangle, minimum width=1.5cm, minimum height=1.5cm, text centered, draw=black, fill=blue!20, font=\footnotesize]
    \tikzstyle{arrow} = [draw, -latex'] 

    \begin{tikzpicture}[node distance=1cm, auto]

        \node [box, fill=red!20] (NS) {NS};
        \node [box, left=of NS, fill=orange!20] (2D) {2D};
        \node [box, right=of NS, fill=orange!20] (3D) {3D};

        \node [box, left=of 2D, yshift=1.5cm, align=center] (Left1) {Flow past an object\\\secref{subsec:2D_cylinder}};
        \node [box, below=of Left1, align=center] (Left2) {Lid-driven cavity \\ flow with obstacles \\ \secref{subsub:ldc}};
        \node [box, below=of Left2, align=center] (Left3) {Flow past an airfoil\\\secref{subsub:airfoil}};

        \node [box, right=of 3D, yshift=2.25cm, align=center] (Right1) {Flow past a sphere\\\secref{subsubsec:Re300},~\secref{subsubsec:Re3700}};
        \node [box, below=of Right1, align=center] (Right2) {Flow past a cylinder\\\secref{subsubsec:Cylinder}};
        \node [box, below=of Right2, align=center] (Right3) {Internal flow through nozzle\\\secref{subsub:nozzle}};
        \node [box, below=of Right3, xshift=-1cm, align=center] (Right4) {Flow through a Gyroid\\\secref{subsubsec:Gyroid}};

        \path [arrow] (NS) -- (2D);
        \path [arrow] (NS) -- (3D);

        \path [arrow] (2D) -- (Left1);
        \path [arrow] (2D) -- (Left2);
        \path [arrow] (2D) -- (Left3);

        \path [arrow] (3D) -- (Right1);
        \path [arrow] (3D) -- (Right2);
        \path [arrow] (3D) -- (Right3);
        \path [arrow] (3D) -- (Right4);
    \end{tikzpicture}
    \caption{Tree diagram for the results section of Octree-SBM for Navier-Stokes.}
    \label{fig:NS}
\end{figure}

\subsection{Two-dimensional Navier-Stokes simulations}

\subsubsection{Unsteady flow past a cylinder}\label{subsec:2D_cylinder}
\begin{figure}[t!]
    \centering
    \includegraphics[width=0.6\linewidth,trim=220 0 0 0,clip]{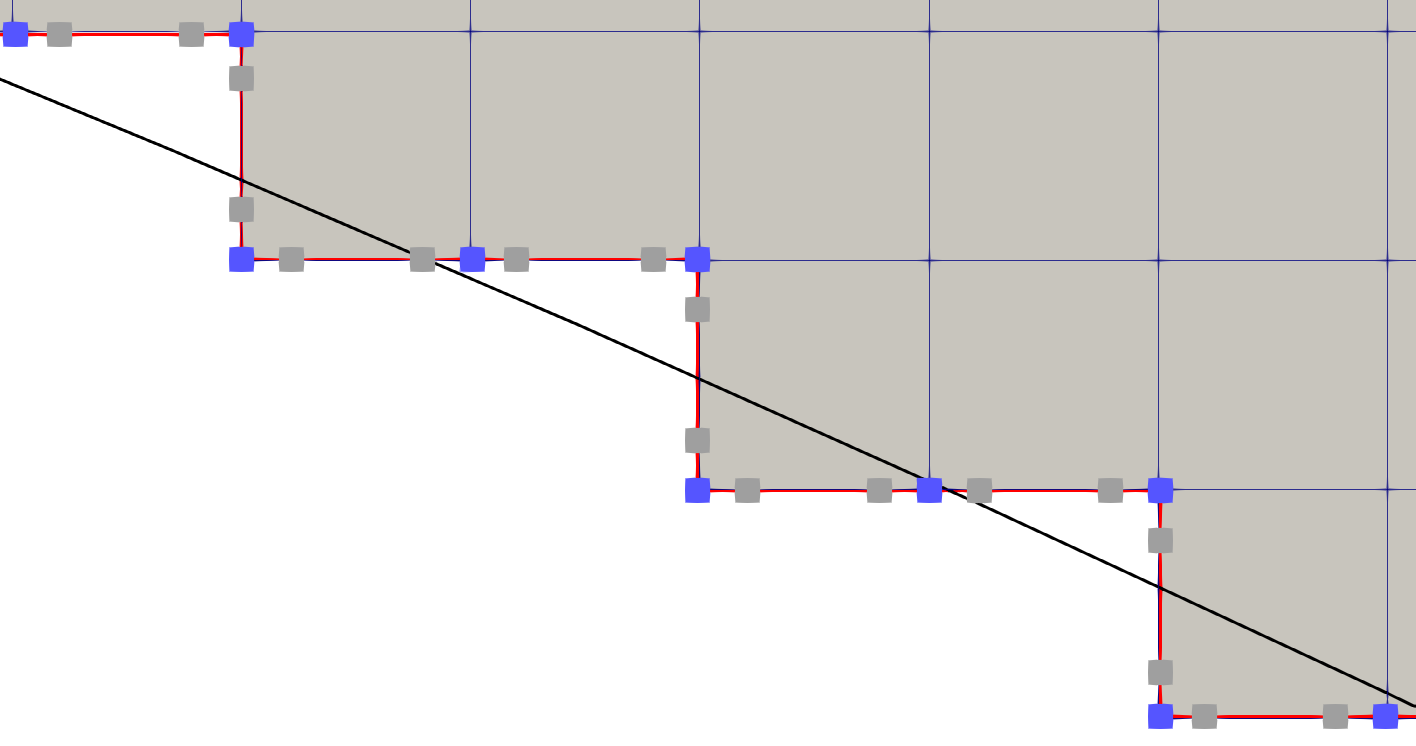}
    \caption{Illustrations of two test cases in \secref{subsec:2D_cylinder}: (a) Strong BC: a no-slip boundary condition is applied strongly at the nodal points (\textcolor{NodalStrongBC}{$\blacksquare$}) along the pixelated surrogate boundary (\textcolor{red}{---}); (b) SBM BC: a shifted boundary condition is applied weakly at the Gauss points (\textcolor{SurrogateGP}{$\blacksquare$}) on the pixelated surrogate boundary (\textcolor{red}{---}).}
    \label{fig:StrongBC}
\end{figure}

\begin{figure}[t!]
    \centering
    \begin{subfigure}{0.5\textwidth}
    \centering
    \includegraphics[width=0.99\linewidth]{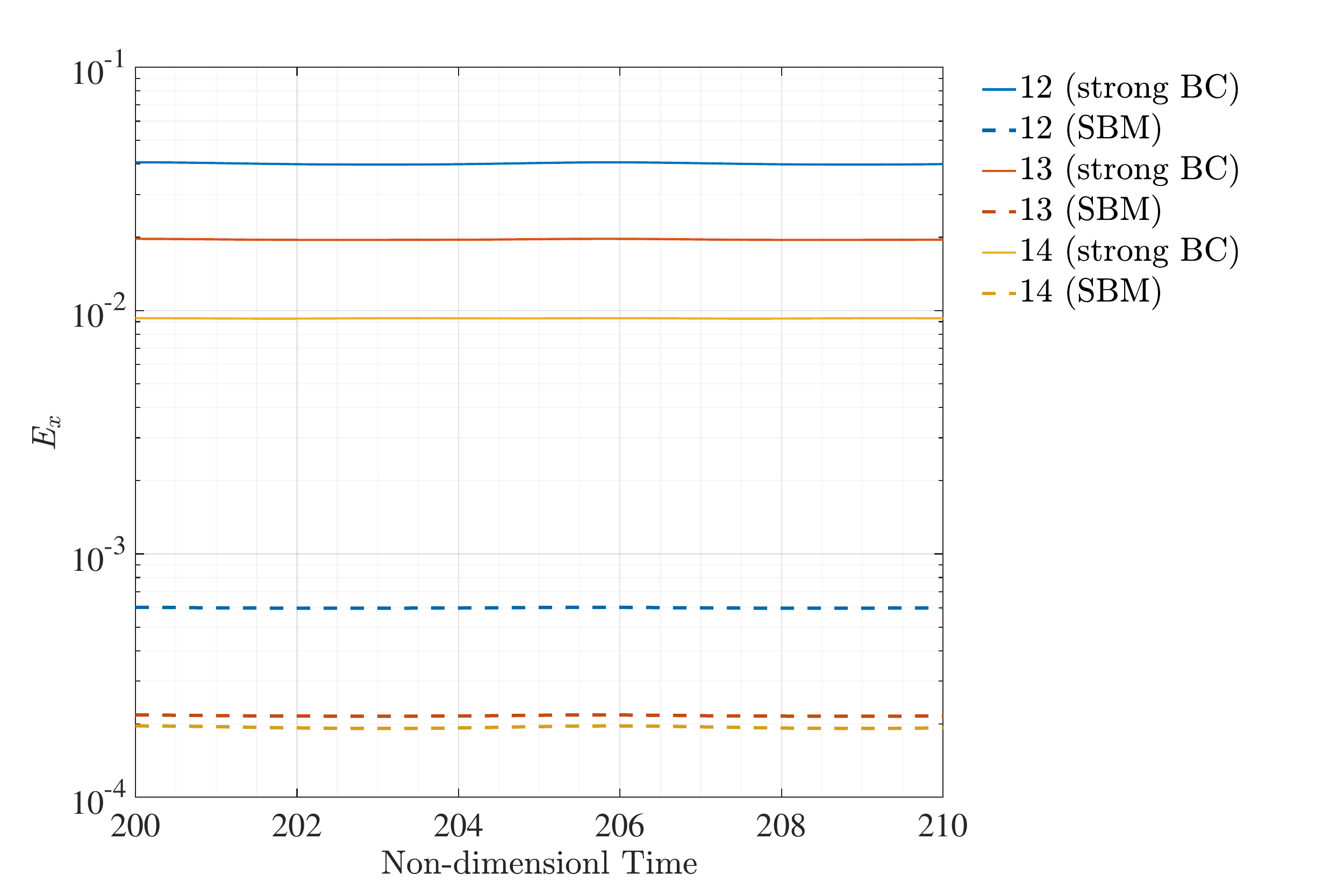}
    \caption{Boundary error of the $x$-component of the velocity.}
    \label{fig:2DCircle_BoundaryErrorX}
    \end{subfigure}%
    \begin{subfigure}{0.5\textwidth}
    \centering
    \includegraphics[width=0.99\linewidth]{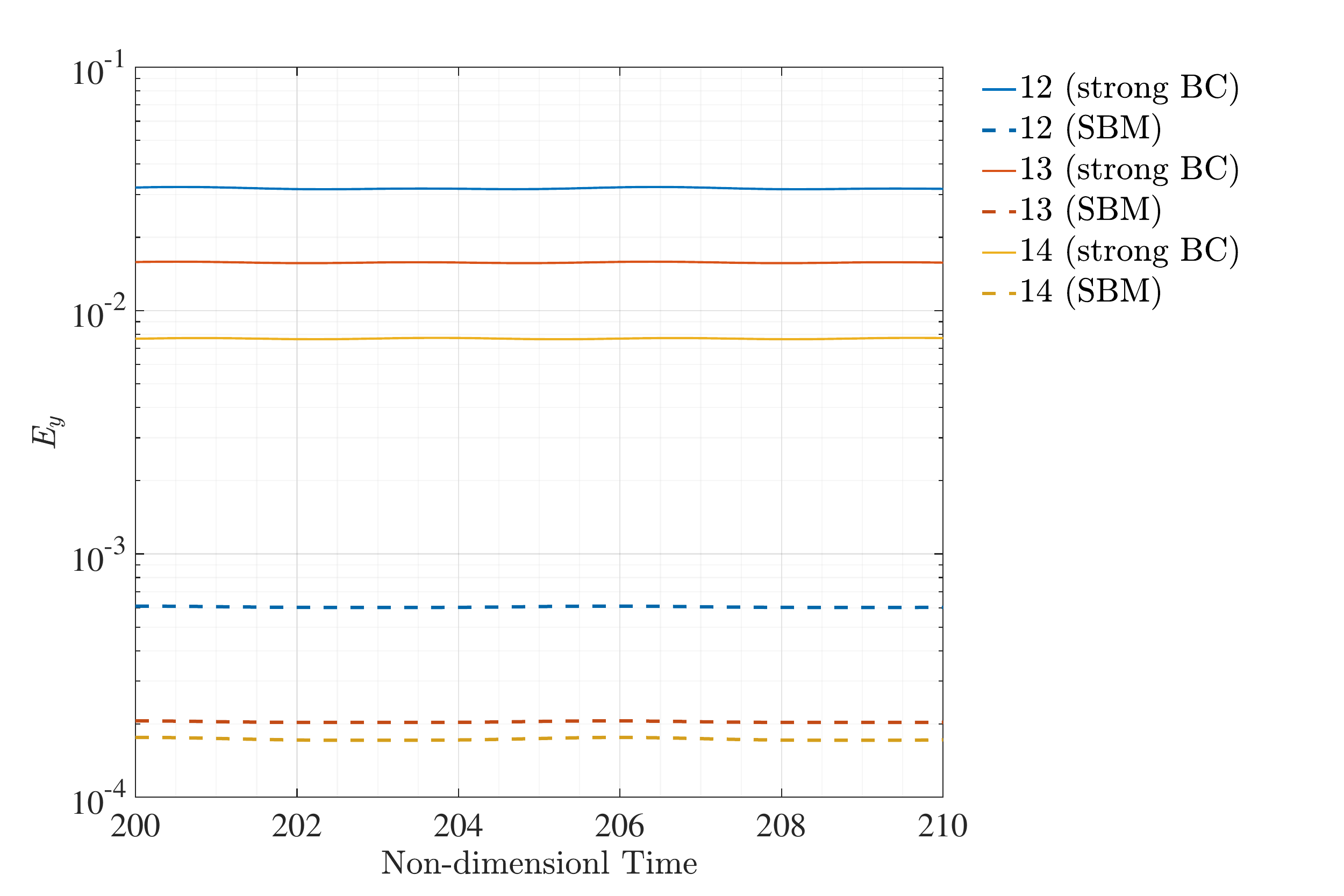}
    \caption{Boundary error of the $y$-component of the velocity.}
    \label{fig:2DCircle_BoundaryErrorY}
    \end{subfigure}%
    \caption{Velocity boundary error comparisons using strong BC and SBM for flow past a cylinder. Note the significant improvement in ensuring no-slip enforcement using the SBM formulation, even at coarse mesh resolution.}
    \label{fig:2DCircle_BoundaryError}
\end{figure}

We evaluate the standard fluid flow problem of flow past a circular geometry at $Re = 100$. The domain is defined as the rectangular region [0, 30] $\times$ [0, 20], within which a circular disk is positioned at the coordinates (10, 10). In the rectangle domain, all walls, excluding the outlet, maintain a consistent non-dimensional freestream velocity of (1, 0, 0). The outlet pressure is set to 0  on the outlet wall. The no-slip boundary condition of the circular obstacle is applied using SBM. The non-dimensional timestep is set to 0.01. We conduct a mesh convergence study as detailed in \tabref{tab:Flow2DCircle}. Furthermore, we engage in a comparative analysis of boundary errors on the true (circular) boundary between the SBM and a strong boundary condition, as illustrated in \figref{fig:2DCircle_BoundaryError}. The term ``strong boundary condition" refers to the direct imposition of the no-slip condition at the nodal points (\textcolor{NodalStrongBC}{$\blacksquare$}) along the surrogate boundary ($\tG$) (\textcolor{red}{---}), as depicted in \figref{fig:StrongBC}. This mirrors the approach adopted in~\citep{tan2023computational}, where a strong Dirichlet boundary condition is applied to  nodal points of octree mesh boundaries. Hence, the geometry appears as ``pixelated" to the fluid: This type of condition may not be preferred in aerospace applications, due to the lack of accuracy in the boundary layer region, but is often utilized in mechanical engineering and in the direct numerical modeling of flow meandering through complex porous media, due to its simplicity.
In contrast, for the SBM, the boundary condition is applied weakly through integration over the Gauss points (\textcolor{SurrogateGP}{$\blacksquare$}), as also  illustrated in \figref{fig:StrongBC}.

Given the no-slip boundary condition, the velocity components in the $x$- and $y$-direction at the boundary should ideally approach zero. The reported boundary error, which quantifies deviations from this ideal behavior, is computed as follows:

In the $x$-direction:
\begin{equation}
    \label{eq:Ex}
    E_x = \sqrt{\int_{\G} (u_x^h)^2 \, d\G}
    = \| u_x \|_{0;\G}
    \, ,
\end{equation}
where $E_x$ represents the boundary error in the $x$-direction, $\G$ denotes the true boundary, and $u_x^h$ is the numerical velocity component in the $x$-direction.
Similarly, in the $y$-direction:
\begin{equation}
    \label{eq:Ey}
    E_y = \sqrt{\int_{\G} (u_y^h)^2 \, d\G}
    = \| u_y \|_{0;\G}
    \, ,
\end{equation}
where $E_y$ denotes the boundary error in the $y$-direction, and $u_y^h$ represents the numerical velocity component in the $y$-direction.
Notably, even when utilizing the coarsest mesh (refine level equal to 12 and mesh size = $\frac{30}{2^{12}}$) with the SBM, the boundary error remains lower than that observed with the most refined mesh (refine level equal to 14 and mesh size = $\frac{30}{2^{14}}$) under a strong Dirichlet boundary condition applied to the surrogate boundary ($\tG$).
The results of~\figref{fig:2DCircle_BoundaryError} are very important, since it is often argued that a pixelated geometry representations are sufficient in some applications (e.g., simulation of Navier-Stokes flow in the porous space of porous media for numerical homogenization purposes).
The SBM offers significantly improved performance over a pixelated geometry, as seen by the fact that SBM errors are about 2 orders of magnitude smaller than the ones with the strong boundary condition on the pixelated geometry. Ultra-fine grids are required for the latter, to match the SBM results on relatively coarse grids. This situation is drastically exacerbated in three dimensions.

\begin{figure}[t!]
    \centering
    \includegraphics[width=0.99\linewidth,trim=0 0 0 0,clip]{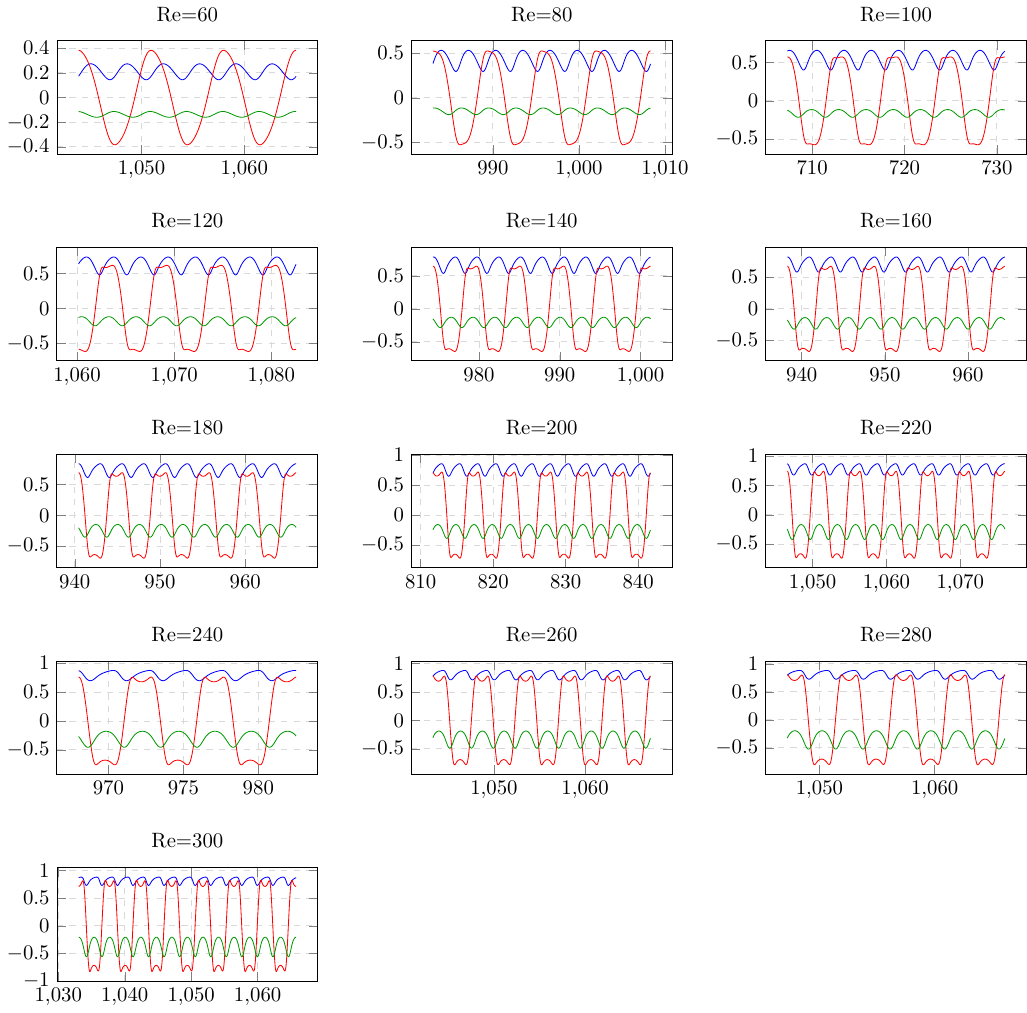}
    \caption{Simulations of flow past a two-dimensional cylinder at different Reynolds numbers. Measurements of flow quantities are taken at a probe positioned downstream from a fixed 2D cylindrical obstacle across various Reynolds numbers. The measurements include: (a) $x$-component of the velocity (\textcolor{blue}{---}), (b) $y$-component of the velocity (\textcolor{red}{---}), and (c) pressure (\textcolor{green!60!black}{---}). The horizontal axes represents a non-dimensional time for all plots.}
    \label{fig:dynamics_reynolds}
\end{figure}

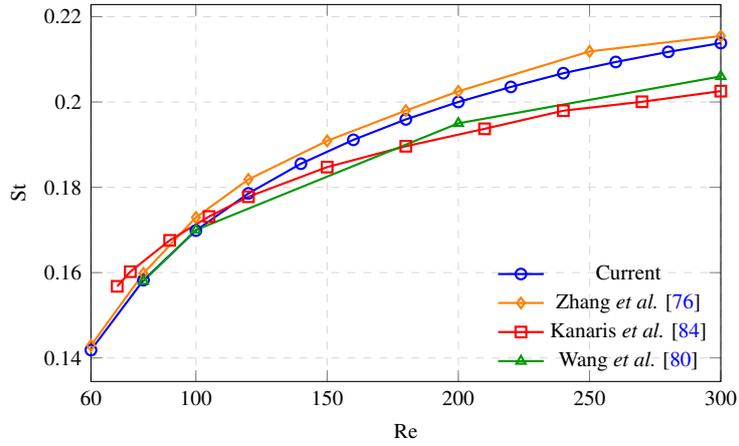
\begin{figure}[t!]
\centering
\begin{tikzpicture}
\begin{axis}[
    width=0.6\linewidth,
    height=0.4\linewidth,
    grid=major,
    grid style={dashed,gray!30},
    xlabel={Re},
    ylabel={St},
    xmin=60, xmax=300,
    xtick={60,100,150,200,250,300},
    xlabel style={font=\fontsize{8}{8}\selectfont, /pgf/number format/assume math mode},
    ylabel style={font=\fontsize{8}{8}\selectfont, /pgf/number format/assume math mode},
    legend style={font=\fontsize{8}{8}\selectfont, at={(1,0)}, anchor=south east, draw=none, fill=none}, 
    tick label style={font=\fontsize{8}{8}\selectfont},
    cycle list name=color list,
    every axis plot/.append style={thick}
]
\addplot+[mark=o, color=blue] table[col sep=comma, x=Re, y=St] {SBM.csv};
\addplot+[mark=diamond, color=orange] table[col sep=comma, x=Re, y=St] {StRe_Zhang_etal.csv};
\addplot+[mark=square, color=red] table[col sep=comma, x=Re, y=St] {StRe_kanaris.csv};
\addplot+[mark=triangle, color=green!60!black] table[col sep=comma, x=Re, y=St] {StRe_Wang_etal.csv};
\addlegendentry{Current}
\addlegendentry{Zhang \textit{et al.} \citep{zhang1995transition}}
\addlegendentry{Kanaris \textit{et al.} \citep{kanaris2011three}}
\addlegendentry{Wang \textit{et al.} \citep{wang2009immersed}}
\end{axis}
\end{tikzpicture}
    \caption{The variation of Strouhal number (the non-dimensional vortex shedding frequency) with respect to Reynolds number for the flow past a two-dimensional cylinder.}
    \label{fig:StRe}
\end{figure}

Maintaining the mesh's highest refinement level at 14, which corresponds to the smallest element size of $\frac{30}{2^{14}}$, we vary the Reynolds number from 60 to 300 in increments of 20. Velocity and pressure measurements are conducted at a point located $3\times D$ downstream from the circular obstacle, where D represents the diameter of the obstacle. Specifically, these measurements are taken at the coordinates (13.5, 10). After the flow stabilizes to a statistically steady state, the recorded data are illustrated in \figref{fig:dynamics_reynolds}. The entire collection of velocity values in the $y$-direction over different times creates a signal in the space-time domain. The signal is analyzed using a discrete Fourier transform (DFT), efficiently calculated using a fast Fourier transform (FFT) algorithm, to identify the vortex shedding frequency. Subsequently, the Strouhal number, representing the dimensionless vortex shedding frequency, is computed and plotted against the Reynolds number in \figref{fig:StRe}, and shows good match with results from literature.

\begin{table}[ht]
    \centering
    \caption{Mesh convergence study and comparisons with literature of flow past cylinder}
    \begin{tabular}{@{}P{6cm}P{2cm}P{2cm}@{}}
        \toprule
        Study           &  $C_d$ &  \textit{St}   \\
        \midrule
         Liu \textit{et al.} \citep{liu1998preconditioned}           &  1.350 & 0.165   \\
         Posdziech \textit{et al.} \citep{posdziech2007systematic}           &  1.31 & 0.163   \\
         Uhlmann \citep{uhlmann2005immersed}  & 1.453 & 0.169 \\
         Wu \textit{et al.} \citep{wu2009implicit}           & 1.364 & 0.163 \\
         Yang \textit{et al.} \citep{yang2009smoothing}  & 1.393 & 0.165 \\
         Rajani \textit{et al.} \citep{rajani2009numerical}    &  1.34 & 0.157   \\
        Kamensky \textit{et al.} \citep{Kamensky:2015ch}           &  1.386 & 0.170  \\
         Main \textit{et al.} (triangular grid) \citep{main2018shifted_2}           &  1.36 & 0.169  \\
         Kang \textit{et al.}  \citep{kang2021variational}           &  1.374 & 0.168  \\
         Current (element size = $30/2^{12}$)           &  1.384 & 0.170   \\
         Current (element size = $30/2^{13}$)           &  1.384 & 0.170    \\
         Current (element size = $30/2^{14}$)           &  1.386 & 0.170   \\

        \bottomrule
    \end{tabular}
    \label{tab:Flow2DCircle}
\end{table}

\begin{figure}[t!]
    \centering
    \begin{subfigure}{0.5\linewidth}
    \centering
        \includegraphics[width=0.99\linewidth,trim=100 0 100 0,clip]{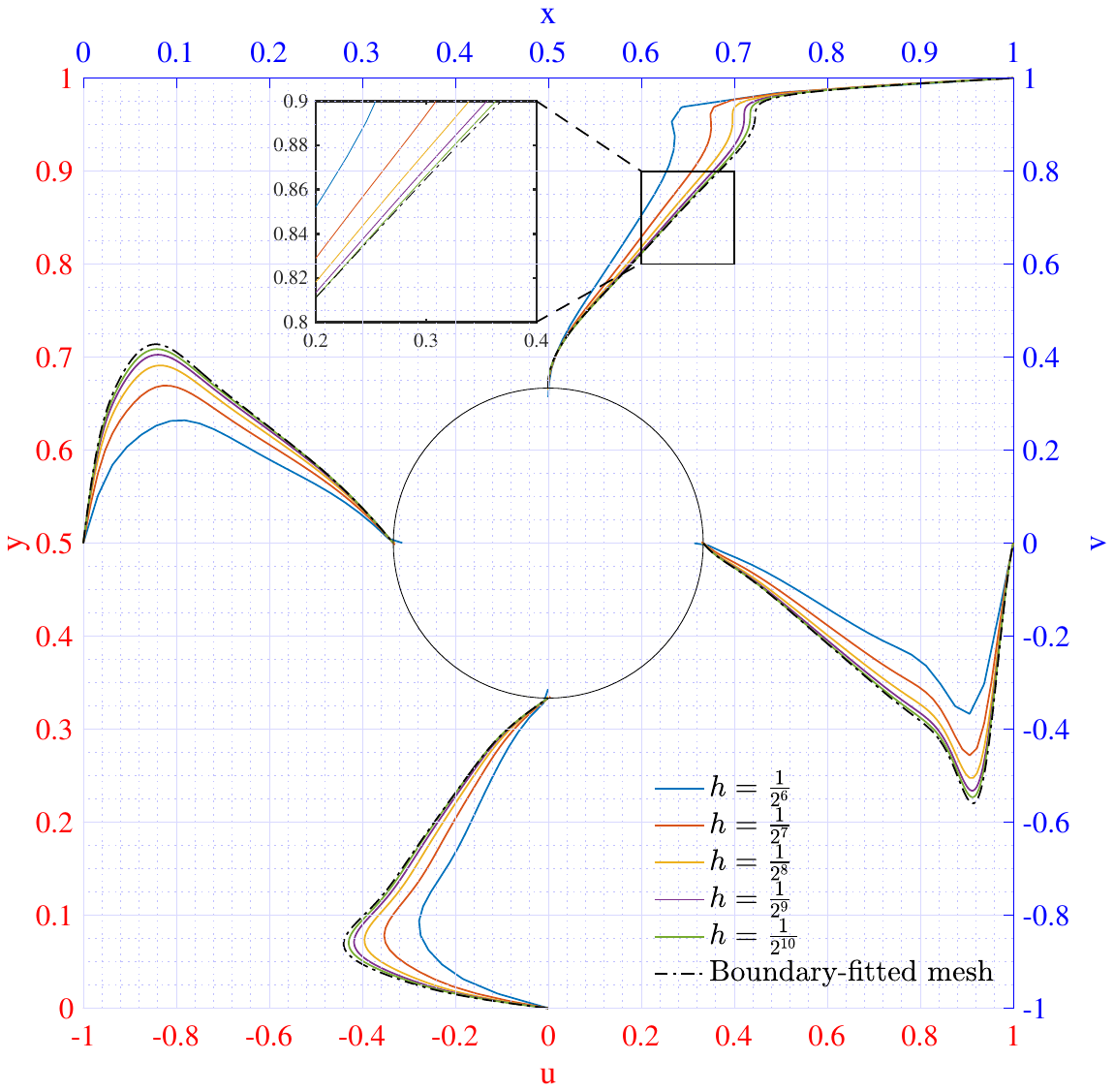}
    \end{subfigure}%
    \caption{Spatial Convergence of lid-driven cavity flow using SBM at a $Re$ of 5000. The figure also plots the velocity computed using a body-fitted mesh as a baseline comparison.}
    \label{fig:Convergence_LDC}
\end{figure}
\subsubsection{Lid-driven cavity flow with obstacles}\label{subsub:ldc}

\begin{figure}[t!]
    \centering
    \begin{subfigure}{0.32\linewidth}
    \centering
    \includegraphics[width=0.99\linewidth,trim= 10 10 10 10,clip]{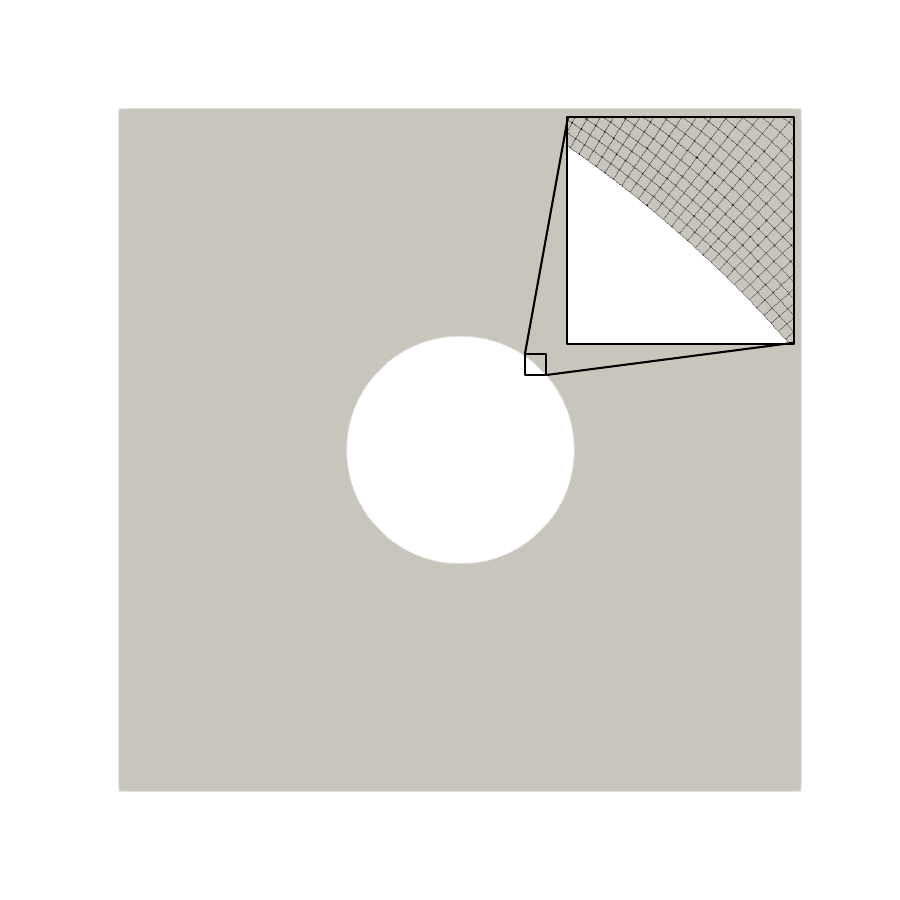}
    \caption{Boundary-fitted quadrilateral mesh.}
    \label{fig:Quad_BoundaryFitted}
    \end{subfigure}%
    \begin{subfigure}{0.32\linewidth}
    \centering
        \includegraphics[width=0.99\linewidth,trim=0 0 0 0,clip]{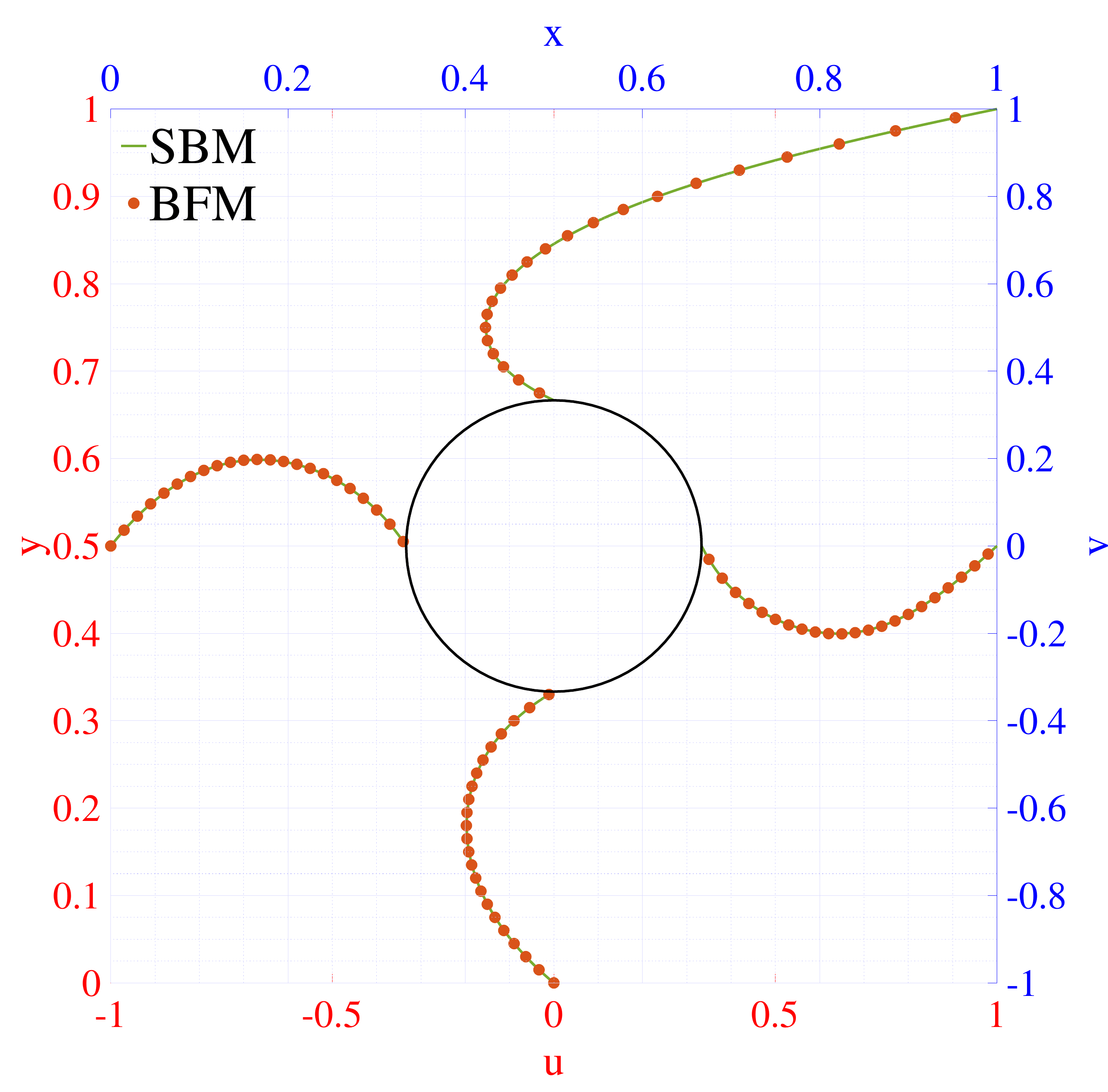}
        \caption{$Re = 100$}
        \label{fig:VelocityProfile100}
    \end{subfigure}%
    \begin{subfigure}{0.32\linewidth}
    \centering
        \includegraphics[width=0.99\linewidth,trim=0 0 0 0,clip]{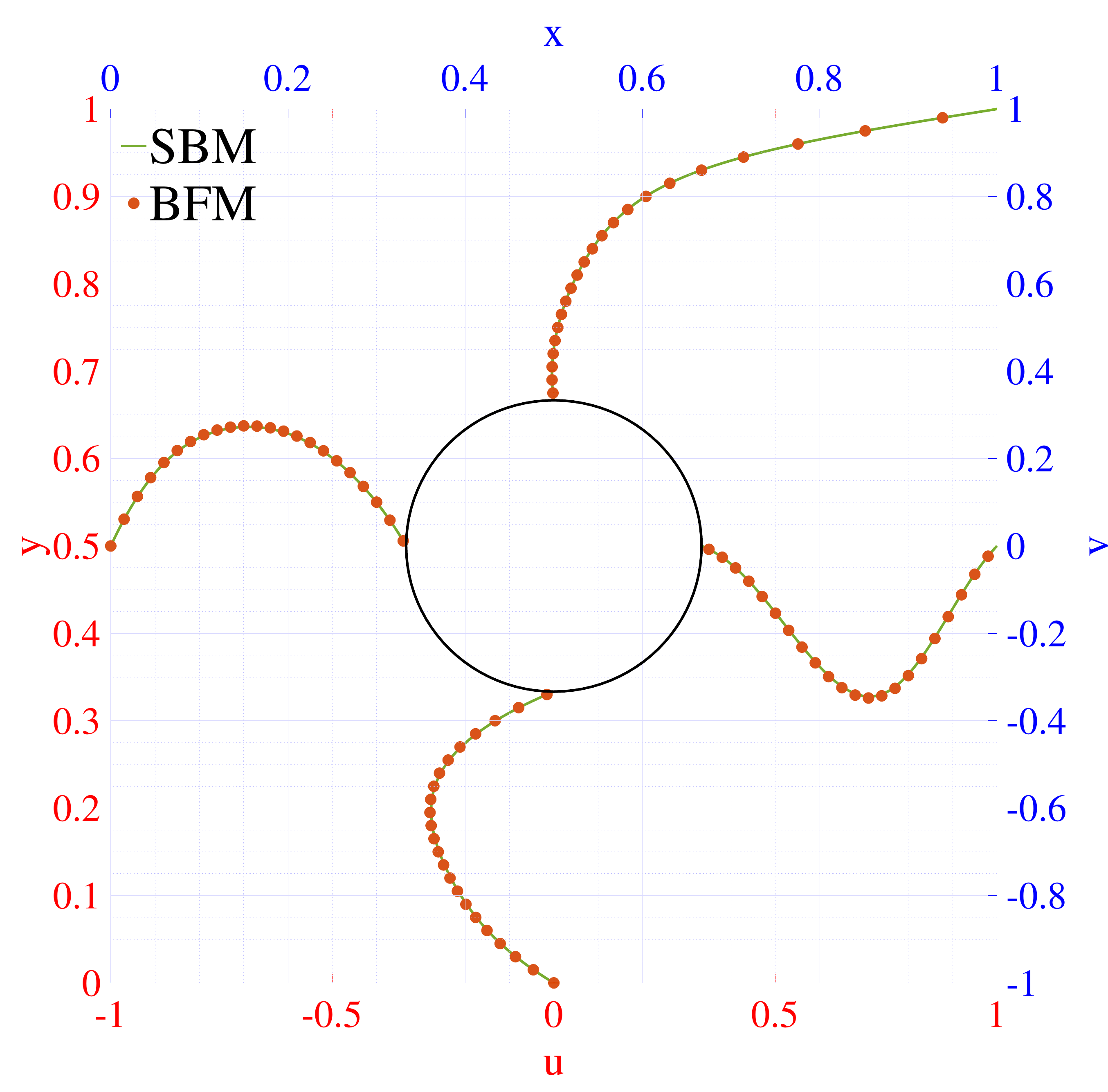}
        \caption{$Re = 400$}
        \label{fig:VelocityProfile400}
    \end{subfigure}%
    \\
        \begin{subfigure}{0.32\linewidth}
        \centering
        \includegraphics[width=0.99\linewidth,trim=0 0 0 0,clip]{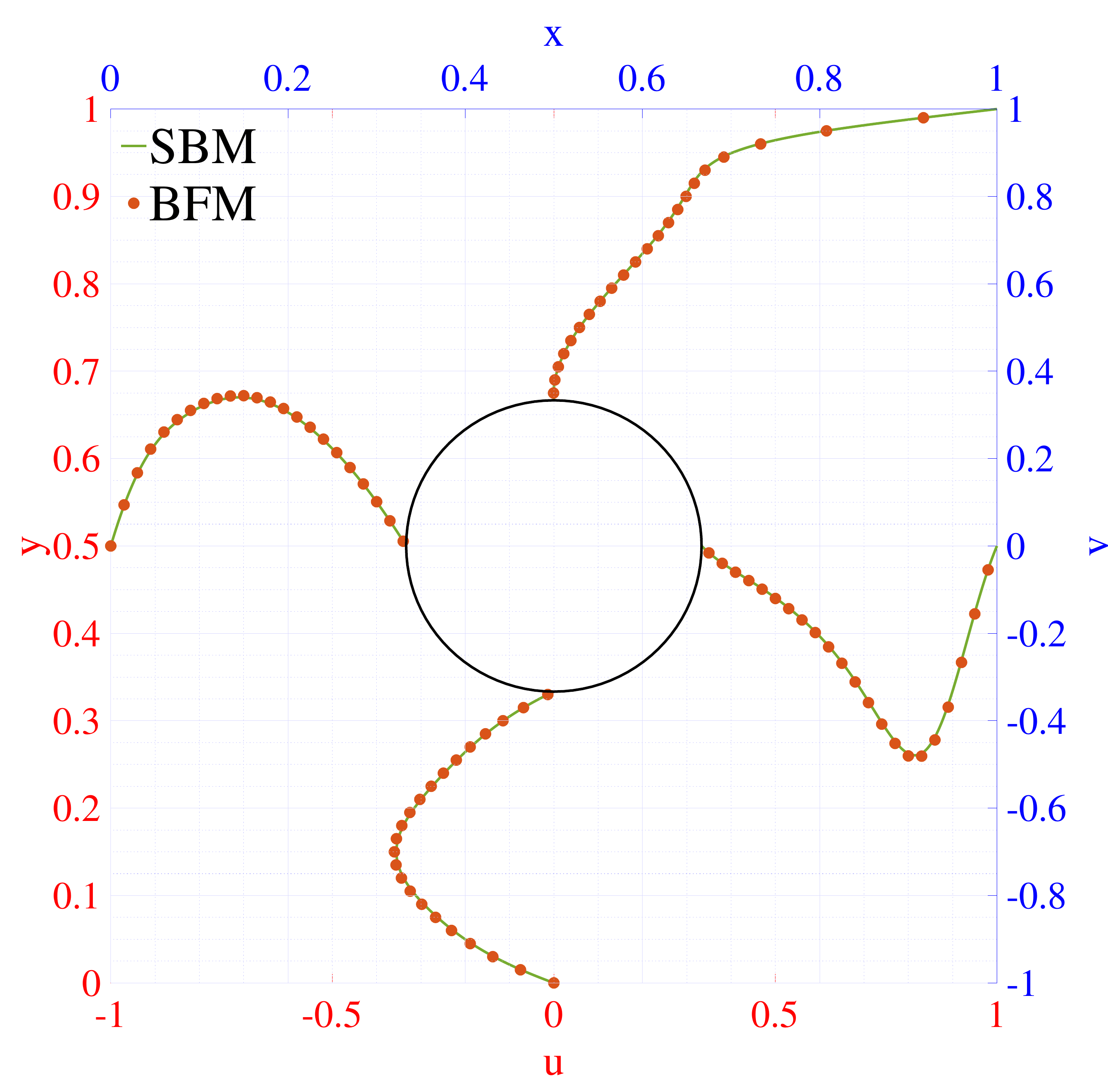}
        \caption{$Re = 1000$}
        \label{fig:VelocityProfile1000}
    \end{subfigure}%
       \begin{subfigure}{0.32\linewidth}
    \centering
        \includegraphics[width=0.99\linewidth,trim=0 0 0 0,clip]{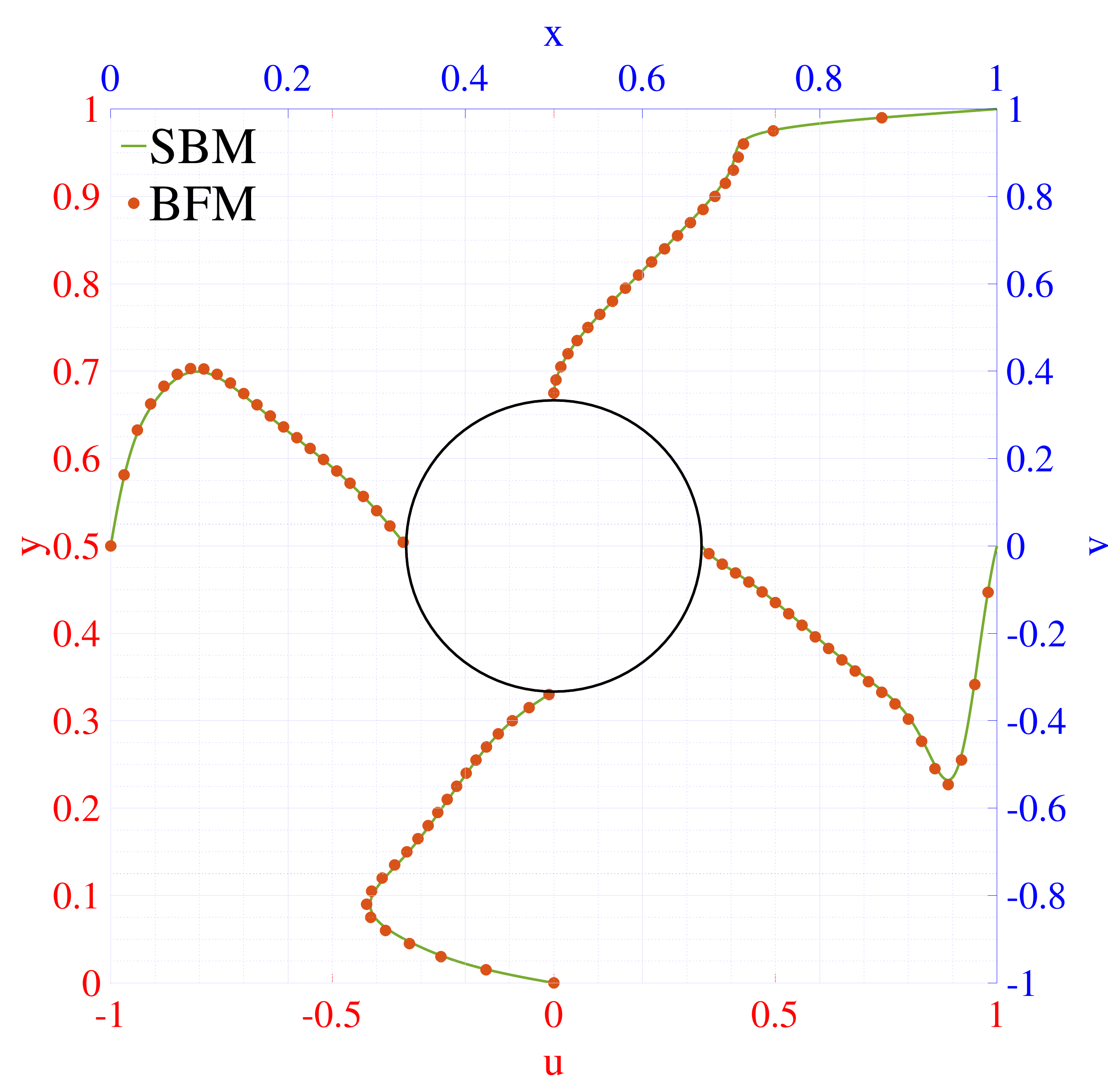}
        \caption{$Re = 3000$}
        \label{fig:VelocityProfile3000}
    \end{subfigure}%
    \begin{subfigure}{0.32\linewidth}
    \centering
        \includegraphics[width=0.99\linewidth,trim=0 0 0 0,clip]{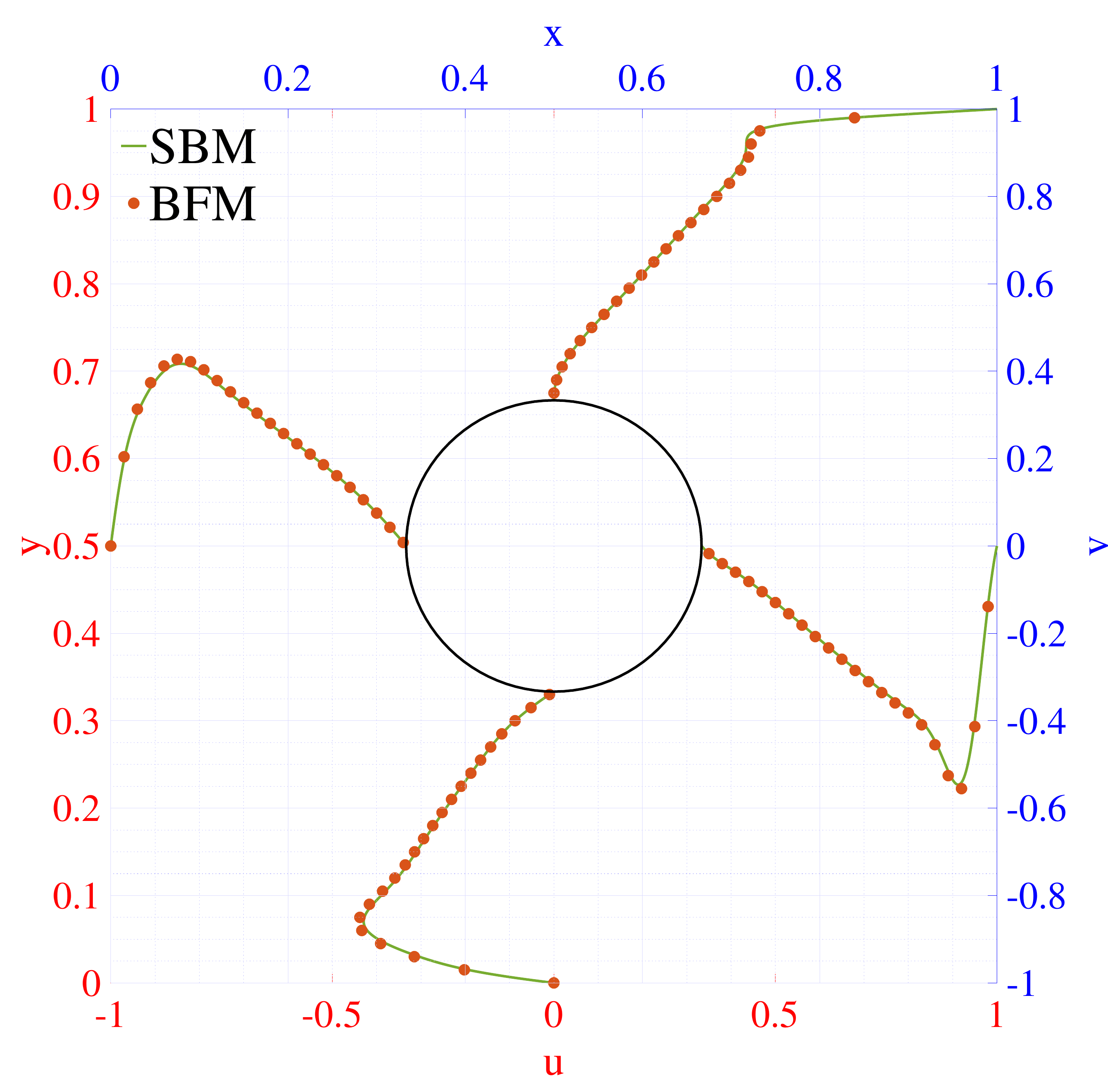}
        \caption{$Re = 5000$}
        \label{fig:VelocityProfile5000}
    \end{subfigure}%
    \caption{We compare the velocity profiles shown in panels (b) through (f) for the lid-driven cavity flow containing a circular disk. The solid line labeled `SBM' represents the velocity profile obtained using the SBM with octree meshes, while the scattered dots labeled `BFM' correspond to the velocity profile obtained using body fitted simulations using a quadrilateral boundary-fitted mesh. Additionally, panel (a) provides a zoomed-in view of the boundary-fitted mesh for greater clarity on its structure. The SBM results closely align with the BFM profiles across a range of Reynolds numbers.}
    \label{fig:VelocityProfile_LDC}
\end{figure}

\begin{figure}[t!]
    \centering
    \begin{subfigure}{0.3\linewidth}
    \centering
        \includegraphics[width=0.99\linewidth,trim=100 100 280 100,clip]{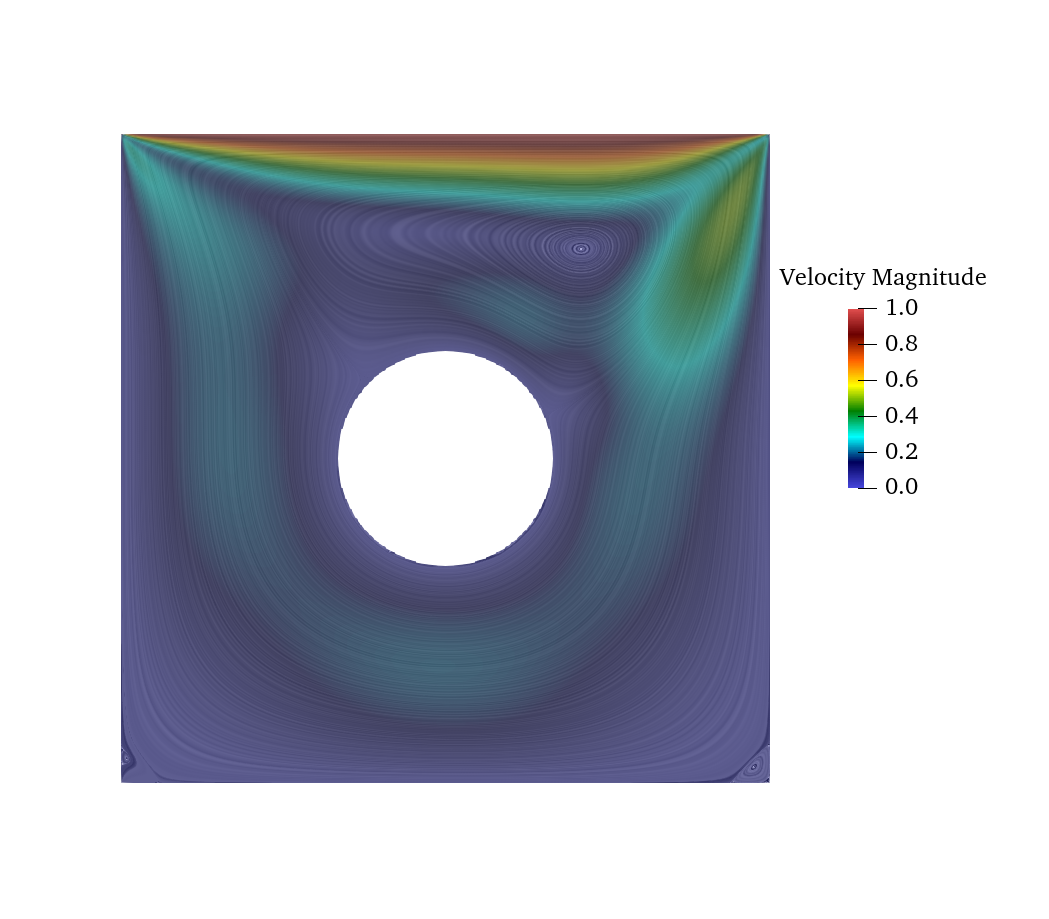}
        \caption{$Re = 100$}
    \end{subfigure}%
    \begin{subfigure}{0.3\linewidth}
    \centering
        \includegraphics[width=0.99\linewidth,trim=100 100 280 100,clip]{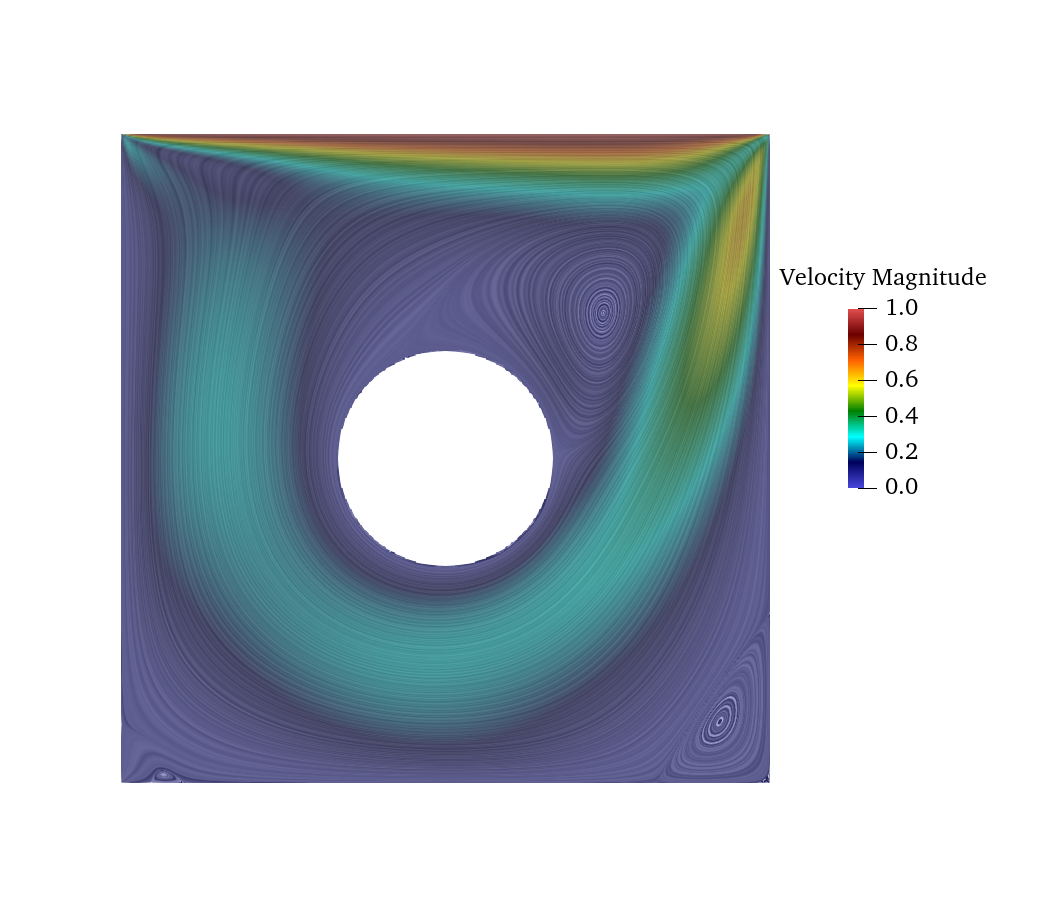}
        \caption{$Re = 400$}
    \end{subfigure}%
        \begin{subfigure}{0.3\linewidth}
        \centering
        \includegraphics[width=0.99\linewidth,trim=100 100 280 100,clip]{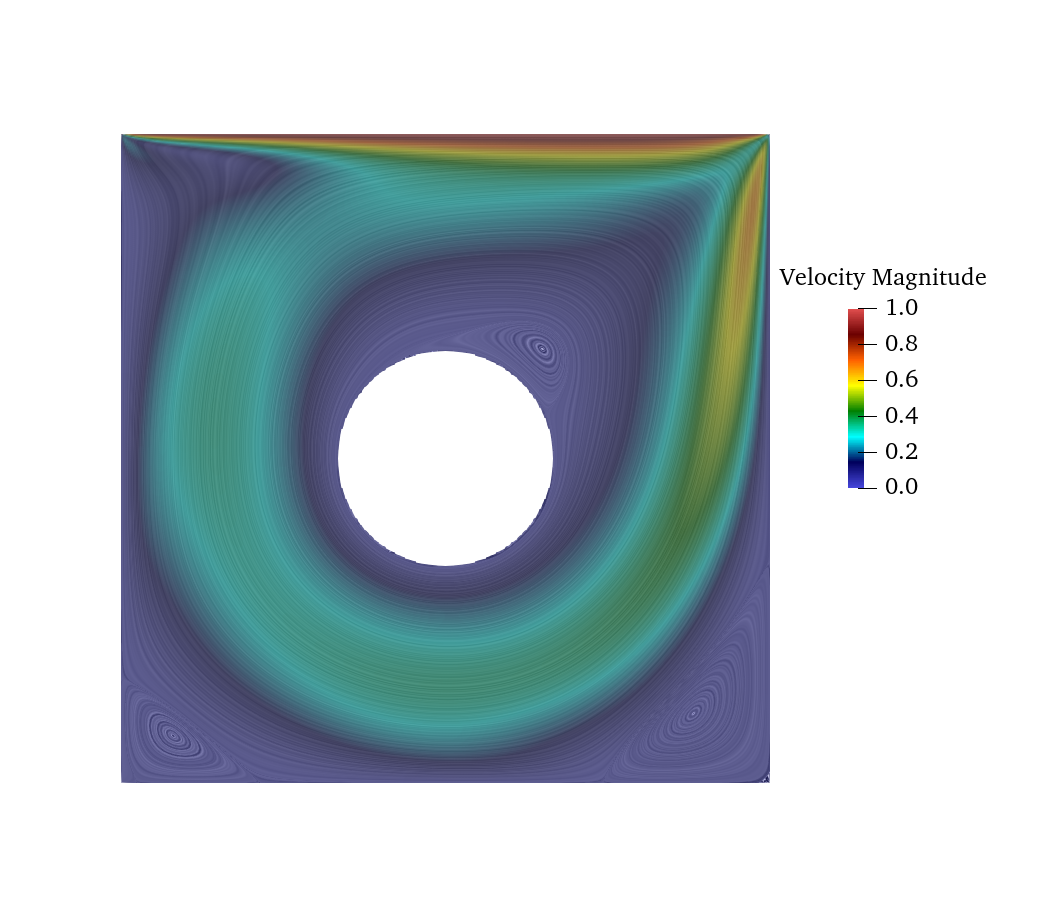}
        \caption{$Re = 1000$}
    \end{subfigure}%
    \\
       \begin{subfigure}{0.3\linewidth}
    \centering
        \includegraphics[width=0.99\linewidth,trim=100 100 280 100,clip]{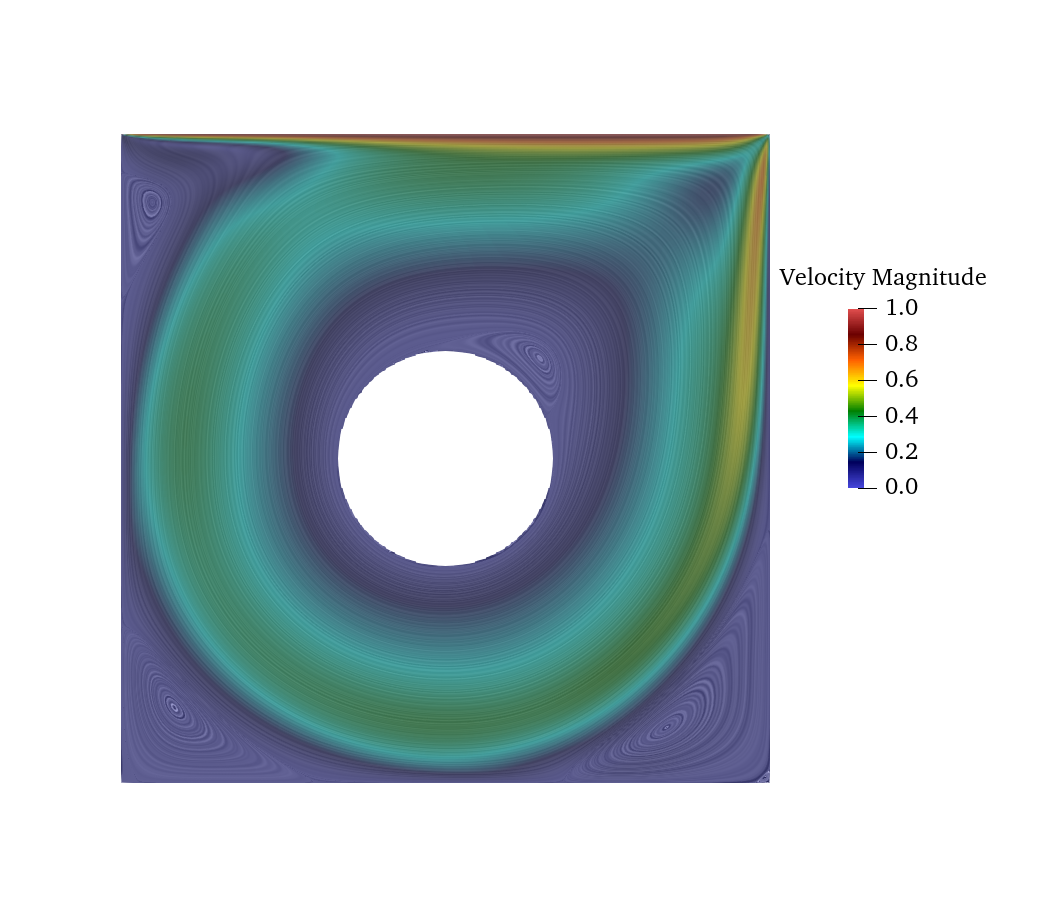}
        \caption{$Re = 3000$}
    \end{subfigure}%
    \begin{subfigure}{0.3\linewidth}
    \centering
        \includegraphics[width=0.99\linewidth,trim=100 100 280 100,clip]{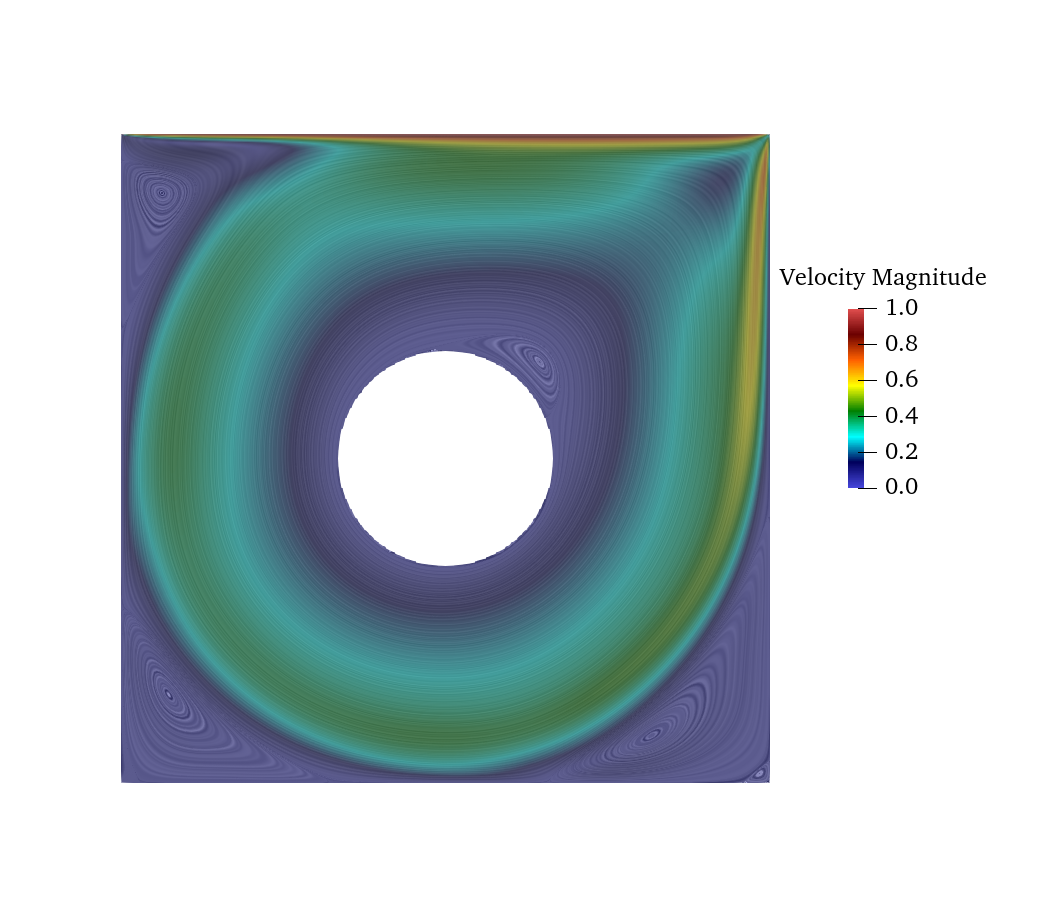}
        \caption{$Re = 5000$}
    \end{subfigure}%
    \begin{subfigure}{0.3\linewidth}
        \centering
        \raisebox{0.4in}{
        \includegraphics[width=0.6\linewidth,trim=775 360 40 250,clip]{LIC_Re5000.png}
        }
    \end{subfigure}%
        \caption{The visualization of LIC (Line Integral Convolution) contours within a lid-driven cavity containing circular obstacles at various Reynolds numbers.}
    \label{fig:LIC_LDC}
\end{figure}

\begin{figure}[t!]
    \centering
    \begin{subfigure}{0.49\textwidth}
    \includegraphics[width=0.99\linewidth,trim=0 0 0 0,clip]{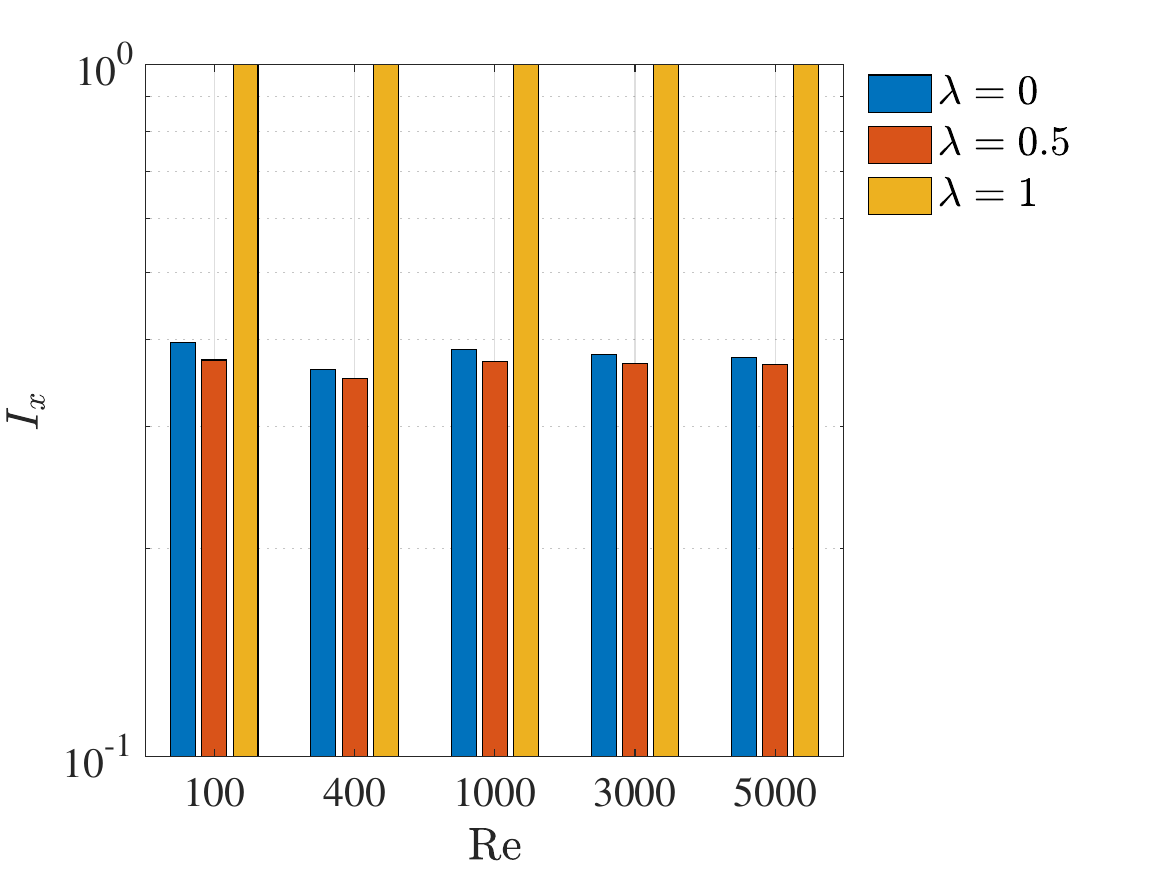}
    \caption{$I_x$}
    \end{subfigure}
    \begin{subfigure}{0.49\textwidth}
    \includegraphics[width=0.99\linewidth,trim=0 0 0 0,clip]{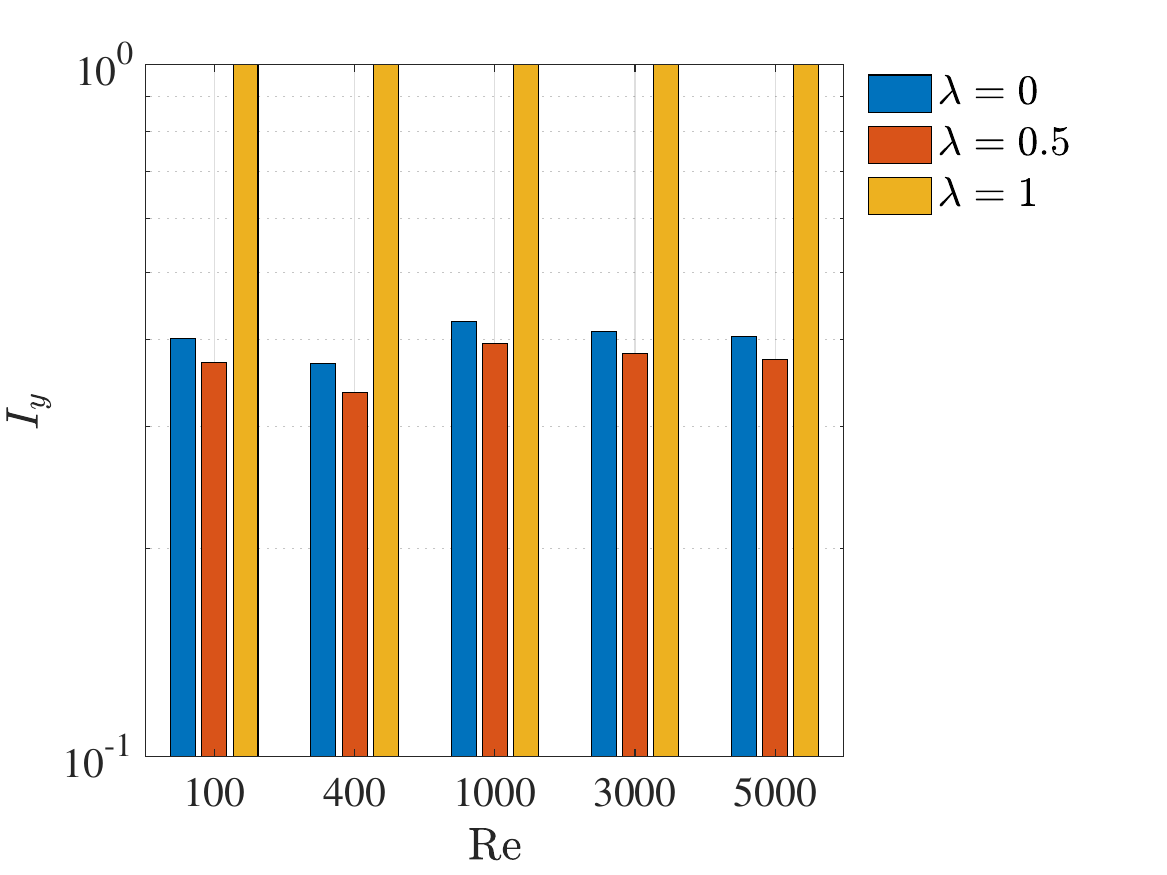}
    \caption{$I_y$}
    \end{subfigure}
    \caption{Comparison of $I_x$ and $I_y$ (\eqnref{eq:lxylambda}) for the lid-driven cavity problem with a circular obstacle, using various values of $\lambda$. The lower values of $I_x$ and $I_y$ for $\lambda = 0.5$ indicate that an optimal surrogate boundary can result in reduced boundary errors.}
    \label{fig:LDC-Ex-Ey}
\end{figure}

For our next benchmark study, the lid-driven cavity flow is often chosen to assess the robustness of the framework. In order to test the problem using the SBM, we select simulations with a disk of diameter $D = \frac{L}{3}$ placed at the center of the lid-driven cavity flow, where $L$ is the box length of the chamber. This setup mirrors the original benchmark by~\citep{huang2020simulation}. A spatial convergence study employs $h = \frac{1}{2^l}$, where $l = 6, 7, 8, 9, 10$, at a Reynolds number of 5000, as illustrated in \figref{fig:Convergence_LDC}. The comparative analysis compares results obtained using the Boundary-Fitted Method (BFM) with a quadrilateral mesh (\figref{fig:Quad_BoundaryFitted}), generated using Gmsh~\citep{geuzaine2009gmsh}. The average mesh size for the boundary-fitted mesh is approximately 0.0014. Observations indicate that the results from the SBM increasingly converge towards those from the BFM as the mesh refines. Following the mesh convergence analysis, tests are conducted across five different Reynolds numbers: 100, 400, 1000, 3000, and 5000 with $h = \frac{1}{2^{10}}$. The outcomes from the SBM are then compared to those from the quadrilateral boundary-fitted meshes to evaluate performance across various flow regimes, as shown in \figref{fig:VelocityProfile_LDC}. The Line Integral Convolution (LIC) contours, once the flow reaches a steady state, are depicted in \figref{fig:LIC_LDC}. To evaluate the impact of selecting $\lambda = 0.5$, we calculate $E_x$ and $E_y$ from \eqnref{eq:Ex} and \eqnref{eq:Ey} for various values of $\lambda$ at different Reynolds numbers, once the flow has stabilized. The mesh size used for all Reynolds numbers is $h = \frac{1}{2^9}$. To compare different surrogate models with the reference case where $\lambda = 1$, we introduce the improvement factors $I_x$ and $I_y$, as defined in equation \eqnref{eq:lxylambda}. Lower values of $I_x$ and $I_y$ indicate more accurate solutions:
\begin{equation}
    I_x = \frac{E_x(\lambda)}{E_x(\lambda = 1)}, \quad
    I_y = \frac{E_y(\lambda)}{E_y(\lambda = 1)}.
    \label{eq:lxylambda}
\end{equation}
\figref{fig:LDC-Ex-Ey} illustrates that $\lambda = 0.5$ yields improved $I_x$ and $I_y$ values across different Reynolds numbers. This strongly suggests that an optimal surrogate boundary using $\lambda = 0.5$ offers significant benefits, not only for PDEs such as the Poisson equation and linear elasticity problems~\citep{yang2024optimal}, but also for the Navier-Stokes equations.

\begin{figure}[t!]
    \centering
    \begin{subfigure}{0.32\linewidth}
    \centering
        \includegraphics[width=0.99\linewidth,trim=0 0 0 0,clip]{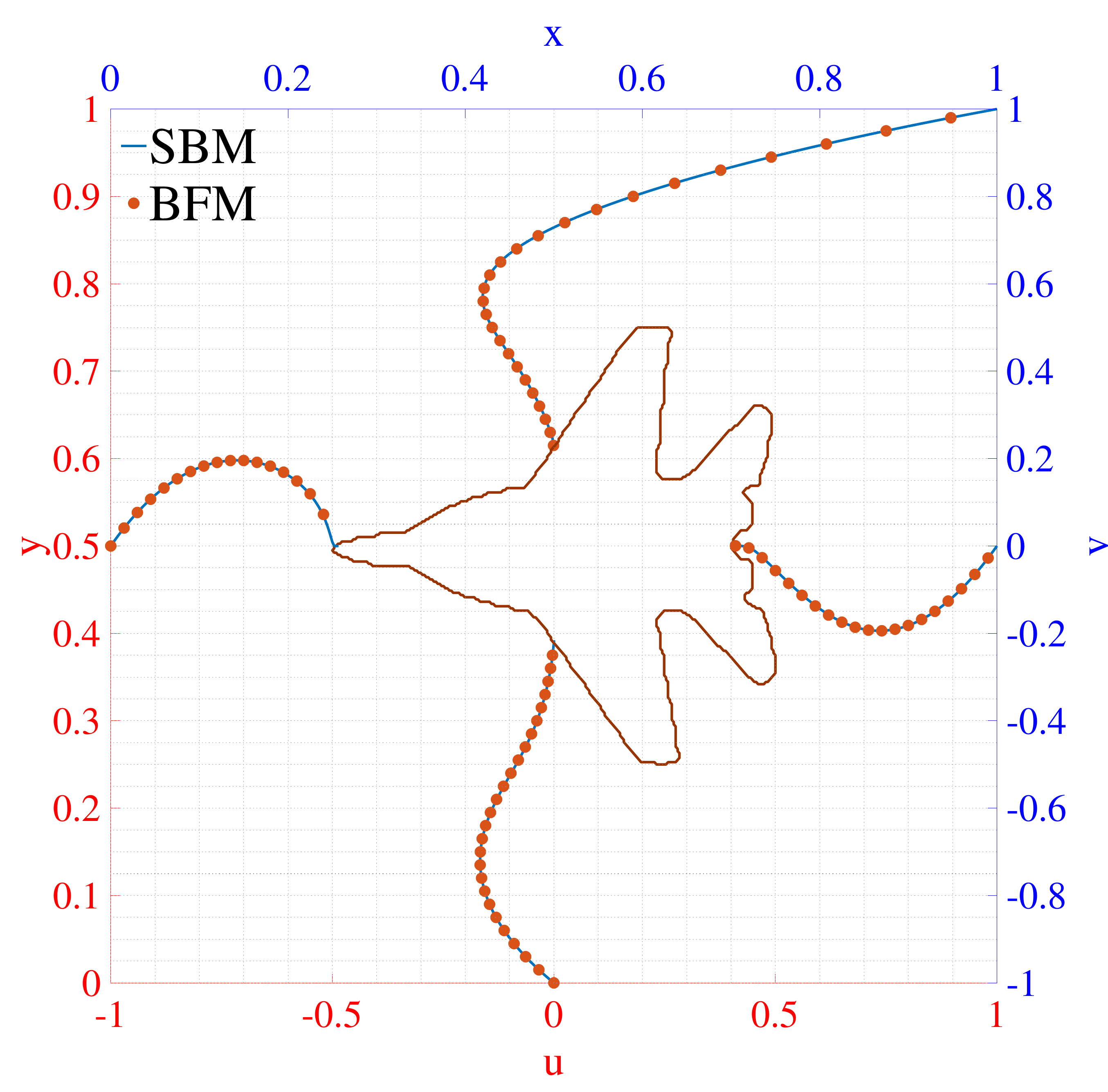}
        \caption{Plane-shaped obstacle, $Re = 50$}
        \label{fig:LDC_plane_Re50}
    \end{subfigure}%
    \begin{subfigure}{0.32\linewidth}
    \centering
        \includegraphics[width=0.99\linewidth,trim=0 0 0 0,clip]{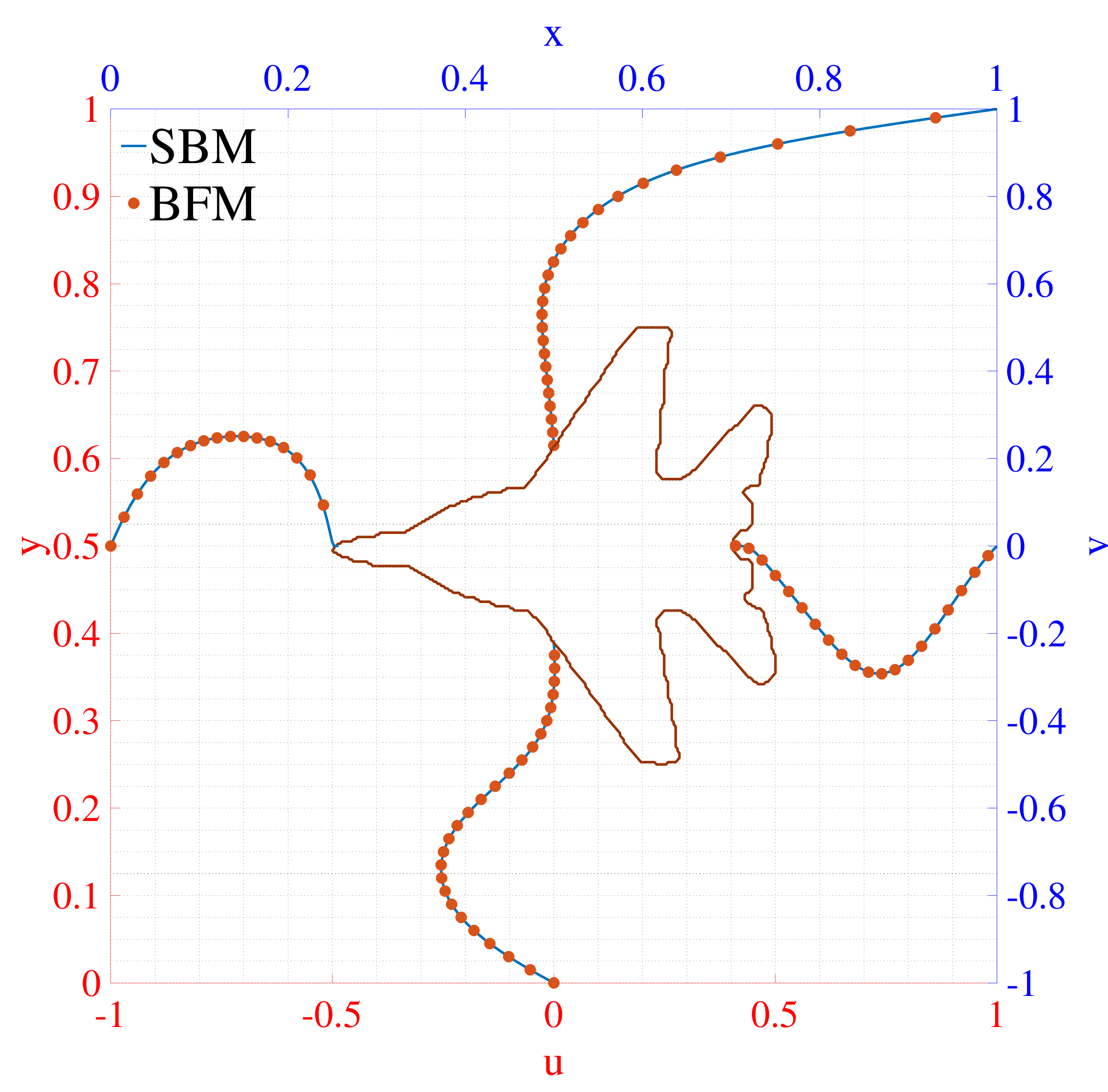}
        \caption{Plane-shaped obstacle, $Re = 500$}
        \label{fig:LDC_plane_Re500}
    \end{subfigure}%
    \begin{subfigure}{0.32\linewidth}
    \centering
        \includegraphics[width=0.99\linewidth,trim=0 0 0 0,clip]{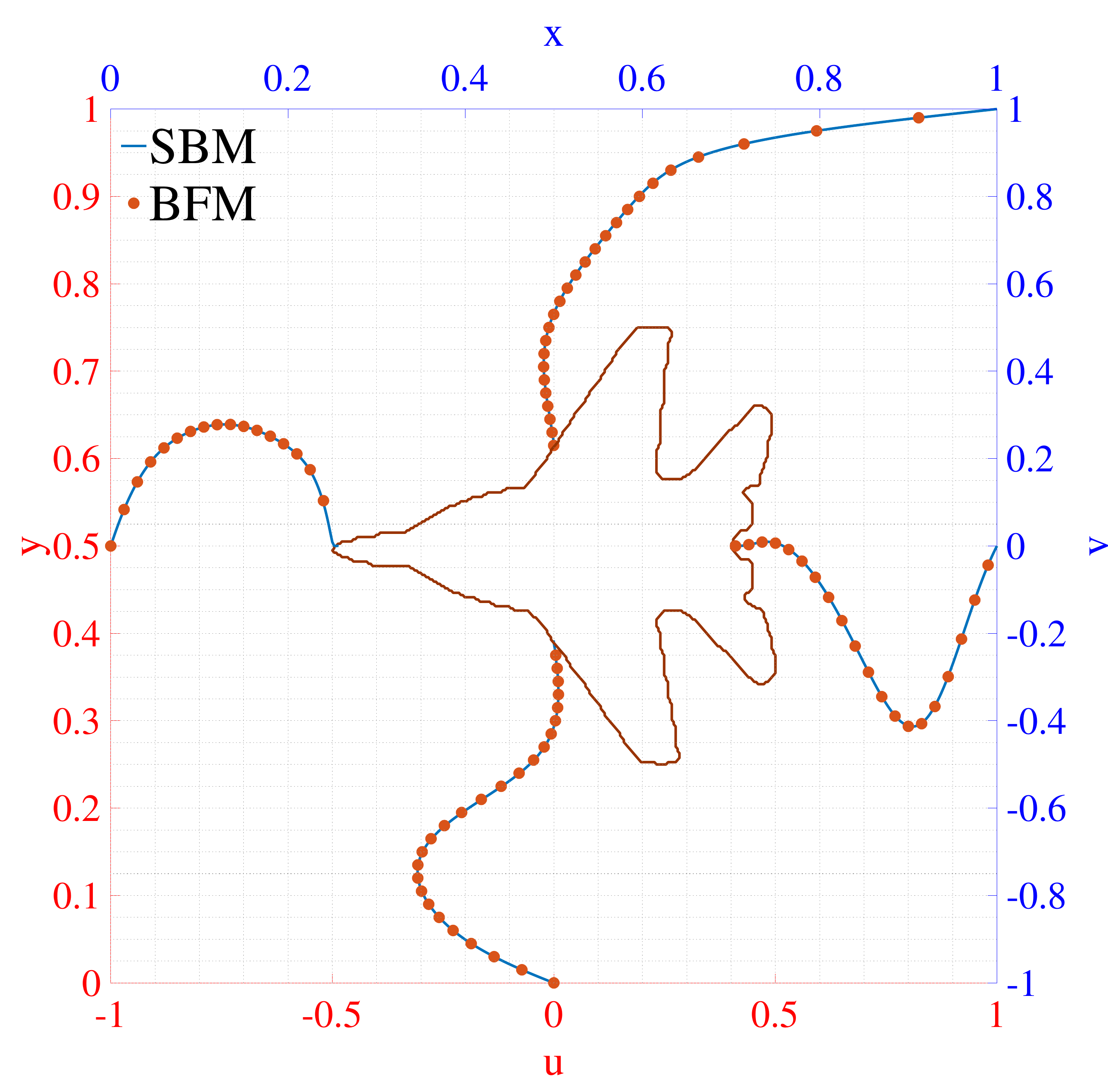}
        \caption{Plane-shaped obstacle, $Re = 1000$}
        \label{fig:LDC_plane_Re1000}
    \end{subfigure}%
    \\
    \begin{subfigure}{0.32\linewidth}
        \centering
        \includegraphics[width=0.99\linewidth,trim=0 0 0 0,clip]{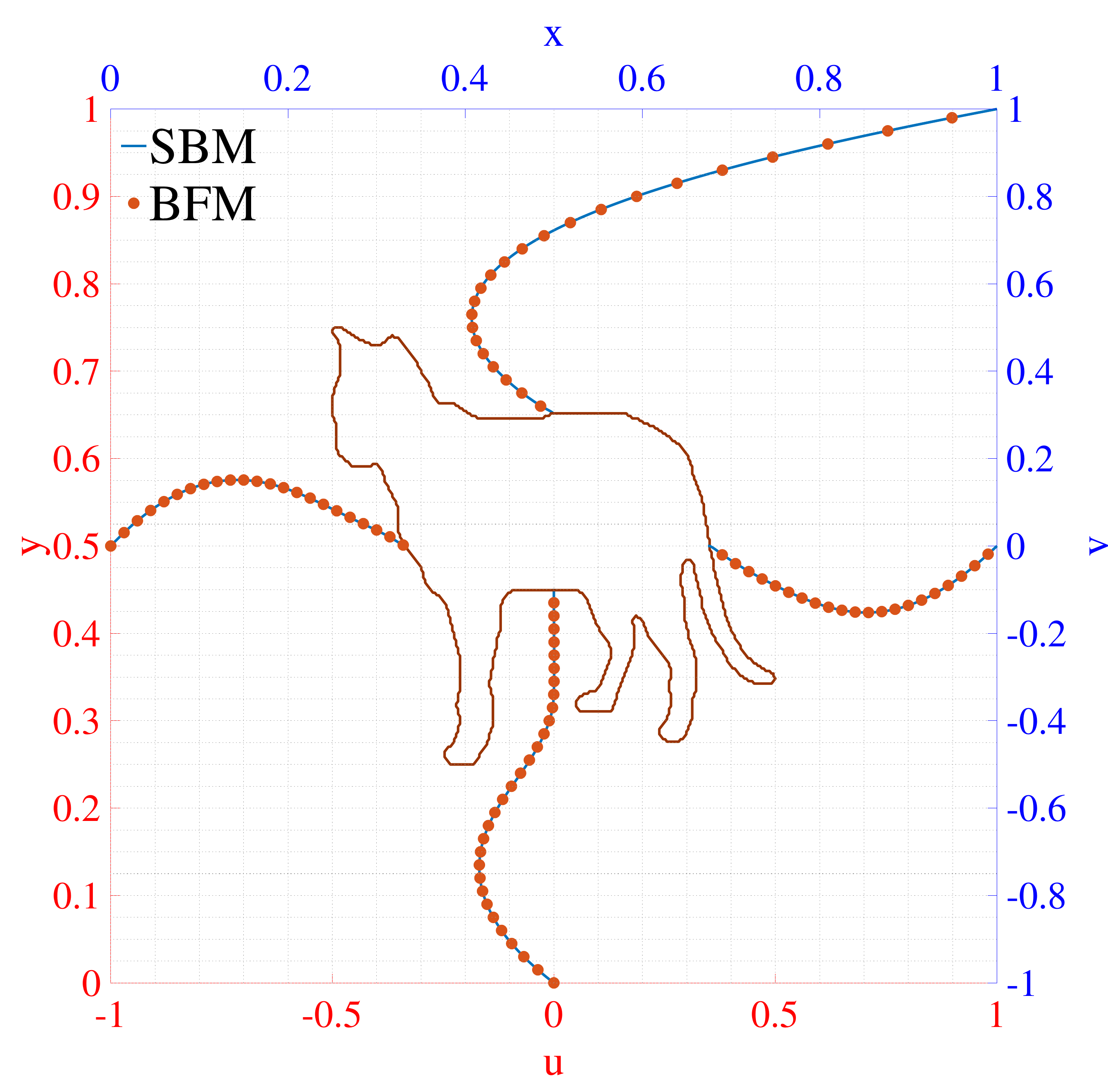}
        \caption{Cat-shaped obstacle, $Re = 50$}
        \label{fig:LDC_cat_Re50}
    \end{subfigure}%
    \begin{subfigure}{0.32\linewidth}
    \centering
        \includegraphics[width=0.99\linewidth,trim=0 0 0 0,clip]{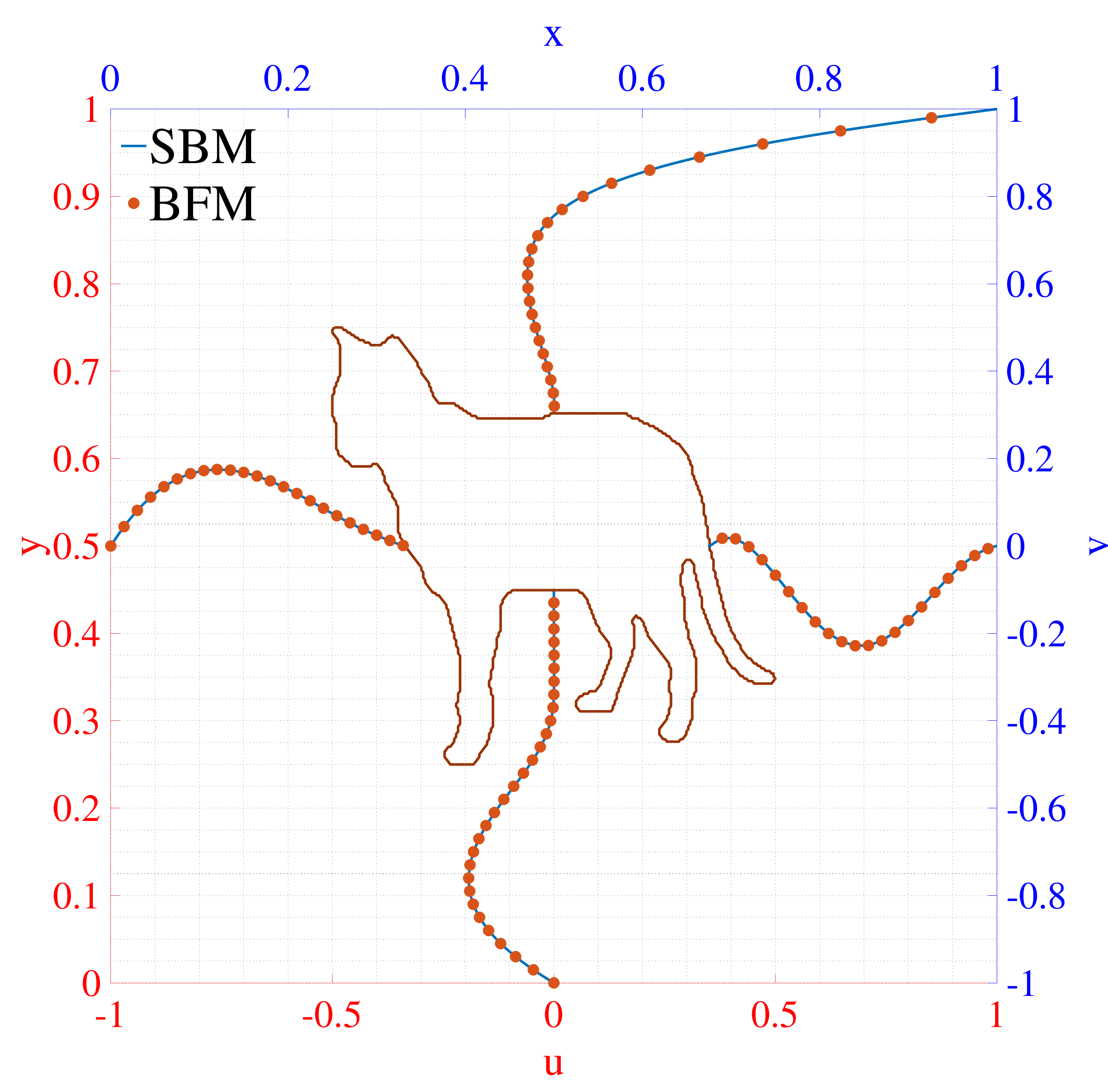}
        \caption{Cat-shaped obstacle, $Re = 500$}
        \label{fig:LDC_cat_Re500}
    \end{subfigure}%
    \begin{subfigure}{0.32\linewidth}
    \centering
        \includegraphics[width=0.99\linewidth,trim=0 0 0 0,clip]{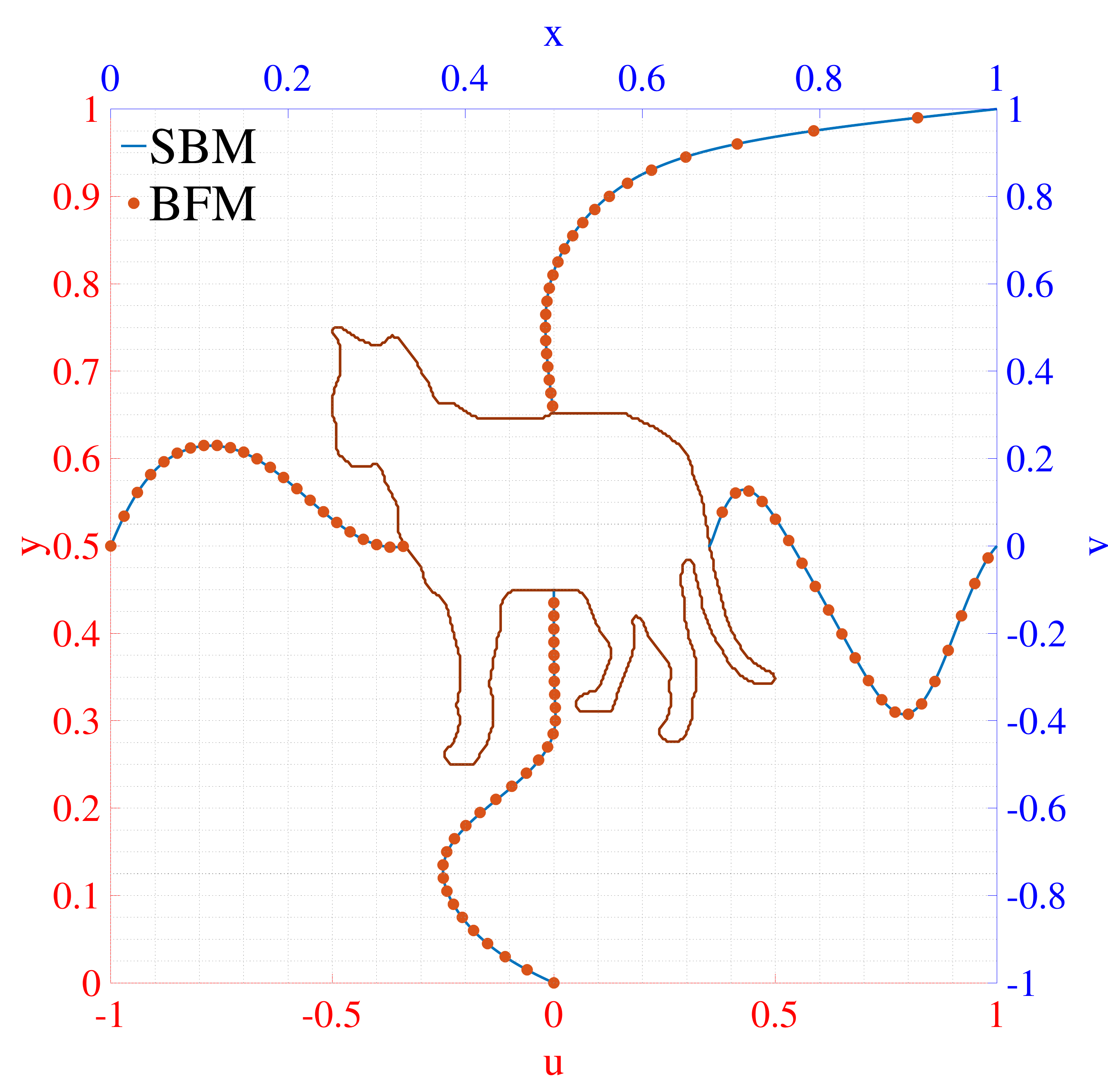}
        \caption{Cat-shaped obstacle, $Re = 1000$}
        \label{fig:LDC_cat_Re1000}
    \end{subfigure}%
    \\
    \begin{subfigure}{0.32\linewidth}
        \centering
        \includegraphics[width=0.99\linewidth,trim=0 0 0 0,clip]{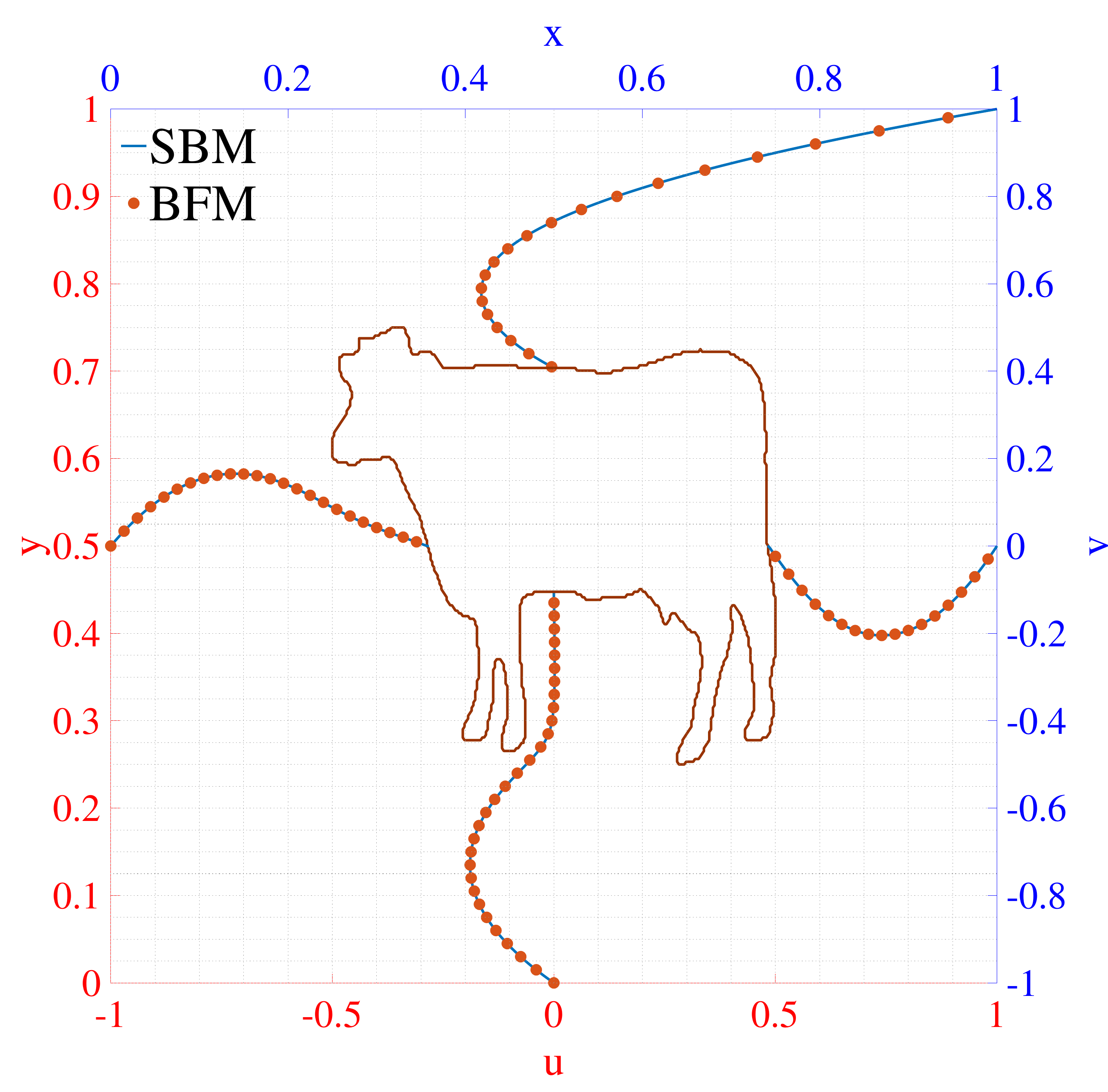}
        \caption{Cattle-shaped obstacle, $Re = 50$}
        \label{fig:LDC_cattle_Re50}
    \end{subfigure}%
    \begin{subfigure}{0.32\linewidth}
    \centering
        \includegraphics[width=0.99\linewidth,trim=0 0 0 0,clip]{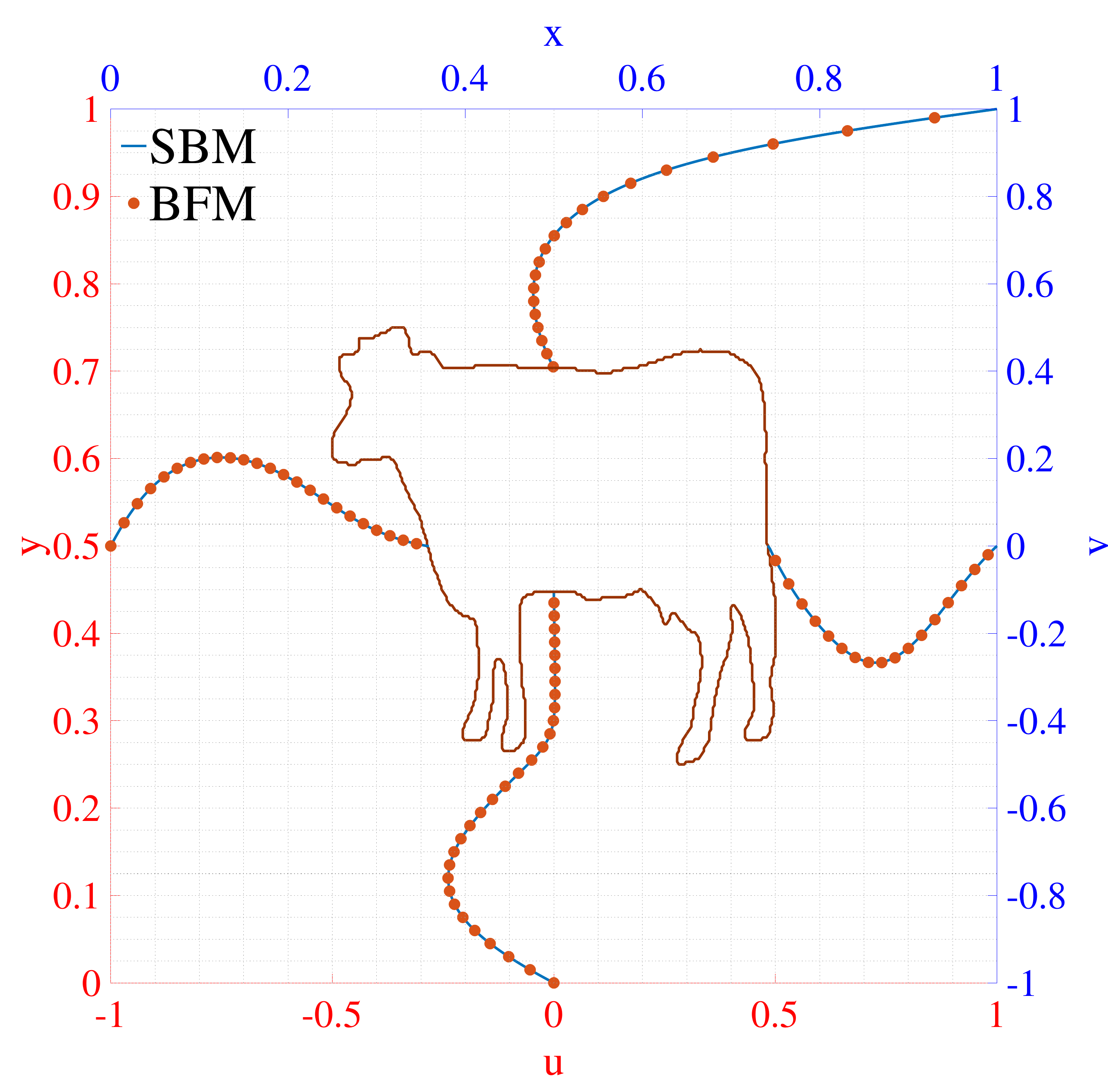}
        \caption{Cattle-shaped obstacle, $Re = 500$}
        \label{fig:LDC_cattle_Re500}
    \end{subfigure}%
    \begin{subfigure}{0.32\linewidth}
    \centering
        \includegraphics[width=0.99\linewidth,trim=0 0 0 0,clip]{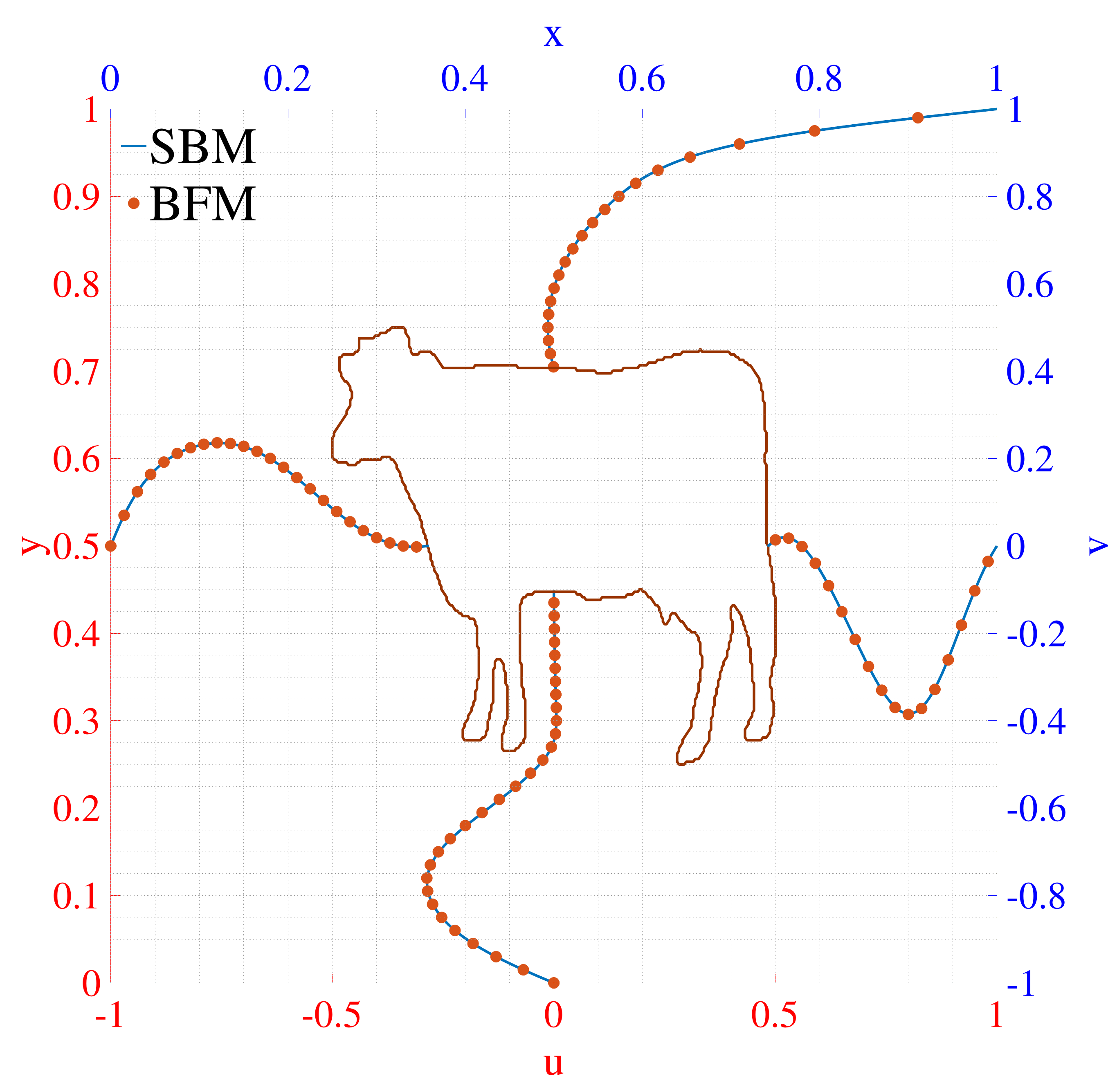}
        \caption{Cattle-shaped obstacle, $Re = 1000$}
        \label{fig:LDC_cattle_Re1000}
    \end{subfigure}%
    \caption{Velocity profiles for lid-driven cavity flow simulations with complex geometries inside the chamber. The first row shows results for a plane-shaped obstacle, the second row features a cat-shaped obstacle, and the third row includes a cattle-shaped obstacle. Simulations were conducted at Reynolds numbers $Re = 50$, $Re = 500$, and $Re = 1000$, demonstrating the influence of obstacle shape and Reynolds number on the flow behavior. SBM and BFM show strong agreement with each other.}
    \label{fig:LDC_ComplexShape1}
\end{figure}

\textbf{Complex shaped obstacles}:  In addition to performing lid-driven cavity flow simulations with a circular disk, we also conducted simulations with more complex geometries placed at the center of the chamber to test SBM's ability to handle complex shapes, as illustrated in \figref{fig:LDC_ComplexShape1}. These non-parametric shapes, sampled from the grayscale dataset of SkelNetOn~\citep{demir2019skelneton,Atienza2019}, were placed inside the chamber. To soften thin features, a Gaussian blur filter with a scale of 2 was applied. The shapes were then resized to fit within the unit square. To validate our results, we compared them with simulations performed using the Boundary-Fitted Method (BFM), where a boundary-fitted mesh was generated with Gmsh. The octree mesh used had a resolution of $1/2^9$, while the boundary-fitted mesh had a comparable resolution of approximately $1/2^9$. The meshes and streamlines for both SBM and BFM are shown in \secref{appendix:LDC-BFM-SBM}. We perform simulations at various Reynolds ($Re = 50,~500,~1000$) for these complex shapes and compare against results obtained using a body fitted mesh, and are shown in \figref{fig:LDC_ComplexShape1}. 
In all cases, the Octree-SBM approach gives identical results to the body-fitted mesh.

\clearpage

\subsubsection{Flow past an airfoil}\label{subsub:airfoil}
\begin{figure}[b!]
    \centering 
    \begin{subfigure}[b]{0.7\linewidth}
        \includegraphics[width=0.99\linewidth,trim=0 20cm 0 0,clip]{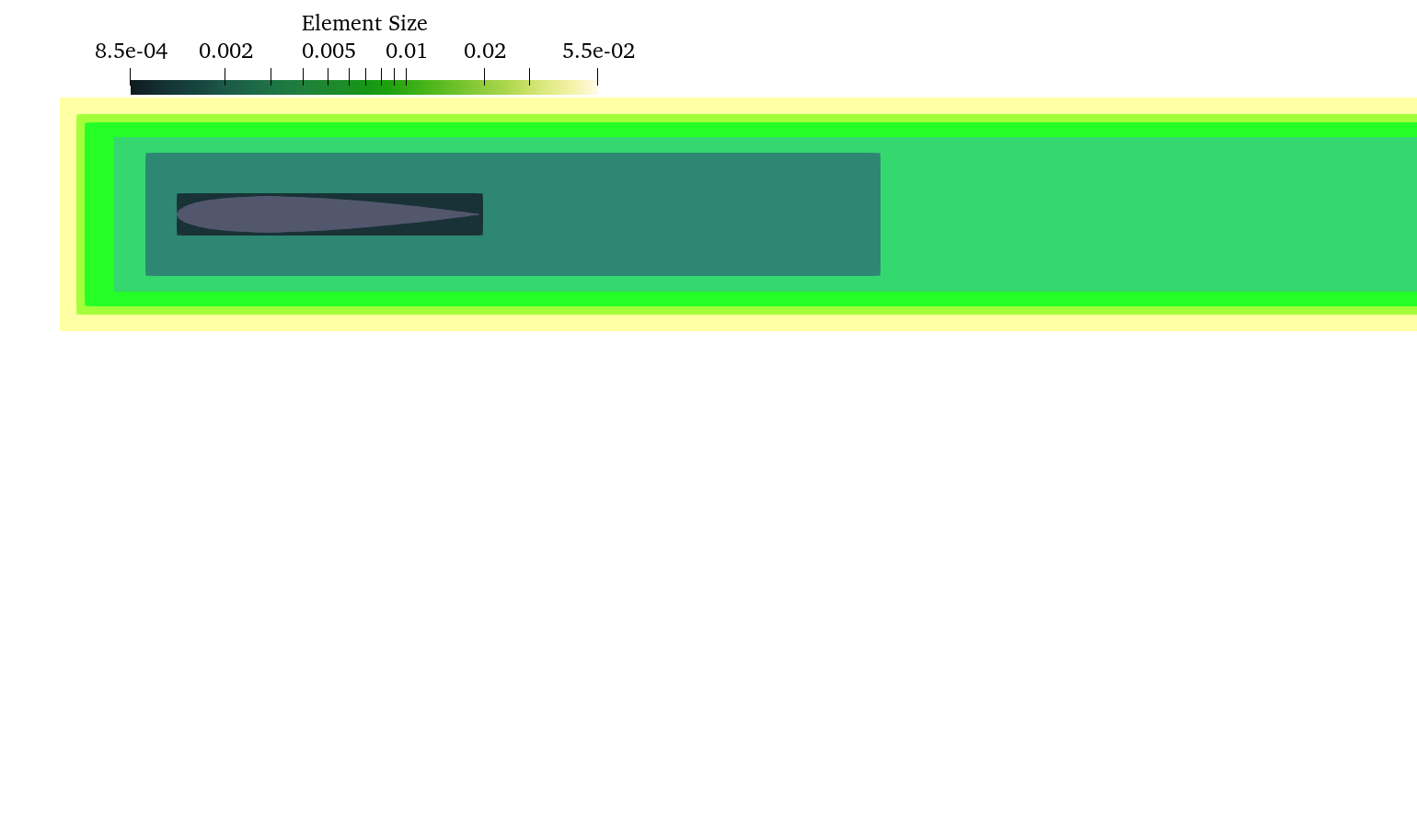}
        \caption{AOA = 0\textdegree}
        \label{fig:aoa0}
    \end{subfigure}
    \begin{subfigure}[b]{0.7\linewidth}
        \includegraphics[width=0.99\linewidth,trim=0 16cm 0 0,clip]{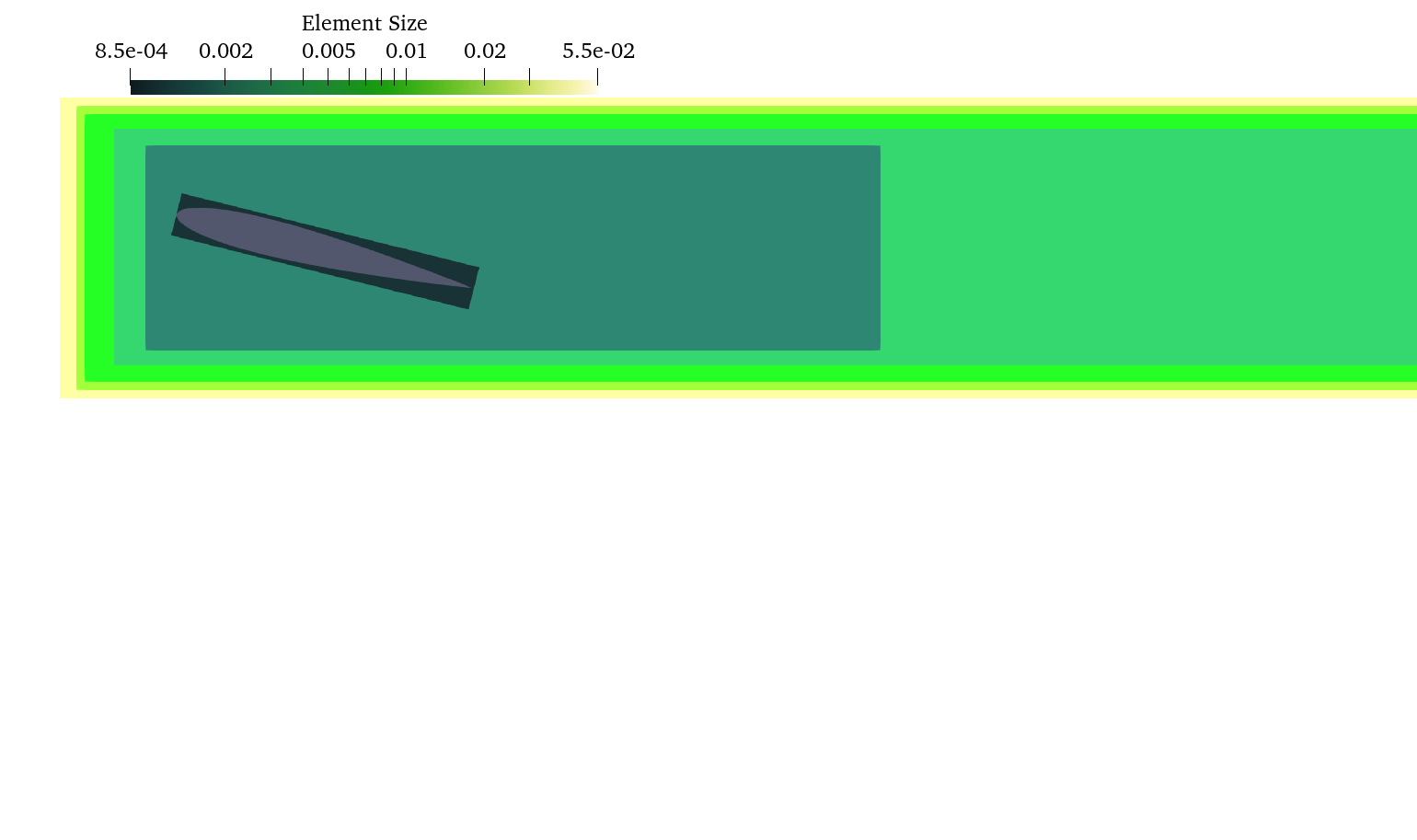}
        \caption{AOA = 14\textdegree}
        \label{fig:aoa14}
    \end{subfigure}
    \begin{subfigure}[b]{0.7\linewidth}
        \includegraphics[width=0.99\linewidth,trim=0 15cm 0 0,clip]{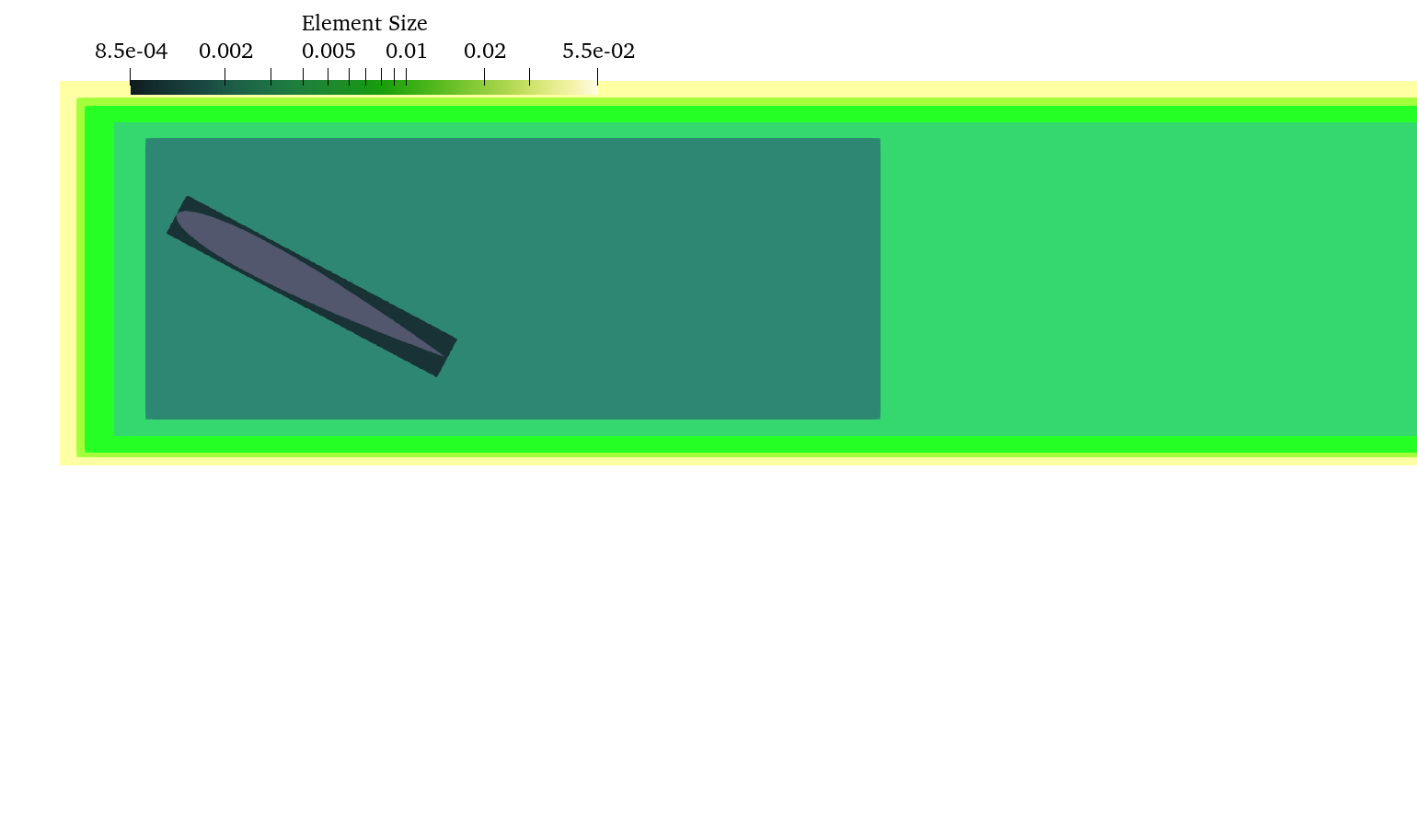}
        \caption{AOA = 28\textdegree}
        \label{fig:aoa28}
    \end{subfigure}
    \caption{The local mesh refinement strategy for the NACA 0012 airfoil at various angles of attack is illustrated. The darker green areas indicate finer mesh regions. This figure clearly shows a box-shaped refinement zone (the darkest green) near the airfoil, which rotates with the airfoil as its angle changes.}
    \label{fig:naca0012_mesh}
\end{figure}

We test flow past a NACA (National Advisory Committee for Aeronautics) 0012 airfoil at $Re = 1000$. The non-dimensional timestep is 0.01. The solution domain is a rectangle region [0, 28] $\times$ [0, 28] with leading edge point placed at (14, 14). To study the aerodynamic behavior under various AOAs (angles of attack), the airfoil is systematically rotated from 0$^{\circ}$ to 28$^{\circ}$ in 2$^{\circ}$ increments, pivoting about its leading edge point against the background octree mesh. The mesh configuration is illustrated in \figref{fig:naca0012_mesh}. Except for the outlet, all other walls are subject to a uniform non-dimensional freestream velocity of (1, 0, 0). To mitigate backflow issues, backflow stabilization techniques, as discussed in \secref{sec:NS_backflow}, are implemented at the outlet. Unlike the outlet boundary condition in \secref{subsec:2D_cylinder}, we do not set the pressure to zero at the outlet, considering the Reynolds number is set to 1000 and the angle of attack (AOA) varies, potentially resulting in strong outlet vortices. Therefore, backflow stabilization is employed to prevent numerical divergence at the outlet.

\begin{figure}[t!]
    \centering
    \begin{tikzpicture}
    \begin{axis}[
        xlabel={AOA (Angle of attack)},
        ylabel={$\frac{C_l}{C_d}$},
        xmin=0, xmax=28,
        grid=both,
        grid style={line width=.1pt, draw=gray!10},
        major grid style={line width=.2pt, draw=gray!50},
        minor tick num=5,
        legend style={at={(0.5,-0.25)}, anchor=north, legend columns=-1},
    ]
    \addplot [solid, color=blue, mark=*] table [x={a}, y={clcd}, col sep=space] {Current.txt};
    \addplot [dashed, color=red, mark=square*] table [x={a}, y={clcd}, col sep=space] {Kurtulus.txt};
    \addplot [dotted, color=brown!60!black, mark=triangle*] table [x={a}, y={clcd}, col sep=space] {Ilio.txt};
    \legend{Current, Kurtulus~\citep{kurtulus2015unsteady}, Ilio \textit{et al.}~\citep{di2018fluid}}
    \end{axis}
    \end{tikzpicture}
    \caption{Aerodynamic efficiency of the NACA 0012 airfoil at $Re = 1000$, showing the lift-to-drag ratio (${C_l}/{C_d}$) as a function of the angle of attack.}
    \label{fig:2Dairfoil_Cl_Cd}
\end{figure}
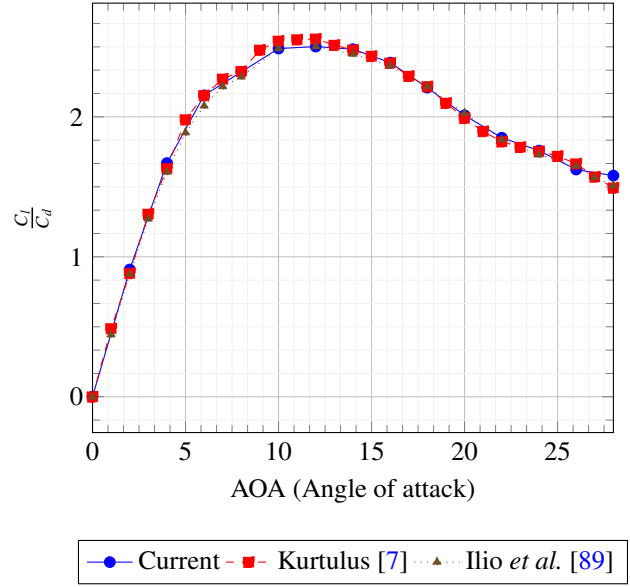

The plot of lift-to-drag ratio ($\frac{C_l}{C_d}$) is shown in \figref{fig:2Dairfoil_Cl_Cd}. The force coefficients are defined as:
\begin{align*}
C_d := \frac{2F_D}{c};C_l := \frac{2F_L}{c},
\end{align*}
where $F_D$, $F_L$ is represent the drag force and lift force applied on the airfoil, respectively, and c is the chord length.
Consistent with the findings of~\citep{kurtulus2015unsteady,di2018fluid}, we conclude that the lift-to-drag ratio ($\frac{C_l}{C_d}$) reaches its maximum between 10 and 12 degrees. This indicates that the airfoil achieves optimal aerodynamic efficiency within this range of angles.

\begin{figure}[t!]
    \centering
    \newcommand{\cropL}{1cm}
    \newcommand{\cropB}{14cm}
    \newcommand{\cropR}{24cm}
    \newcommand{\cropT}{7cm}
    \begin{tikzpicture}
        \newcounter{imageCounter}
        \setcounter{imageCounter}{0}
        \foreach \angle in {0,2,...,28}
        {
            \stepcounter{imageCounter}
            \pgfmathsetmacro{\x}{mod(\theimageCounter - 1, 3) * 4.5} 
            \pgfmathsetmacro{\y}{-int((\theimageCounter - 1) / 3) * 3.5} 

            \node at (\x, \y) {
                \includegraphics[width=4.5cm, trim=\cropL{} \cropB{} \cropR{} \cropT{}, clip]{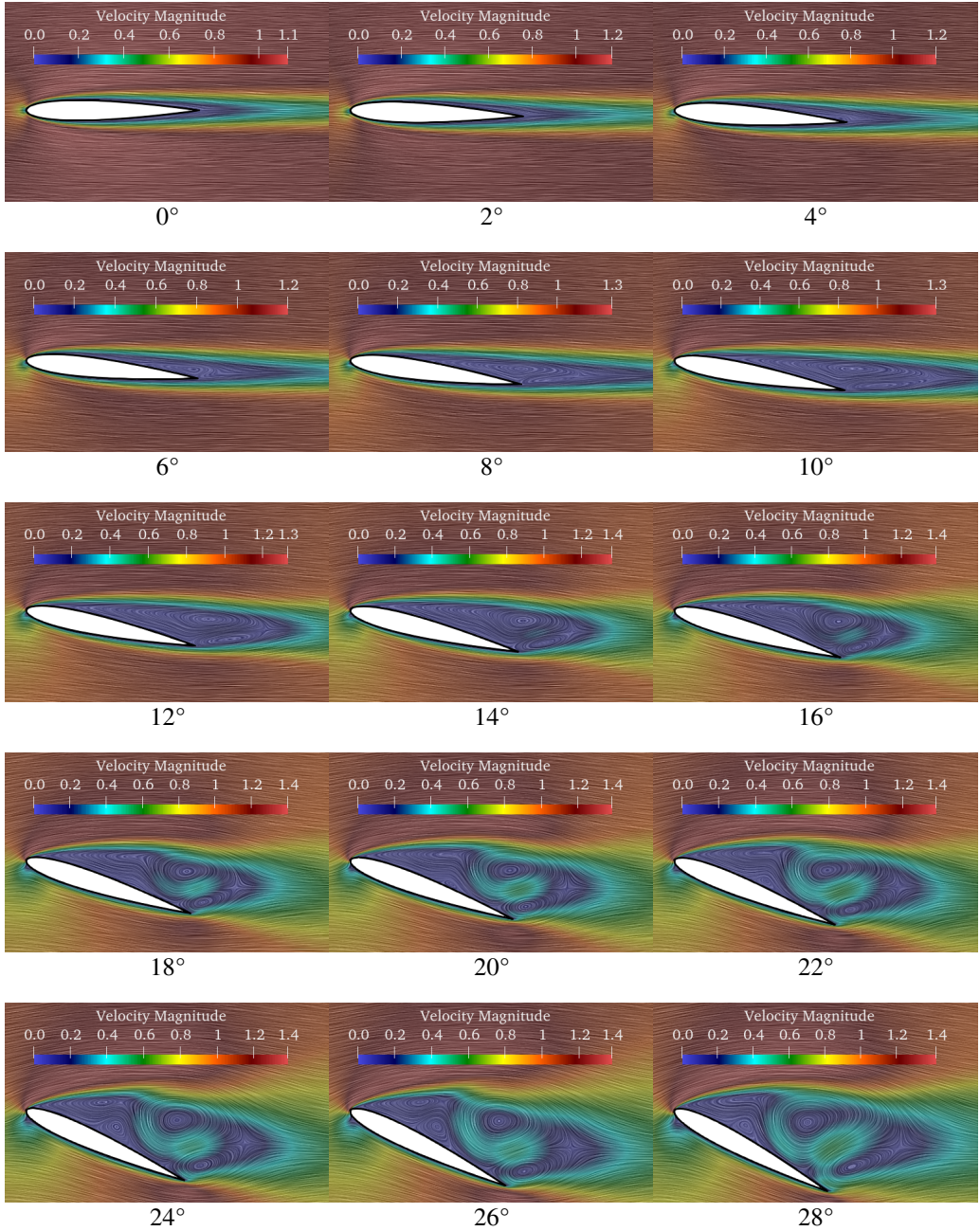}
            };
            \node at (\x, \y - 1.6) {\angle\textdegree}; 
        }
    \end{tikzpicture}
    \caption{Line Integral Convolution (LIC) visualization of flow past the NACA 0012 airfoil at various angles of attack, with contours colored by velocity magnitude.}
    \label{fig:LIC_airfoil_2D}
\end{figure}

\begin{figure}[t!]
    \centering
    \includegraphics[width=0.99\linewidth,trim=0 0 0 0,clip]{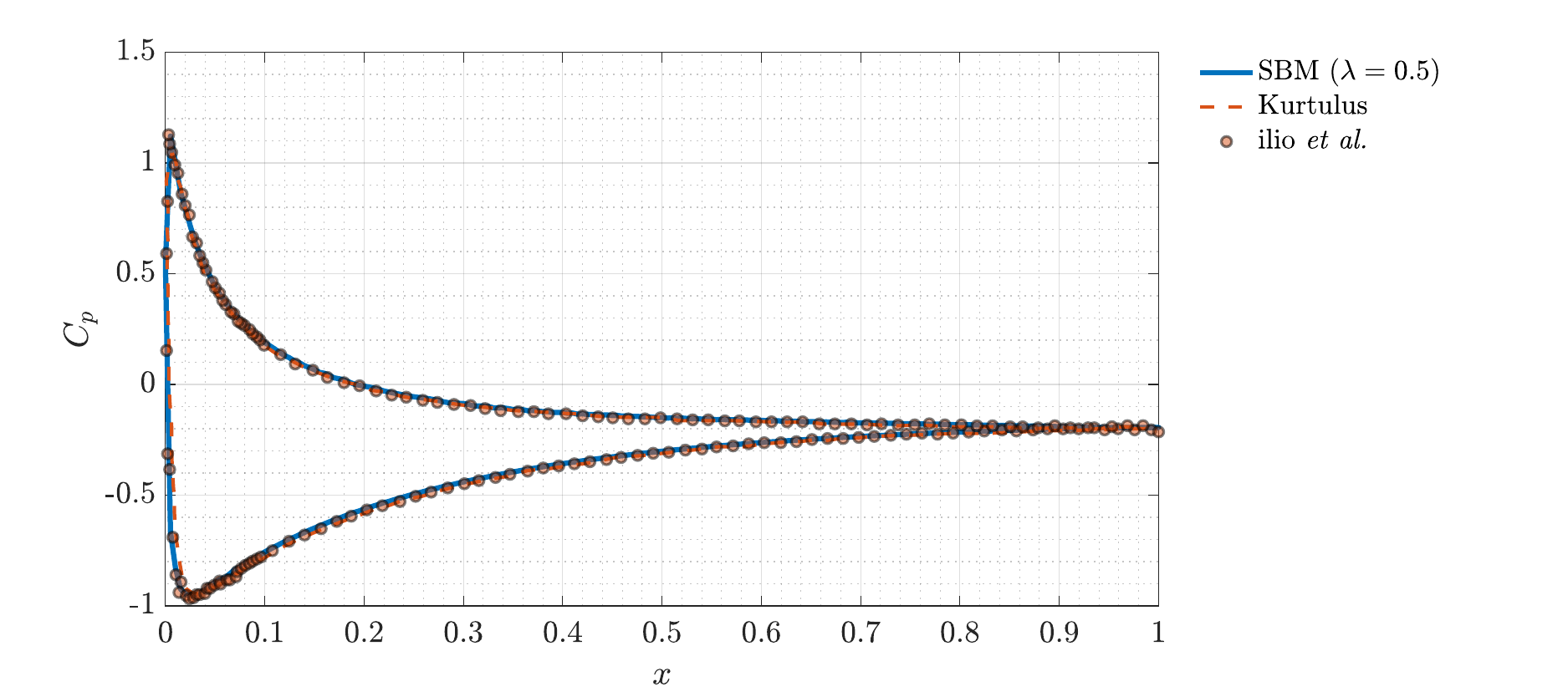}
    \caption{Comparison of the mean pressure coefficient profile over the surface of the NACA 0012 airfoil at an angle of attack of 8\textdegree\ using $\lambda = 0.5$, with values from the literature.}
    \label{fig:PressureOverAirfoil_lambda0p5}
\end{figure}

The flow patterns around the NACA 0012 airfoil are depicted through Line Integral Convolution (LIC) contours in \figref{fig:LIC_airfoil_2D}.
\figref{fig:2Dairfoil_Cl_Cd} demonstrates SBM's ability to accurately compute the lift-to-drag ratio, an integrated value derived by summing the contributions at Gauss points along the true boundary ($\G$). To further highlight SBM's local performance on individual elements, we extracted the pressure coefficient along $\G$ of the airfoil using the methodology outlined in \secref{sec:Eval_quantity}. We selected an angle of attack of 8\textdegree\ to illustrate the pressure coefficient distribution around the airfoil, as presented in other studies~\cite{kurtulus2015unsteady,di2018fluid}. For the pressure calculation, we averaged the values from ten points along the segmented surface region to determine the mean pressure at each position, enabling a detailed assessment of SBM's accuracy in capturing local quantities at the elemental level. Our results for the pressure coefficient with $\lambda = 0.5$ show strong agreement with the literature~\cite{kurtulus2015unsteady,di2018fluid}, as plotted in \figref{fig:PressureOverAirfoil_lambda0p5}.

\begin{figure}[t!]
    \centering
    \begin{subfigure}[b]{0.56\linewidth}
        \includegraphics[width=0.99\linewidth,trim=0 0 0 0,clip]{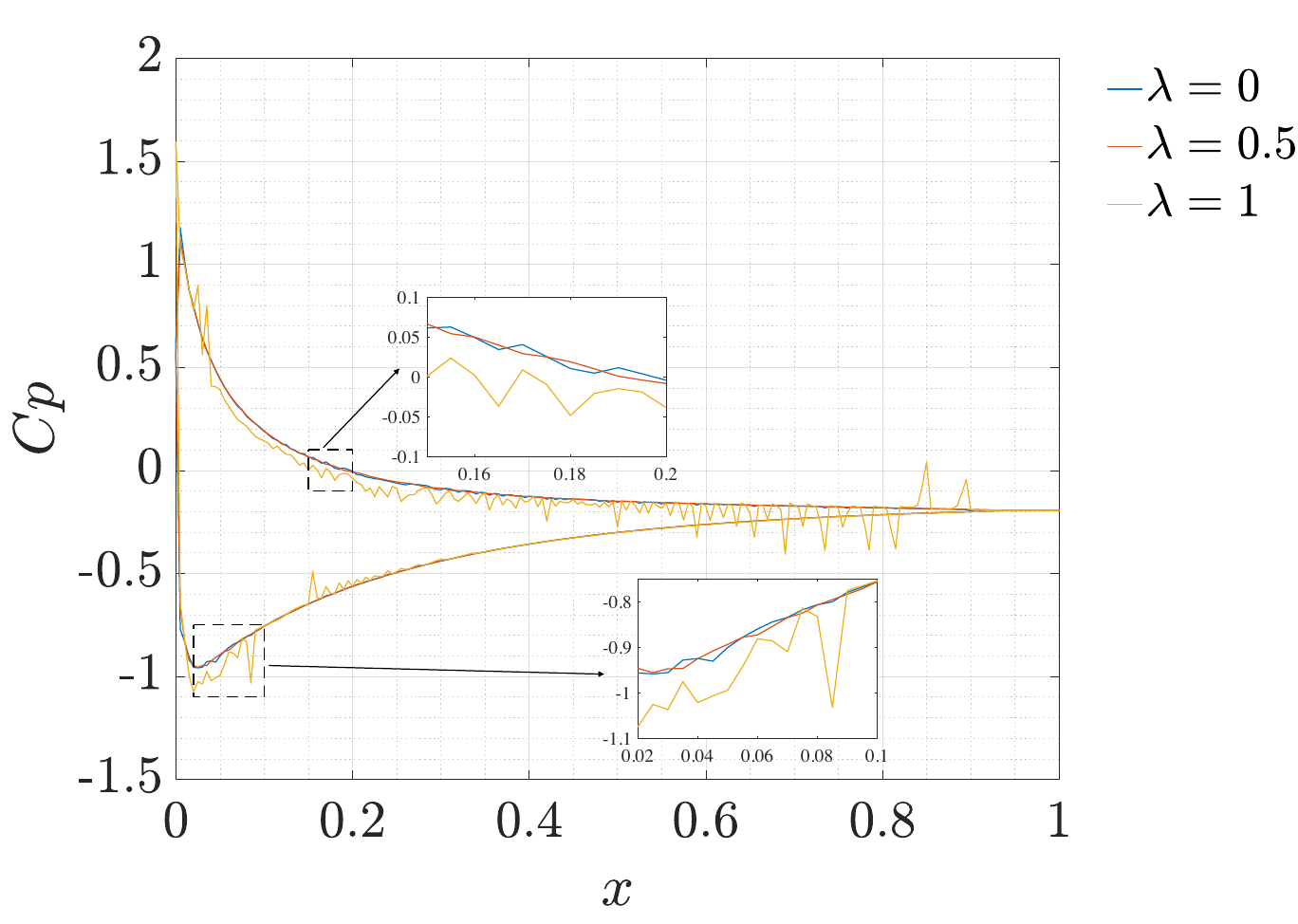}
        \caption{Mean pressure coefficient profile over the surface of the NACA 0012. Diverse $\lambda$ studies are provided to show the benefit of selecting $\lambda = 0.5$.}
        \label{fig:PressureOverAirfoil}
    \end{subfigure}
    \hfill
    \begin{subfigure}[b]{0.40\linewidth}
        \includegraphics[width=1.07\linewidth,trim=0 0 0 0,clip]{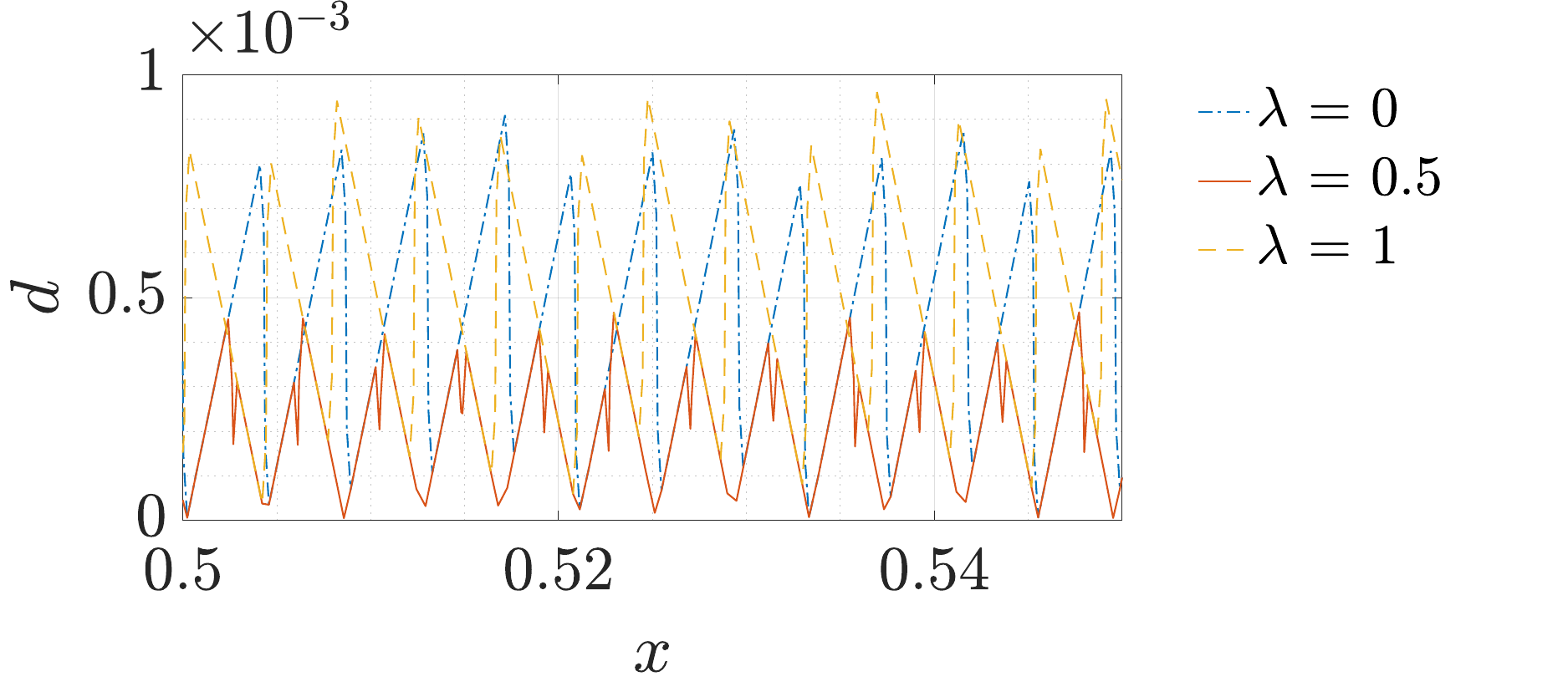}
        \caption{Distance function ($d$) from Gauss points on surrogate boundary to the true boundary.}
        \label{fig:DistanceOverAirfoil}
    \end{subfigure}
    \caption{Mean pressure coefficient profile and distance function variation over the surface of the NACA 0012 airfoil at an angle of attack of 8\textdegree. We present results for three $\lambda$ values. Panel (a) shows that $\lambda = 0.5$ produces a smoother pressure coefficient curve, while panel (b) demonstrates that $\lambda = 0.5$ reduces the variation in the distance between the surrogate boundary and the true boundary.}
    \label{fig:combinedFigures}
\end{figure}

Additionally, we evaluated the effect of varying $\lambda$ values to support the choice of $\lambda = 0.5$ for this study. \figref{fig:PressureOverAirfoil} presents the pressure coefficients over the airfoil for $\lambda = 0$, $\lambda = 0.5$, and $\lambda = 1$. The pressure coefficients obtained with $\lambda = 1$ show significant oscillations along the airfoil's surface, while $\lambda = 0$ results in milder oscillations. In contrast, $\lambda = 0.5$ yields the smoothest pressure profile, making it the most suitable for accurate local assessments. The smoother profile observed for $\lambda = 0.5$ can be attributed to the behavior of the distance function ($d$), which measures the distance from Gauss points on the surrogate boundary ($\tG$) to the true boundary ($\G$). For both $\lambda = 0$ and $\lambda = 1$, the oscillation amplitude of this distance is greater than that for $\lambda = 0.5$, as shown in \figref{fig:DistanceOverAirfoil}. Moreover, it is observed that pressure tends to increase in regions where the distance is larger.

\subsection{Three-dimensional Navier-Stokes}
\subsubsection{Flow past a sphere at $Re = 300$}
\label{subsubsec:Re300}
\begin{figure}[t!]
    \centering
    \begin{subfigure}[b]{0.25\linewidth}
        \includegraphics[width=0.99\linewidth,trim=0 0 0 0,clip]{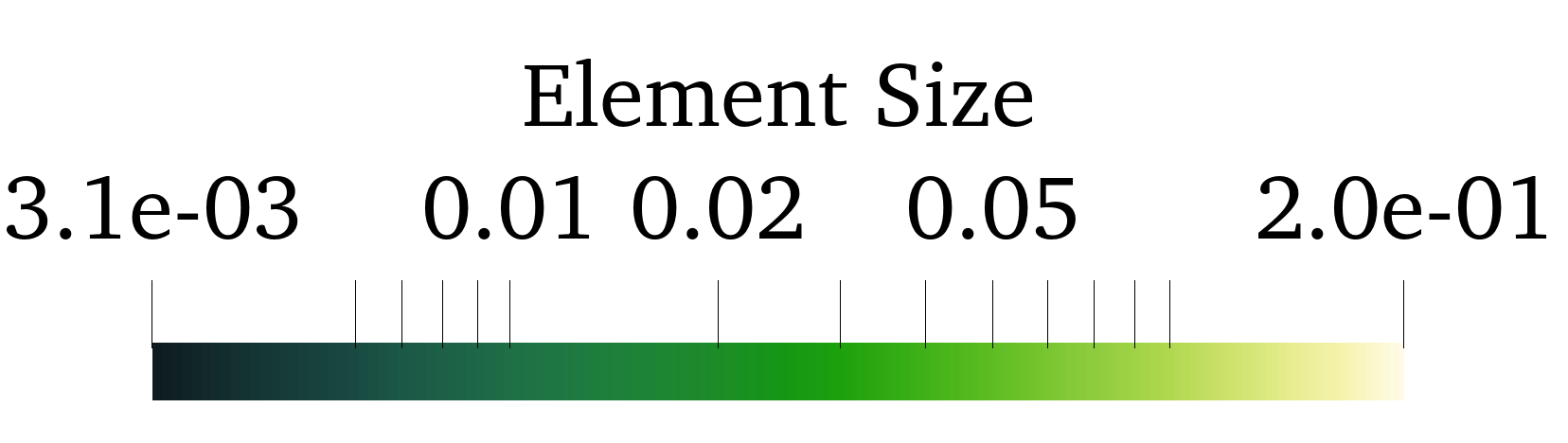}
    \end{subfigure}\\
    \begin{subfigure}[b]{0.99\linewidth}
        \centering
        \includegraphics[width=0.5\linewidth,trim=0 0 0 0,clip]{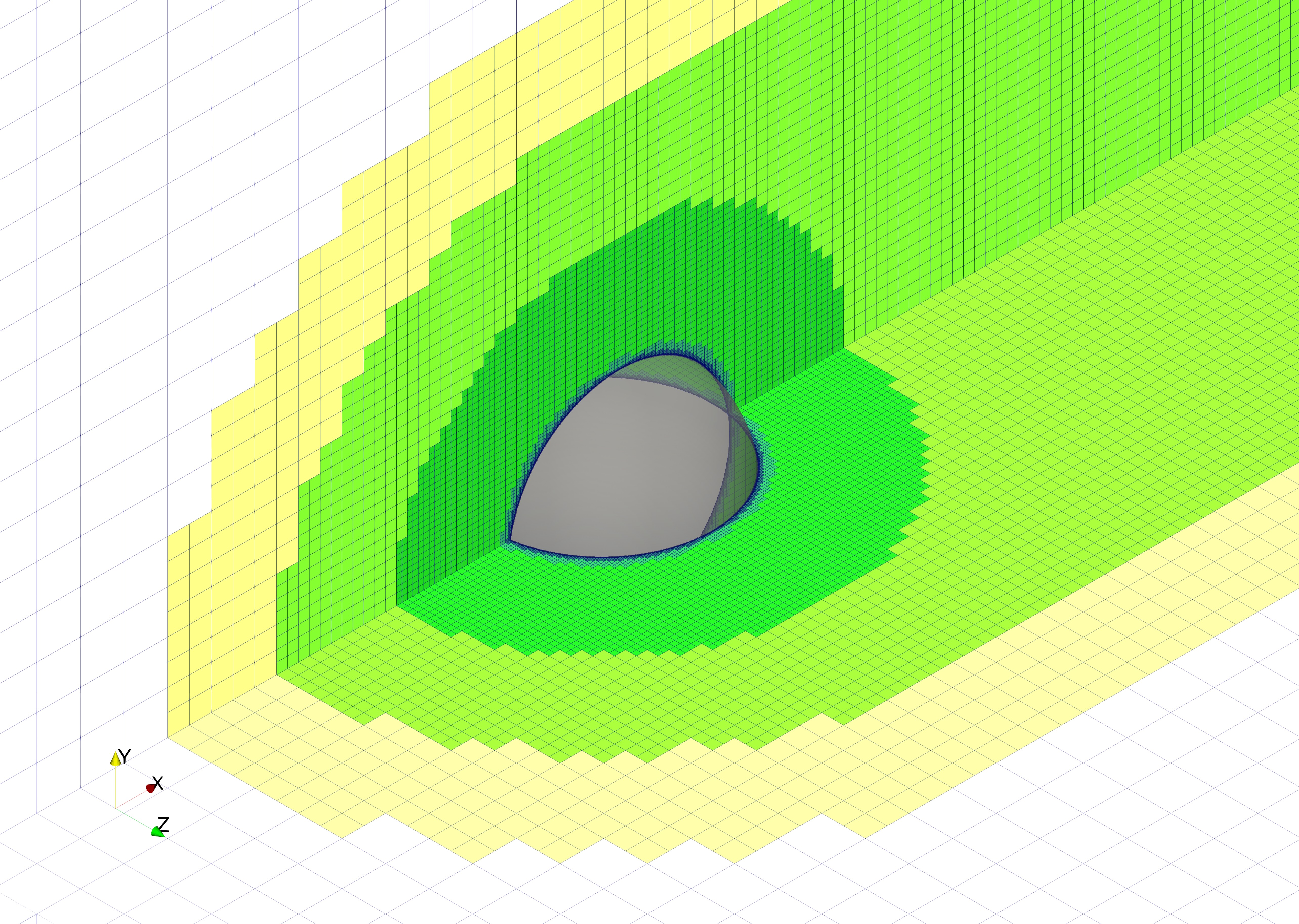}
        \caption{Slices showing the refined region near the sphere and the downstream flow area.}
    \end{subfigure}
    \begin{subfigure}[b]{0.7\linewidth}
        \includegraphics[width=0.99\linewidth,trim=0 0 0 0,clip]{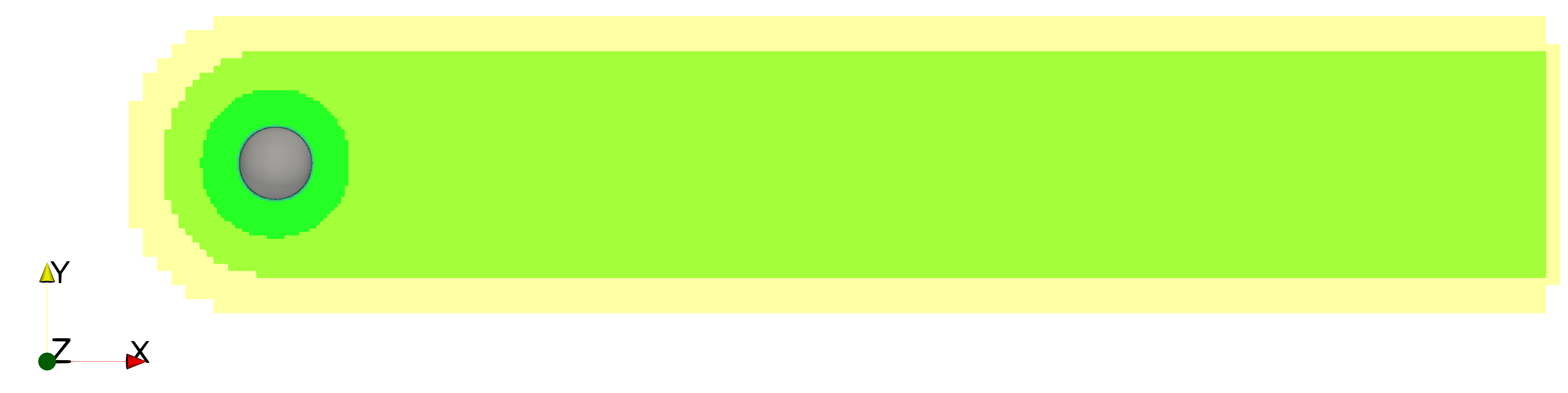}
        \caption{Slice view illustrating the refinement in the region capturing the wake generated by the flow past a sphere.}
    \end{subfigure}
    \hfill
    \begin{subfigure}[b]{0.29\linewidth}
        \includegraphics[width=0.99\linewidth,trim=0 0 0 0,clip]{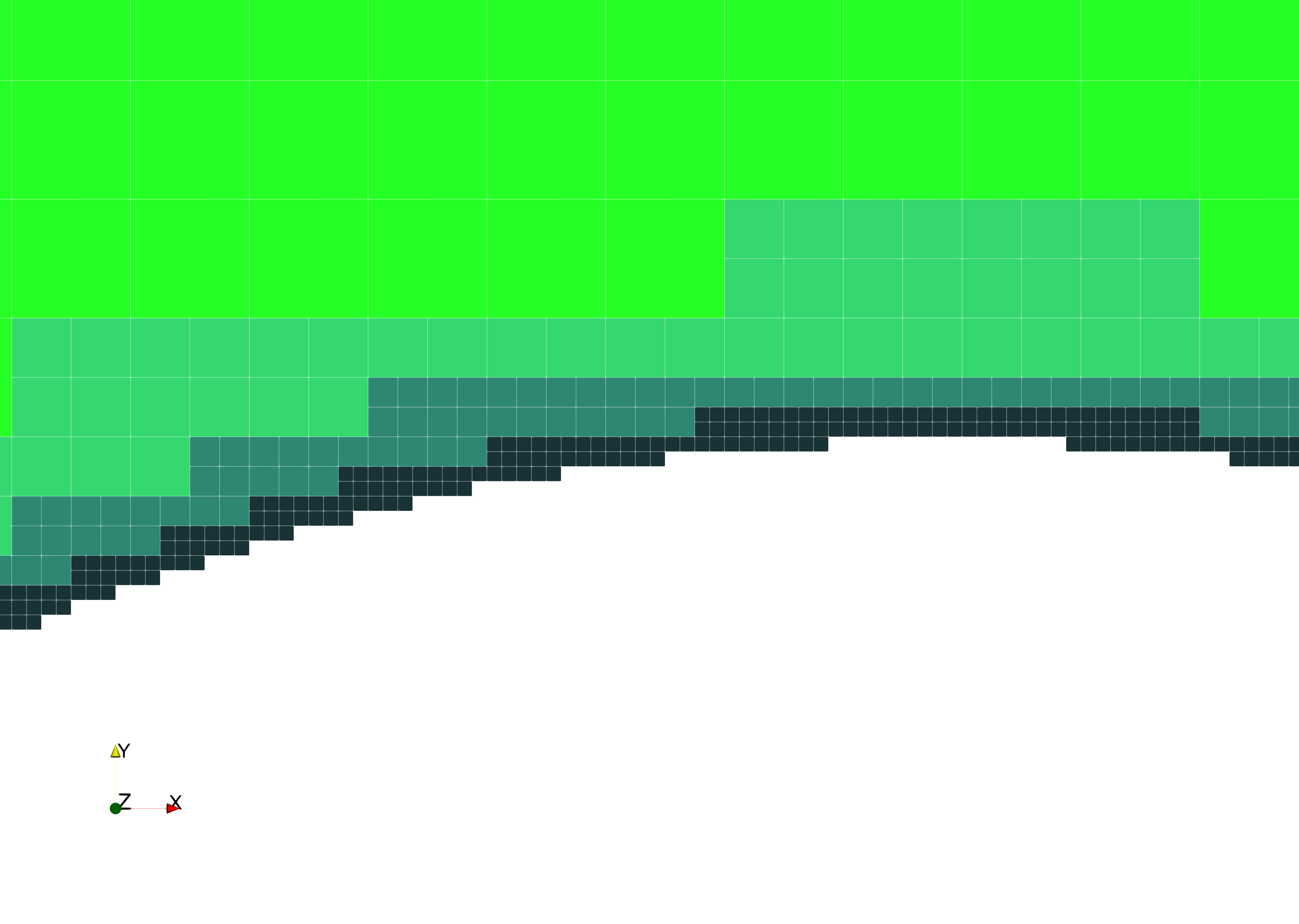}
        \caption{Slice view highlighting the refined region in proximity to the sphere.}
    \end{subfigure}
    \caption{Flow past a sphere: Illustrating the locally refined octree mesh.}
    \label{fig:Mesh_flow_pass_sphere}
\end{figure}

To test fluid flow past a sphere at a $Re$ of 300, we set up a computational domain identical in dimensions to the one described by~\citep{angelidis2016unstructured}: a rectangular region spanning [0, 25] $\times$ [0, 10] $\times$ [0, 10], with a sphere of radius 0.5 centered at the coordinates (3, 5, 5). Our mesh refinement strategy closely follows that of~\citep{angelidis2016unstructured}, involving several layers of progressively coarser mesh surrounding the sphere and along the flow path. Specifically, we utilize three concentric spheres centered at (3, 5, 5), with the innermost sphere having a radius of 1 and a refinement level of 10 (yielding an element size of $\frac{25}{2^{10}}$), the next sphere with a radius of 1.5 and a refinement level of 9 ($\frac{25}{2^9}$ element size), and the outer sphere with a radius of 2 and a refinement level of 8 ($\frac{25}{2^8}$ element size). Additionally, we define two cylindrical refinement regions aligned with the flow direction from (3, 5, 5) to (20.5, 5, 5), the first cylinder with a radius of 1.5 and refinement level 9 ($\frac{25}{2^9}$ element size), and the second with a radius of 2 and refinement level 8 ($\frac{25}{2^8}$ element size). Close to the sphere, to ensure detailed capture of the flow dynamics, we achieve a finer mesh with a refinement level of 13, resulting in an element size of $\frac{25}{2^{13}}$ and approximately three elements covering the sphere's boundary. The mesh refinement and mesh size of this problem is shown in \figref{fig:Mesh_flow_pass_sphere}. The non-dimensional timestep we use to perform the simulation is 0.025. For the outlet, we apply backflow stabilization, while all other walls are subjected to a uniform non-dimensional freestream velocity of (1, 0, 0).

\begin{figure}[t!]
    \centering
    \begin{subfigure}[b]{0.5\linewidth}
        \includegraphics[width=0.99\linewidth,trim=0 0 0 0,clip]{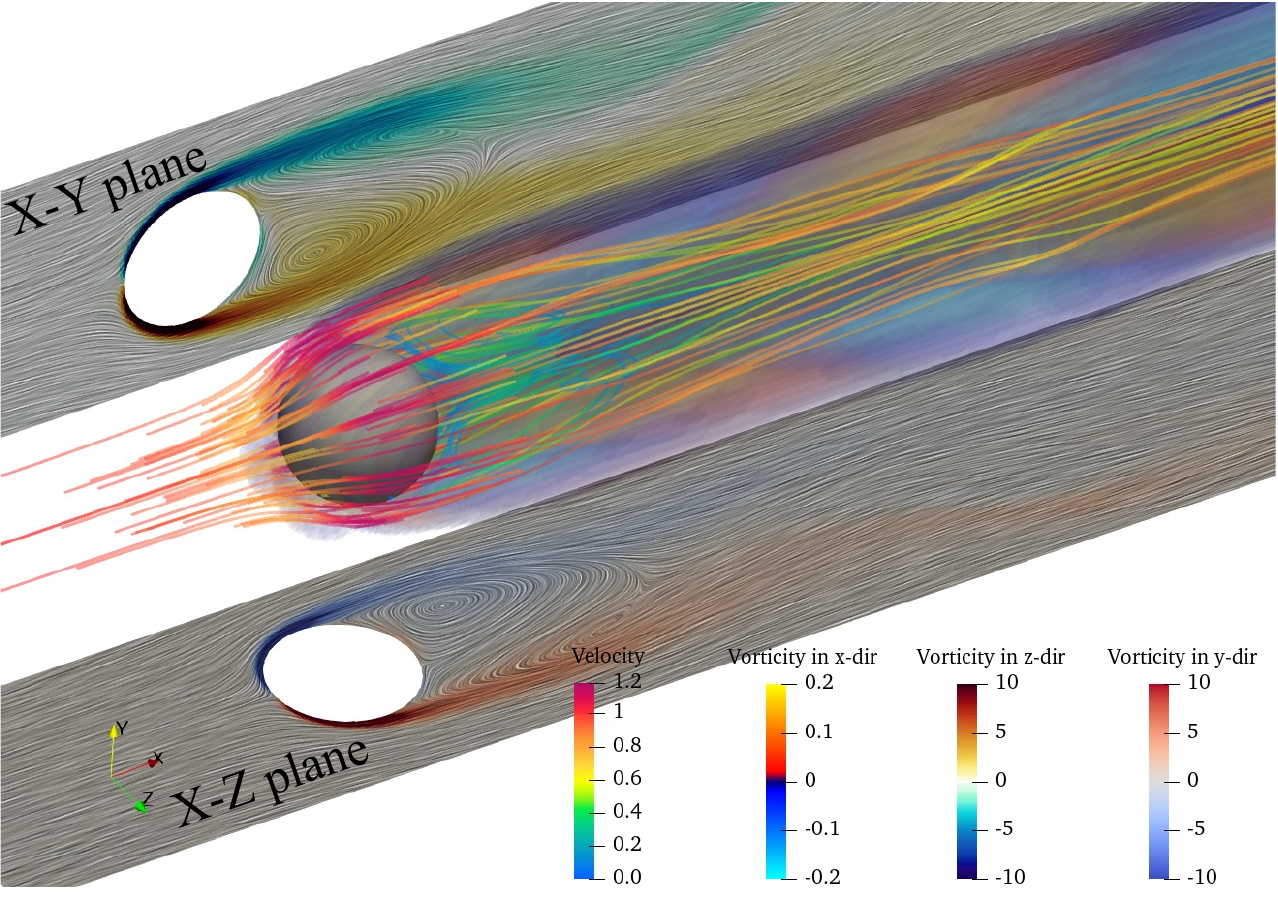}
    \end{subfigure}
    \caption{Visualization of the instantaneous flow with a $Re$ of 300. The vorticity component, $\omega_{z}$, is calculated and visualized in the region downstream of the sphere. The component $\omega_{y}$ is projected onto the $x$-$z$ plane and depicted using LIC contours. Similarly, $\omega_{z}$ is projected onto the $x$-$y$ plane and illustrated with LIC contours. Additionally, streamlines around the sphere are visualized and colored according to the magnitude of velocity, offering a comprehensive view of the flow dynamics.}
    \label{fig:flow_visualization}
\end{figure}

\begin{figure}[t!]
    \centering
    \begin{subfigure}{0.45\textwidth}
    \centering
    \includegraphics[width=0.8\linewidth,trim=0 180 0 200,clip]{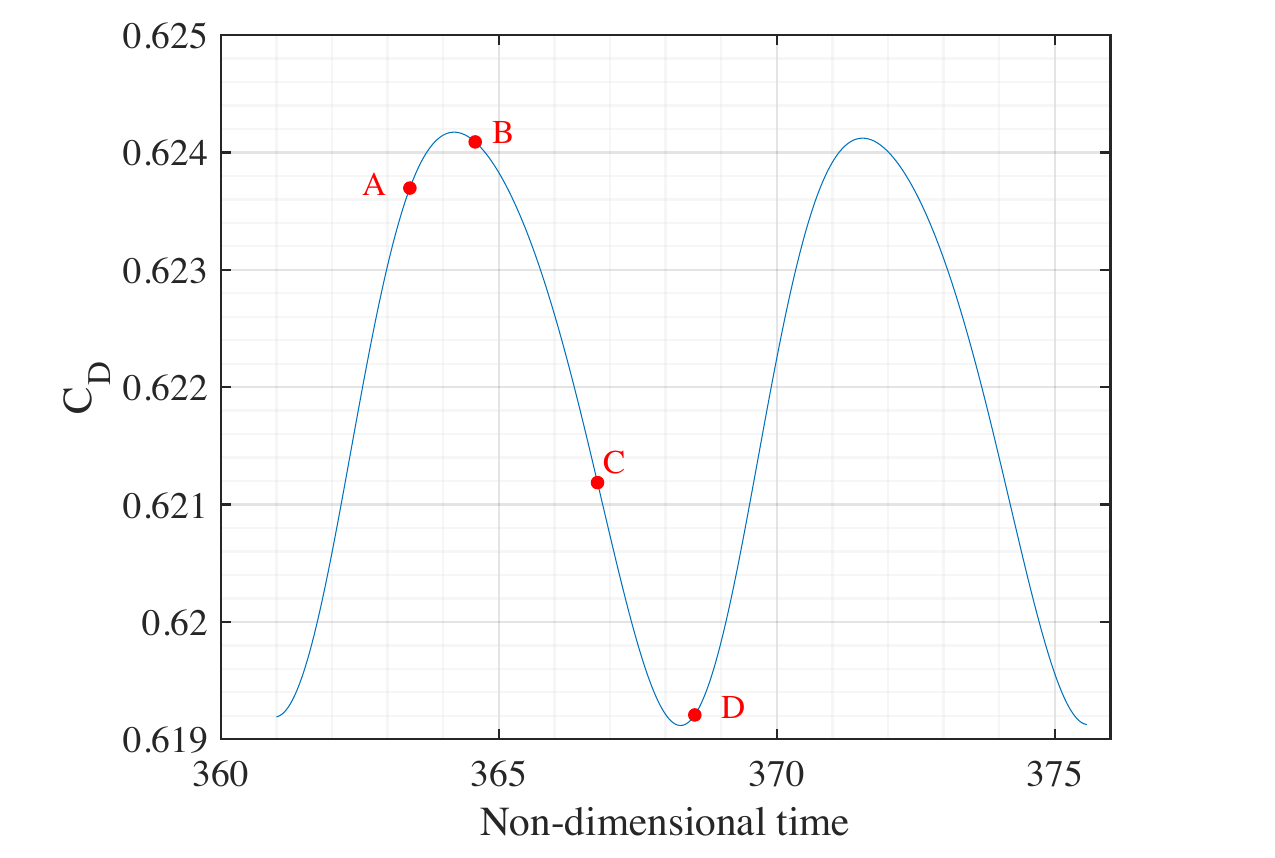}
    \caption{Time history for the drag coefficient for $Re$ = 300.}
    \label{fig:Cd_history}
    \end{subfigure}
\qquad
    \begin{subfigure}{0.45\textwidth}
    \centering
    \includegraphics[width=0.99\linewidth]{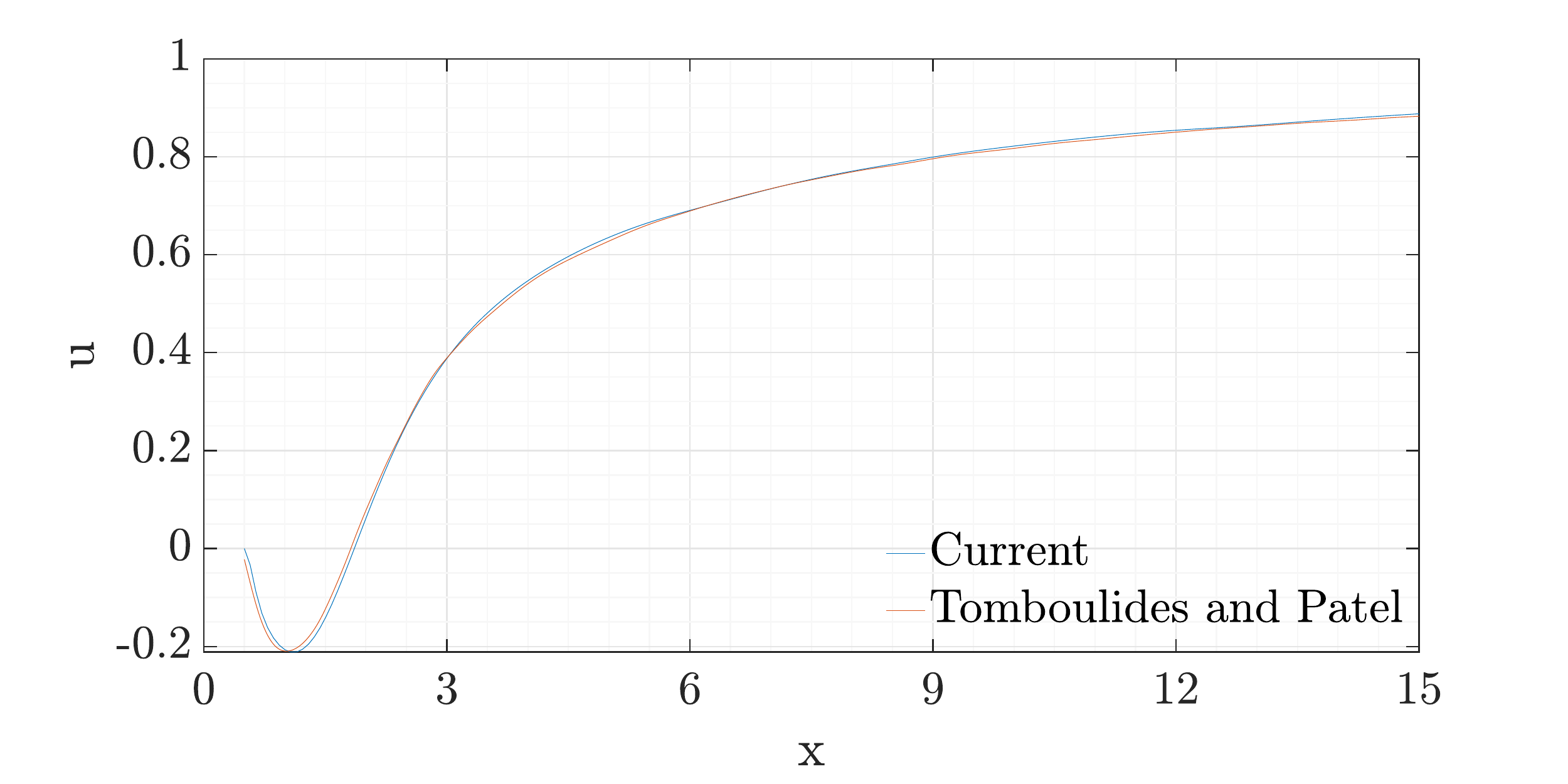}
    \caption{Time-averaged streamwise $x$-component of the velocity profile.}
    \label{fig:TimeAverage_Re300}
    \end{subfigure}
\\
    \begin{subfigure}{0.45\textwidth}
    \centering
    \includegraphics[width=0.99\linewidth]{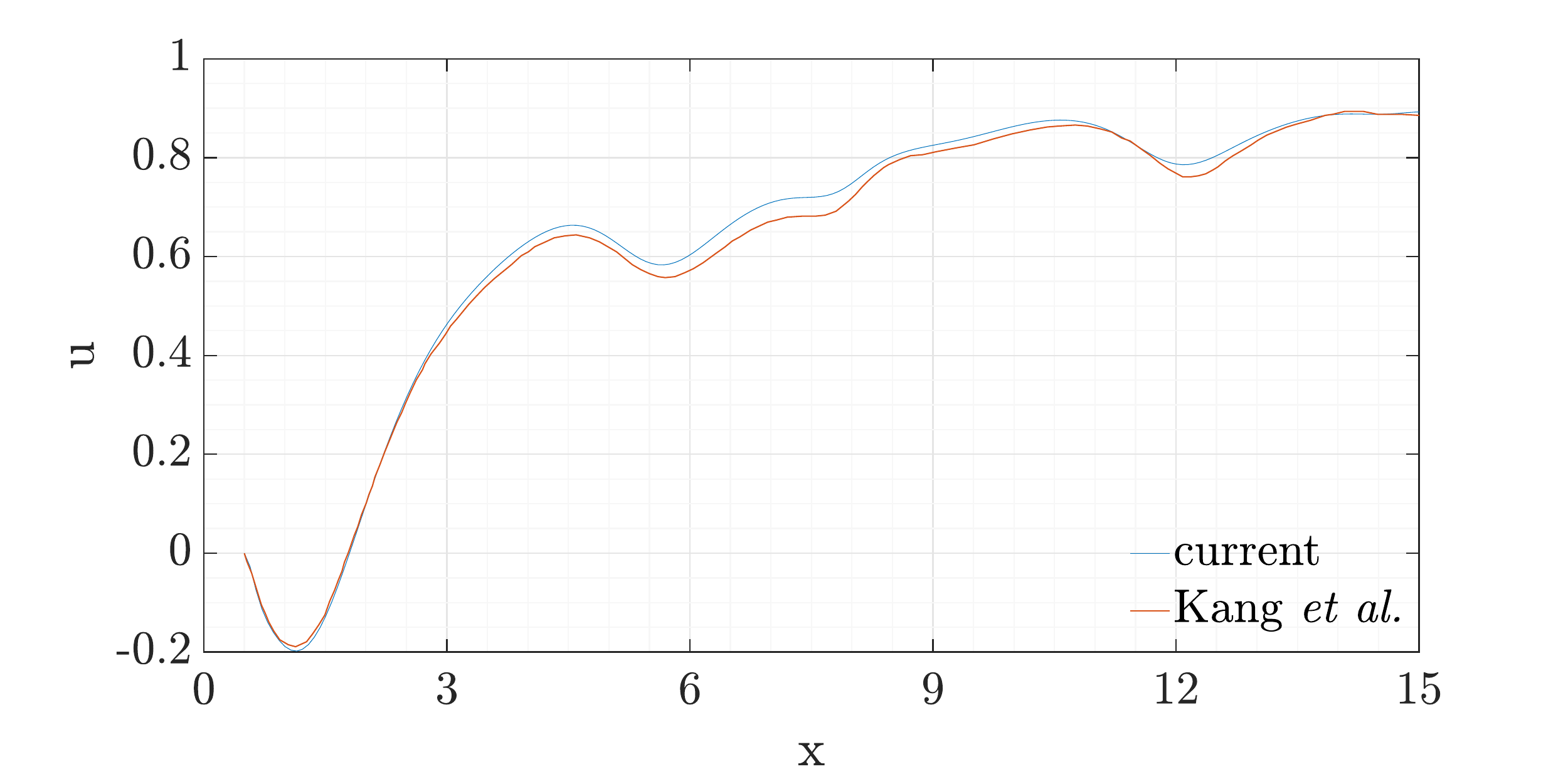}
    \caption{Streamwise $x$-component of the velocity profile at time point A.}
    \label{fig:A}
    \end{subfigure}%
\qquad
    \begin{subfigure}{0.45\textwidth}
    \centering
    \includegraphics[width=0.99\linewidth]{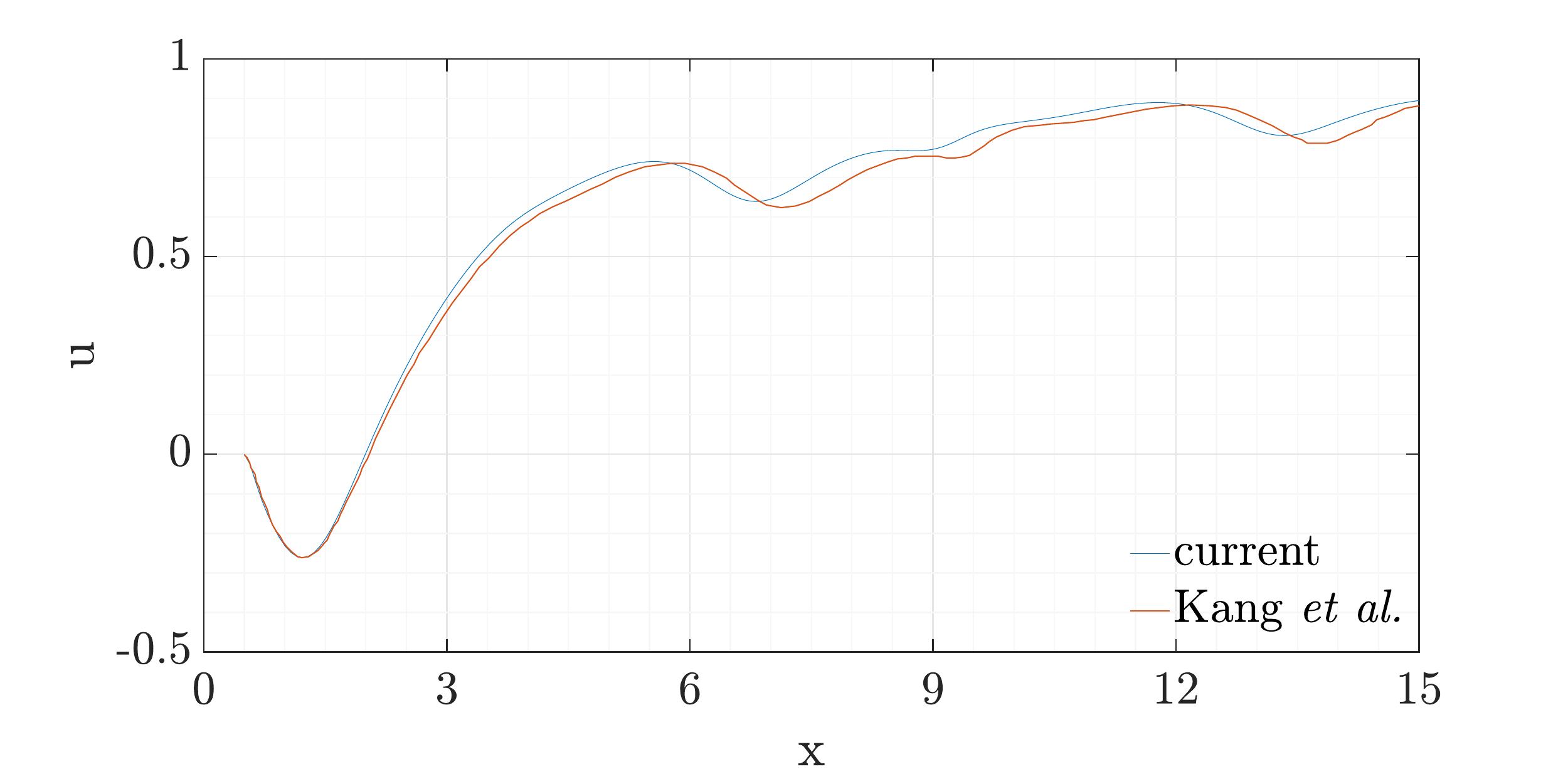}
    \caption{Streamwise $x$-component of the velocity profile at time point B.}
    \label{fig:B}
    \end{subfigure}%
\\
    \begin{subfigure}{0.45\textwidth}
    \centering
    \includegraphics[width=0.99\linewidth]{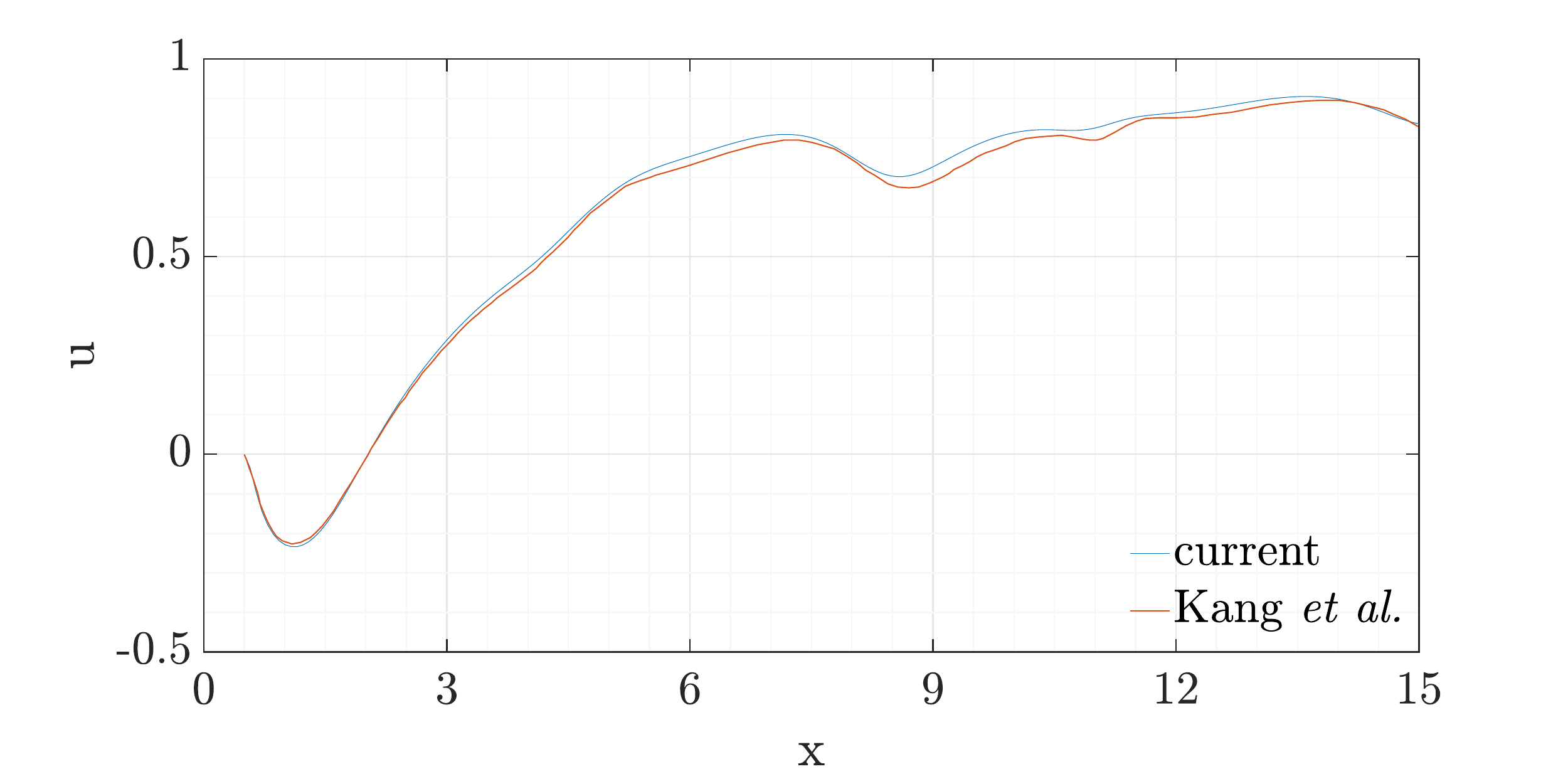}
    \caption{Streamwise $x$-component of the velocity profile at time point C.}
    \label{fig:C}
    \end{subfigure}%
\qquad
    \begin{subfigure}{0.45\textwidth}
    \centering
    \includegraphics[width=0.99\linewidth]{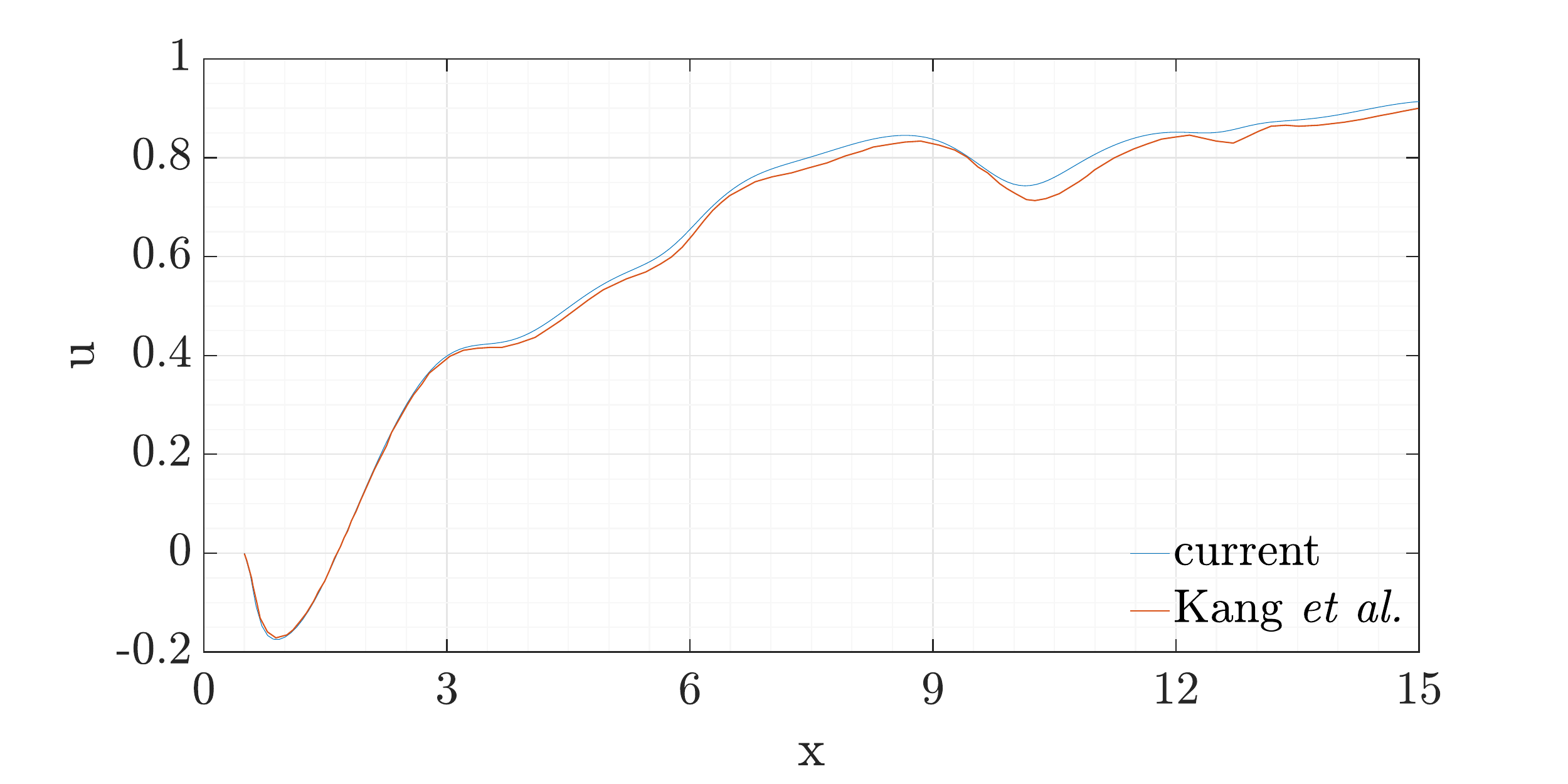}
    \caption{Streamwise $x$-component of the velocity profile at time point D.}
    \label{fig:D}
    \end{subfigure}%
    \caption{The streamwise velocity profile and drag history of flow past a sphere at $Re$ = 300. Panel(a) shows the time history of the drag coefficient. A, B, C, and D in Panel (a) are the four times we show instaneous streamwise velocity profile in Panel (c), (d), (e) and compare with~\cite{kang2021variational}, and (f). Panel (b) illustrate the time-averaged streamwise velocity comparing with~\cite{tomboulides_orszag_2000}.}
    \label{fig:streamwiest_velocity}
\end{figure}

Compared with the literature, the computed time-averaged drag coefficient and Strouhal number are shown in the \tabref{table:Re300_Cd_St}, and match well. \figref{fig:Cd_history} shows the history of the drag coefficient changing with the non-dimensional time. Following \citet{kang2021variational}, we selected four specific time points to compare the streamwise velocity profiles in the $x$-direction along the line extending from the center of the sphere with \citep{kang2021variational} in \figref{fig:A}, \figref{fig:B}, \figref{fig:C}, and \figref{fig:D}. We also compare our time-averaged velocity profile with \citet{tomboulides_orszag_2000} in \figref{fig:TimeAverage_Re300}. Note that the variable x, as shown in \figref{fig:TimeAverage_Re300}, \figref{fig:A}, \figref{fig:B}, \figref{fig:C}, and \figref{fig:D} is measured downstream from the centroid of the sphere, which aligns with that provided in the works of \citep{tomboulides_orszag_2000,kang2021variational}. The instantaneous and time-averaged velocity profiles demonstrate that our framework is in good agreement with results from literature.
The instantaneous flow visualization of streamline and vorticity is shown in \figref{fig:flow_visualization}.

\begin{table}[ht]
    \centering
    \caption{Summary of drag coefficients and Strouhal numbers from various studies for flow past a sphere at Re = 300.}
    \begin{tabular}{@{}P{5cm}P{1.5cm}P{1.5cm}@{}}
        \toprule
        \textbf{Study} & $C_d$ & \textit{St} \\
        \midrule
        Roos and Willmarth \citep{roos1971some} (interpolated experiment value) & 0.629 & - \\
        Le Clair \textit{et al.} \citep{le1970numerical} & 0.632 & - \\
        Johnson and Patel \citep{Johnson1999FlowPA}         & 0.656 & 0.137 \\
        Marella \textit{et al.} \citep{Marella2005}             & 0.621 & 0.133 \\
        Vanella \textit{et al.} \citep{vanella2010direct}            & 0.634 & 0.132 \\
        Wang and Zhang  \citep{wang2011immersed}           & 0.680 & 0.135 \\
        Angelidis \textit{et al.} \citep{angelidis2016unstructured}           & 0.665 & 0.132 \\
        Kang \textit{et al.}  \citep{kang2021variational}              & 0.663 & 0.134 \\
        Current                    & 0.622 & 0.137 \\
        \bottomrule
    \end{tabular}
    \label{table:Re300_Cd_St}
\end{table}

\clearpage

\subsubsection{Flow past a sphere at $Re = 3700$}\label{subsubsec:Re3700}
To evaluate the accuracy of the current framework, we increase the $Re$ to 3700. The boundary conditions we apply here are identical to those in \secref{subsubsec:Re300}. We expand the solution domain from \secref{subsubsec:Re300} to [0, 25] $\times$ [0, 15.625] $\times$ [0, 15.625], and position the sphere at (6, 7.8125, 7.8125). The mesh refinement strategy remains essentially the same as in \secref{subsubsec:Re300}. We employ three concentric spheres centered at (6, 7.8125, 7.8125) with refinement levels of 8, 9, and 10, respectively. Additionally, two cylindrical refinement regions are applied beyond the sphere in the wake area, with refinement levels of 8 and 9. Close to the sphere, we utilize a finer mesh with a refinement level of 13. The non-dimensional timestep used for performing the simulation is 0.01.

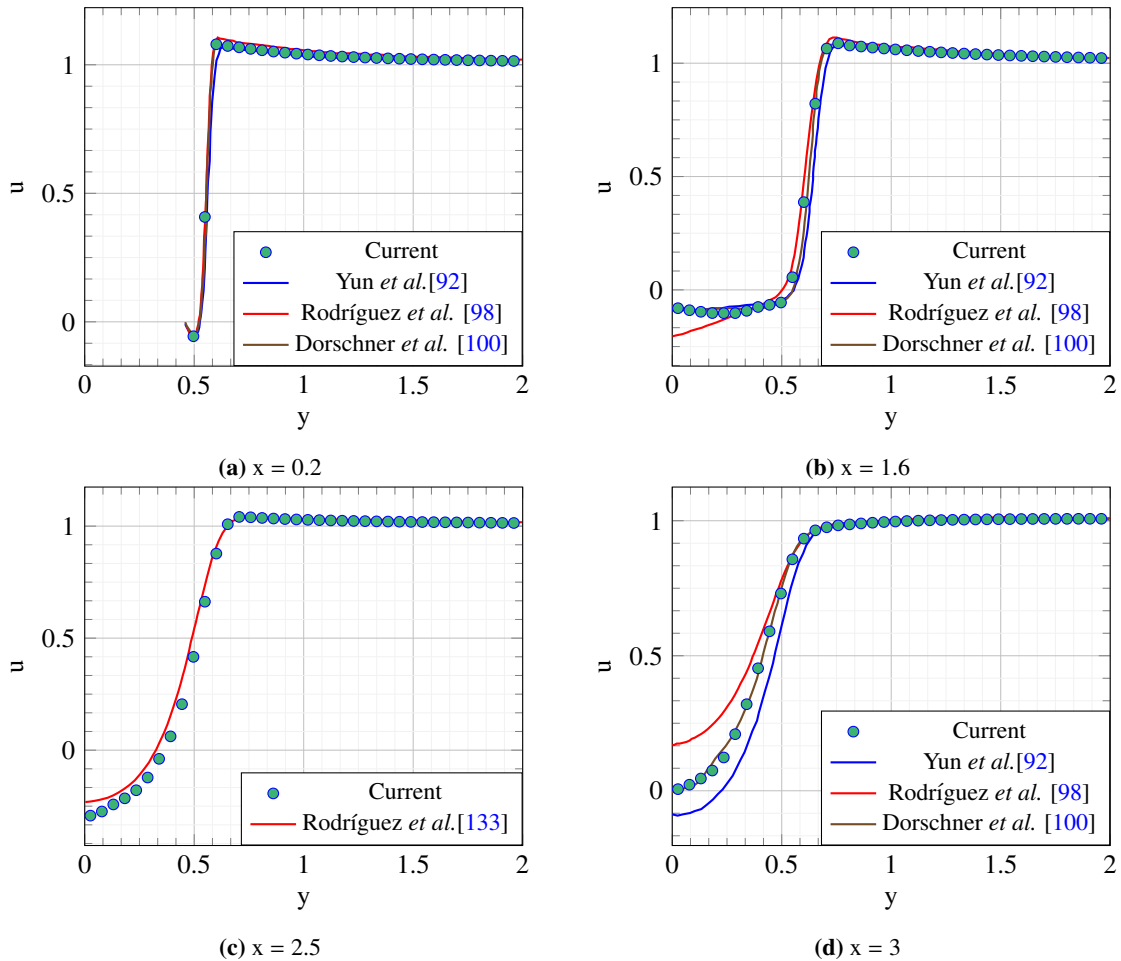
\begin{figure}[b!]
\centering
\begin{subfigure}{.47\linewidth}
\centering
\begin{tikzpicture}
    \begin{axis}[
        width = 0.95\linewidth,
        xlabel={y},
        ylabel={u},
        grid=both,
        grid style={line width=.1pt, draw=gray!10},
        major grid style={line width=.2pt, draw=gray!50},
        minor tick num=5,
        xmin=0, xmax=2, 
        legend style={
            at={(1,0)},           
            anchor=south east,    
            font=\small
        },
        legend entries={Current, Yun \textit{et al.}\citep{yun2006vortical}, Rodr{\'\i}guez \textit{et al.} \citep{rodriguez2011direct}, Dorschner \textit{et al.} \citep{dorschner2016grid}}
    ]
    \addplot +[only marks, mark=*, mark options={fill=lightprofgreen}] table [
        col sep=comma,
        x index=0, 
        y index=1  
    ] {current_0p2.csv}; 
    \addplot [no markers,color=blue, thick] table [
        col sep=comma,
        x index=0, 
        y index=1  
    ] {Yun_0p2.csv}; 
    \addplot [no markers,color=red, thick] table [
        col sep=comma,
        x index=0, 
        y index=1  
    ] {Re3700_0p2.csv}; 
    \addplot [no markers,color=brown!60!black, thick] table [
        col sep=comma,
        x index=0, 
        y index=1  
    ] {Dorschner_0p2.csv}; 
    \end{axis}
\end{tikzpicture}
\caption{x = 0.2}
\label{subfig:0p2}
\end{subfigure}%
\begin{subfigure}{.47\linewidth}
\centering
\begin{tikzpicture}
    \begin{axis}[
        width = 0.95\linewidth,
        xlabel={y},
        ylabel={u},
        grid=both,
        grid style={line width=.1pt, draw=gray!10},
        major grid style={line width=.2pt, draw=gray!50},
        minor tick num=5,
        xmin=0, xmax=2, 
        legend style={
            at={(1,0)},           
            anchor=south east,    
            font=\small
        },
        legend entries={Current, Yun \textit{et al.}\citep{yun2006vortical}, Rodr{\'\i}guez \textit{et al.} \citep{rodriguez2011direct}, Dorschner \textit{et al.} \citep{dorschner2016grid}}
    ]
    \addplot +[only marks, mark=*, mark options={fill=lightprofgreen}] table [
        col sep=comma,
        x index=0, 
        y index=1  
    ] {current_1p6.csv}; 
    \addplot [no markers,color=blue, thick] table [
        col sep=comma,
        x index=0, 
        y index=1  
    ] {Yun_1p6.csv}; 
    \addplot [no markers,color=red, thick] table [
        col sep=comma,
        x index=0, 
        y index=1  
    ] {Re3700_1p6.csv}; 
    \addplot [no markers,color=brown!60!black, thick] table [
        col sep=comma,
        x index=0, 
        y index=1  
    ] {Dorschner_1p6.csv}; 
    \end{axis}
\end{tikzpicture}
\caption{x = 1.6}
\label{subfig:1p6}
\end{subfigure}
\begin{subfigure}{.47\linewidth}
\centering
\begin{tikzpicture}
    \begin{axis}[
        width = 0.95\linewidth,
        xlabel={y},
        ylabel={u},
        grid=both,
        grid style={line width=.1pt, draw=gray!10},
        major grid style={line width=.2pt, draw=gray!50},
        minor tick num=5,
        xmin=0, xmax=2, 
        legend style={
            at={(1,0)},           
            anchor=south east,    
            font=\small
        },
        legend entries={Current, Rodr{\'\i}guez \textit{et al.}\citep{rodriguez2013flow}}
    ]
    \addplot +[only marks, mark=*, mark options={fill=lightprofgreen}] table [
        col sep=comma,
        x index=0, 
        y index=1  
    ] {current_2p5.csv}; 
    \addplot [no markers,color=red, thick] table [
        col sep=comma,
        x index=0, 
        y index=1  
    ] {Re3700_2p5.csv}; 
    \end{axis}
\end{tikzpicture}
\caption{x = 2.5}
\label{subfig:2p5}
\end{subfigure}%
\begin{subfigure}{.47\linewidth}
\centering
\begin{tikzpicture}
    \begin{axis}[
        width = 0.95\linewidth,
        xlabel={y},
        ylabel={u},
        grid=both,
        grid style={line width=.1pt, draw=gray!10},
        major grid style={line width=.2pt, draw=gray!50},
        minor tick num=5,
        xmin=0, xmax=2, 
        legend style={
            at={(1,0)},           
            anchor=south east,    
            font=\small
        },
        legend entries={Current, Yun \textit{et al.}\citep{yun2006vortical}, Rodr{\'\i}guez \textit{et al.} \citep{rodriguez2011direct}, Dorschner \textit{et al.} \citep{dorschner2016grid}}
    ]
    \addplot +[only marks, mark=*, mark options={fill=lightprofgreen}] table [
        col sep=comma,
        x index=0, 
        y index=1  
    ] {current_3p0.csv}; 
    \addplot [no markers,color=blue, thick] table [
        col sep=comma,
        x index=0, 
        y index=1  
    ] {Yun_3p0.csv}; 
    \addplot [no markers,color=red, thick] table [
        col sep=comma,
        x index=0, 
        y index=1  
    ] {Re3700_3p0.csv}; 
    \addplot [no markers,color=brown!60!black, thick] table [
        col sep=comma,
        x index=0, 
        y index=1  
    ] {Dorschner_3p0.csv}; 
    \end{axis}
  \end{tikzpicture}
  \caption{x = 3}
  \label{subfig:3p0}
\end{subfigure}
\caption{Flow past a sphere at $Re = 3700$: Mean velocity profiles in the $x$-direction at four different locations, compared with data from the literature.}
\label{fig:Mean_velocity_Re3700}
\end{figure}

\begin{figure}[t!]
    \centering
    \begin{subfigure}[b]{0.99\linewidth}
        \includegraphics[width=0.9\linewidth,trim=0 0 0 0,clip]{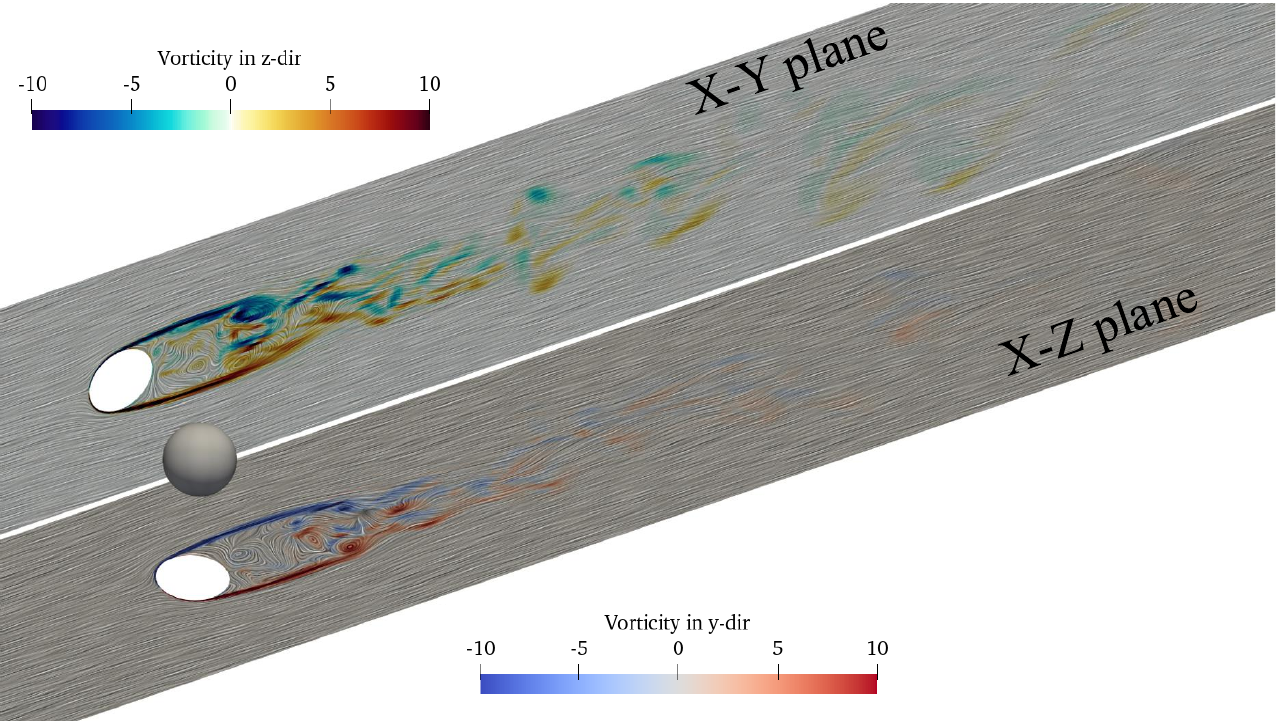}
        \caption{The flow past the sphere on the x-z plane is projected to display $\omega_{y}$. Similarly, the flow around the sphere on the x-y plane is projected to show $\omega_{z}$.}
        \label{fig:flow_visualization_Re3700_vorticity}
    \end{subfigure}
    \begin{subfigure}{0.99\textwidth}
        \centering
        \includegraphics[width=0.9\textwidth,trim=0 1000 0 0,clip]{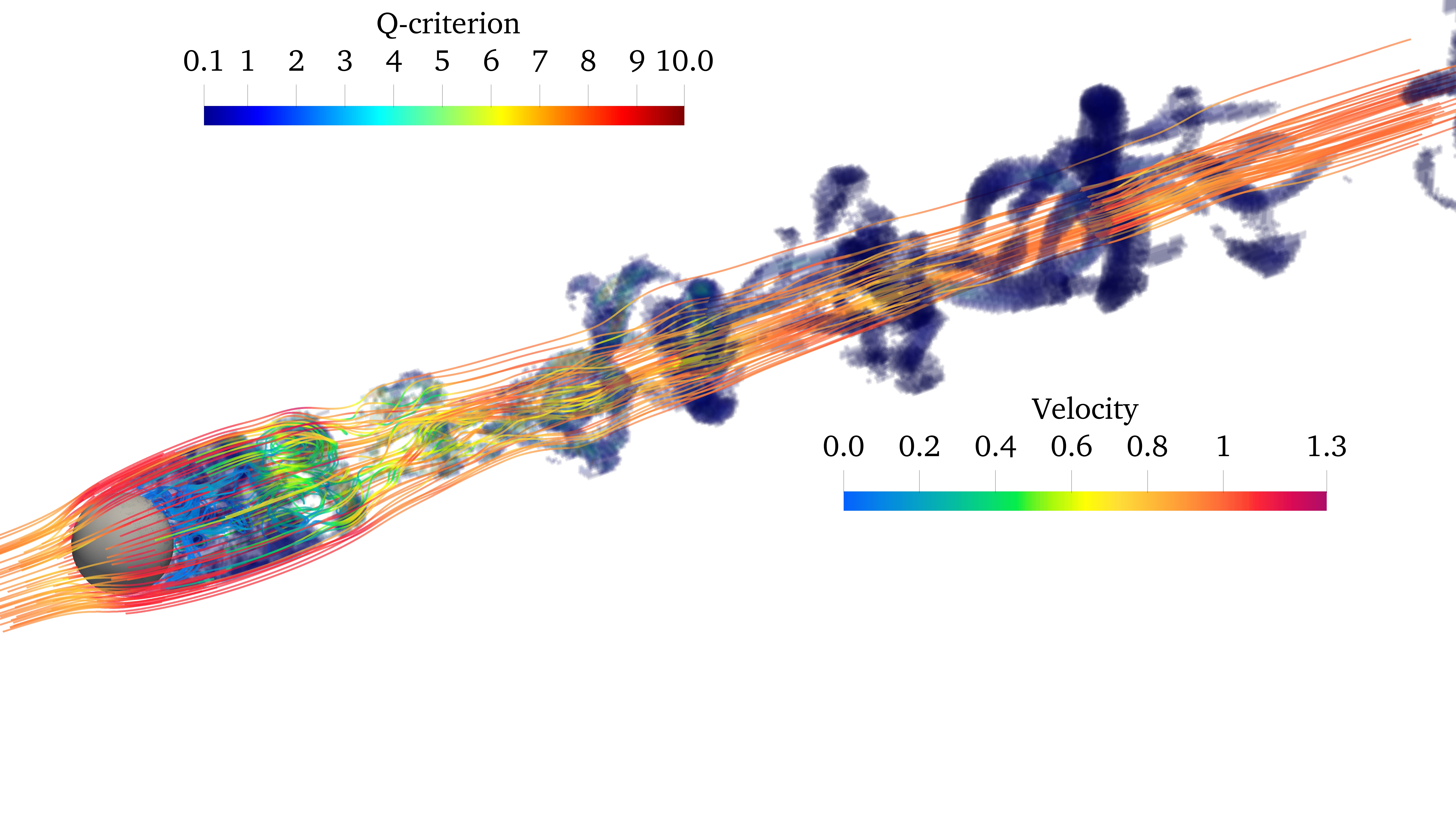}
        \caption{We depicted the Q-criteria, ranging from 0.1 to 10, with streamlines. The streamlines are colored according to the magnitude of the flow velocity.}
        \label{fig:flow_visualization_Re3700_qcriteria}
    \end{subfigure}
    \caption{Instantaneous flow visualization for flow past a sphere at $Re = 3700$, showcasing the vorticity distribution (top) and Q-criteria with streamlines (bottom).}
    \label{fig:flow_visualization_Re3700}
\end{figure}

The mean velocity profiles at different downstream distances from the sphere's center (denoted as x) are compared with the literature in \figref{fig:Mean_velocity_Re3700}. Our results show good agreement with the literature. To evaluate the accuracy of the drag force captured at the sphere's boundary, we compare the drag coefficient obtained from our simulations with those reported in the literature, as shown in \tabref{table:Cd_Re3700}. Our results for $C_d$ are close to those from Large Eddy Simulation (LES), which is expected given that the residual-based Variational Multiscale (VMS) formulation used in our framework is similar to an LES approach. Instantaneous flow visualizations are shown in \figref{fig:flow_visualization_Re3700} to provide insights into the mechanisms of the flow past the sphere. As the flow past the sphere, it stretches and wraps around, forming a horseshoe-shaped vortex in the wake region shown on \figref{fig:flow_visualization_Re3700_qcriteria}. We can successfully capture this structure in our simulations, provided that the refinement level in the wake region is sufficient.

\begin{table}[t!]
    \centering
    \caption{Summary of Results from Various Studies for flow past a sphere at $Re = 3700$}
    \begin{tabular}{@{}P{5cm}P{1.5cm}@{}}
        \toprule
        \textbf{Study} & $C_d$ \\
        \midrule
        Schlichting (exp.) \citep{schlichting2016boundary}  & 0.39 \\
        Rodr{\'\i}guez \textit{et al.} (DNS) \citep{rodriguez2011direct} & 0.394 \\
        Yun \textit{et al.} (LES) \citep{yun2006vortical} & 0.355 \\
        Park \textit{et al.} (LES) \citep{park2006dynamic} & 0.359 \\
        Current                    &  0.359\\
        \bottomrule
    \end{tabular}
    \label{table:Cd_Re3700}
\end{table}

\clearpage

\subsubsection{Flow past a three-dimensional cylinder at $Re = 3900$}\label{subsubsec:Cylinder}
To further evaluate our framework, we selected a canonical problem of flow past a cylinder at $Re = 3900$. The solution domain is \([0, 25.12] \times [0, 3.14] \times [0, 25.12]\), with the cylinder positioned at \((5, 0, 12.56)\) and its central axis along the $y$-direction. For the inlet, we apply a uniform non-dimensional freestream velocity of (1, 0, 0). For the outlet, we apply backflow stabilization. For the other walls, we apply periodic boundary conditions.

The current study's time-averaged drag coefficients compared with those reported in the literature are shown in \tabref{table:Cd_Cylinder_Re3900}. Additionally, the streamwise velocity field is averaged at two positions in the wake region, and these results are compared with those found in the literature in \figref{fig:Mean_velocity_Cylinder_Re3900}.

\begin{table}[h!]
    \centering
    \caption{$C_d$ from Various Studies for flow past a cylinder at $Re = 3900$}
    \begin{tabular}{@{}P{5cm}P{1.5cm}@{}}
        \toprule
        \textbf{Study} & $C_d$ \\
        \midrule
        Lourenco and Shih (exp.)~\citep{lourenco1994characteristics} & 0.99 \\
        Norberg (exp.)~\citep{norberg1994experimental} & 0.98 \\
        Beaudan (LES)~\citep{beaudan1995numerical} & 0.92 \\
        Franke and Frank (LES)~\citep{franke2002large}  & 0.978 \\
        Ma \textit{et al.} (DNS)~\citep{ma2000dynamics} & 0.96 \\
        Meyer \textit{et al.} (LES)~\citep{meyer2010conservative} & 1.05 \\
        Lysenko \textit{et al.} (LES) ~\citep{lysenko2012large} & 0.97 \\
        Current                    &  0.965\\

        \bottomrule
    \end{tabular}
    \label{table:Cd_Cylinder_Re3900}
\end{table}

\begin{figure}[h!]
\centering
\begin{subfigure}{1.0\textwidth}
\centering
\begin{tikzpicture}
    \begin{axis}[
        width=10cm, 
        height=5cm, 
        xlabel={z},
        ylabel={u},
        grid=both,
        grid style={line width=.1pt, draw=gray!10},
        major grid style={line width=.2pt, draw=gray!50},
        minor tick num=5,
        xmin=-2, xmax=2,
        ymin=-0.4, ymax=1.6,
        legend style={draw=none} 
    ]
    \addplot [no markers,color=black, thick] table [
        col sep=comma,
        x index=0,
        y index=1
    ] {current_0p58.csv};
    \addplot [no markers,color=blue, thick] table [
        col sep=comma,
        x index=0,
        y index=1
    ] {Re3900_0p58_Meyer.csv};
    \end{axis}
\end{tikzpicture}
  \caption{x = 0.58}
  \label{subfig:0p58}
\end{subfigure}%
\\
\begin{subfigure}{1.0\textwidth}
\centering
\begin{tikzpicture}
    \begin{axis}[
        width=10cm,
        height=5cm,
        xlabel={z},
        ylabel={u},
        grid=both,
        grid style={line width=.1pt, draw=gray!10},
        major grid style={line width=.2pt, draw=gray!50},
        minor tick num=5,
        xmin=-2, xmax=2,
        ymin=-0.4, ymax=1.6,
        legend style={draw=none} 
    ]
    \addplot [no markers,color=black, thick] table [
        col sep=comma,
        x index=0,
        y index=1
    ] {current_1p06.csv};
    \addplot [no markers,color=blue, thick] table [
        col sep=comma,
        x expr=\thisrowno{0},
        y expr=\thisrowno{1} + 1 
    ] {Re3900_1p06_Meyer.csv};
    \addplot [no markers,dashed,color=red, thick] table [
        col sep=comma,
        x index=0,
        y index=1
    ] {Main_1p06.csv};
    \addplot [
        only marks,
        mark=*,
        color=brown!60!black,
        thick,
        mark size=1
    ] table [
        col sep=comma,
        x index=0,
        y index=1
    ] {Re3900_1p06_Parnaudeau_exp.csv};
    \end{axis}
\end{tikzpicture}
  \caption{x = 1.06}
  \label{subfig:1p06}
\end{subfigure}
\centering
\\[.3cm]
\begin{tikzpicture}
    \begin{axis}[
        width=14cm,
        height=7cm,
        xmin=0, xmax=1,
        ymin=0, ymax=1,
        axis lines=none,
        hide x axis,
        hide y axis,
        legend style={
            at={(0.5,-0.3)}, 
            anchor=south,
            legend columns=4,
            legend cell align=left,
            /tikz/every even column/.append style={column sep=0.5cm} 
        }
    ]

    \addlegendimage{no markers, color=black, thick}
    \addlegendentry{Current}

    \addlegendimage{no markers, color=blue, thick}
    \addlegendentry{Meyer \textit{et al.}\citep{meyer2010conservative}}

    \addlegendimage{no markers, dashed, color=red, thick}
    \addlegendentry{Main \textit{et al.}\citep{Main2018TheSB}}

    \addlegendimage{only marks, mark=*, color=brown!60!black, thick}
    \addlegendentry{Experimental (Parnaudeau \textit{et al.})\citep{parnaudeau2008experimental}}

    \end{axis}
\end{tikzpicture}

\caption{Mean velocity profiles in the $x$-direction at two locations, compared with data from the literature.}
\label{fig:Mean_velocity_Cylinder_Re3900}
\end{figure}

\begin{figure}[t!]
    \centering
    \begin{subfigure}[b]{0.9\linewidth}
        \includegraphics[width=0.99\linewidth,trim=0 0 0 0,clip]{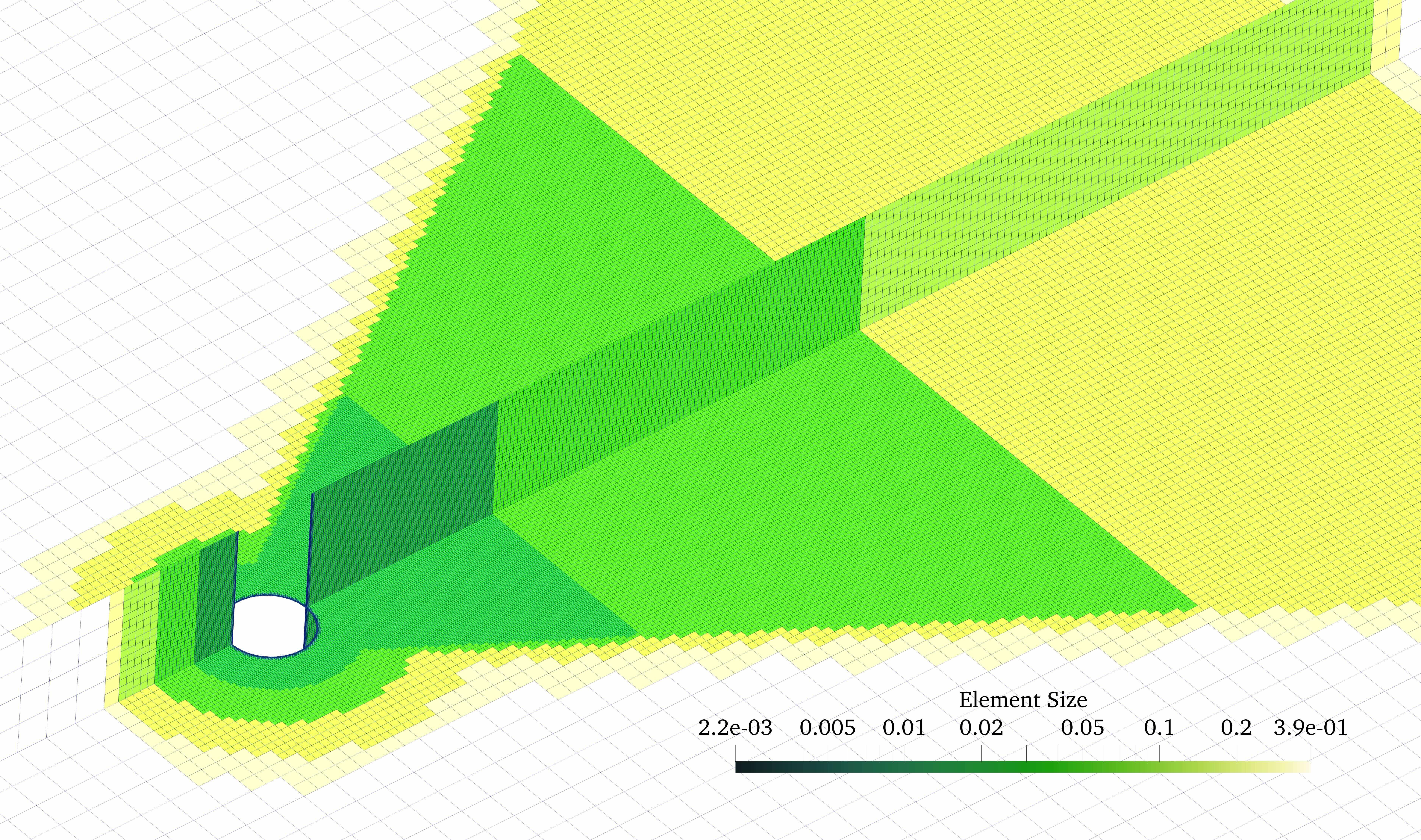}
        \caption{Local refinement of the octree mesh.}
        \label{fig:Mesh_flow_pass_cylinder}
    \end{subfigure}
    \begin{subfigure}[b]{0.9\linewidth}
        \centering
        \includegraphics[width=0.99\linewidth,trim=0 0 0 0,clip]{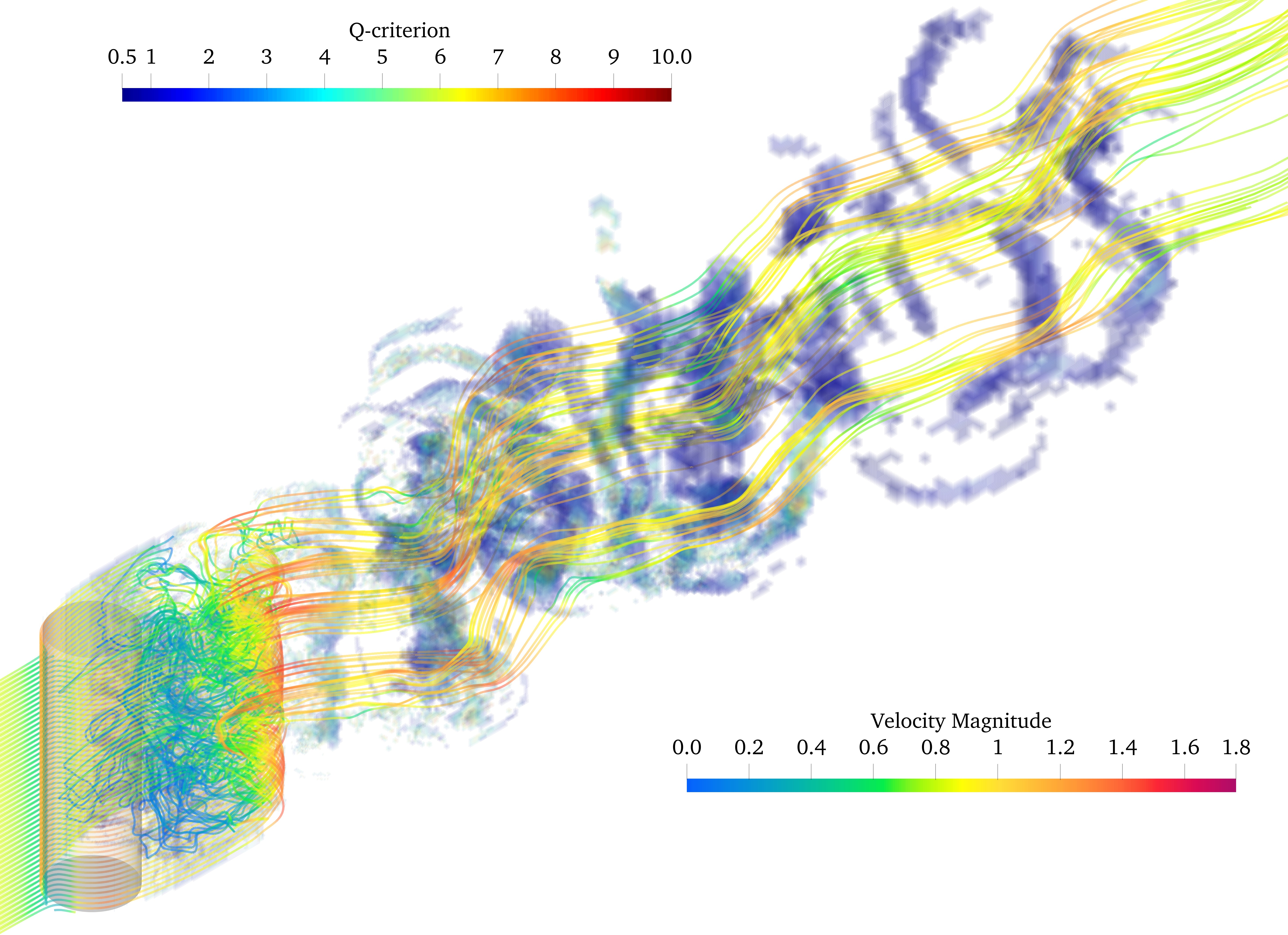}
        \caption{Instantaneous flow visualization using Q-criteria, ranging from 0.5 to 10. Streamlines are colored according to the flow's velocity magnitude.}
        \label{fig:flow_visualization_Re3900_qcriteria}
    \end{subfigure}
    \caption{Local mesh refinement and instantaneous flow visualization for the flow past a cylinder.}
    \label{fig:cylinder_case}
\end{figure}

\clearpage

The mesh refinement regions, shown in \figref{fig:Mesh_flow_pass_cylinder}, include cylindrical refinement regions centered along the same line as the main cylinder geometry. The refinement levels are as follows: a small cylinder with a radius of 0.51 at level 12, a medium cylinder with a radius of 1 at level 10, a larger cylinder with a radius of 1.5 at level 9, and the largest cylinder with a radius of 2 at level 8. Additionally, three trapezoidal regions in the wake are refined to levels 8, 9, and 10. As depicted in \figref{fig:flow_visualization_Re3900_qcriteria}, instantaneous flow visualization illustrates the flow physics of the fluid past a cylinder in three dimensions.

\subsubsection{Internal Flow through a nozzle}\label{subsub:nozzle}
\begin{figure}
    \centering
    \begin{subfigure}{0.45\textwidth}
        \centering
        \includegraphics[width=\linewidth,trim=0 50 50 0,clip]{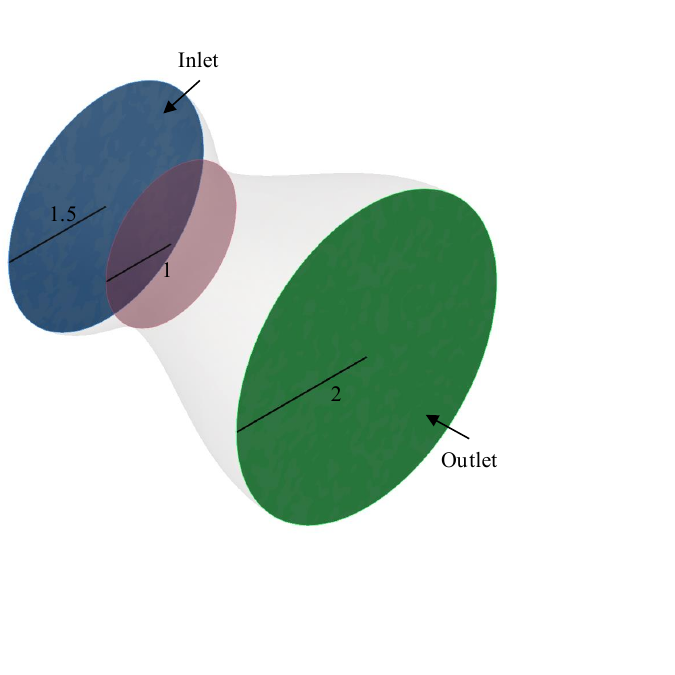}
        \caption{Schematic of the nozzle domain. A parabolic velocity profile is imposed at the inlet, while a zero-pressure condition is applied at the outlet. The Shifted Boundary Method (SBM) enforces no-slip Dirichlet boundary conditions on all other surfaces.}
        \label{fig:solving-domain-nozzle1}
    \end{subfigure}
    \hspace{5pt}
    \begin{subfigure}{0.45\textwidth}
        \centering
        \includegraphics[width=\linewidth,trim=0 50 50 0,clip]{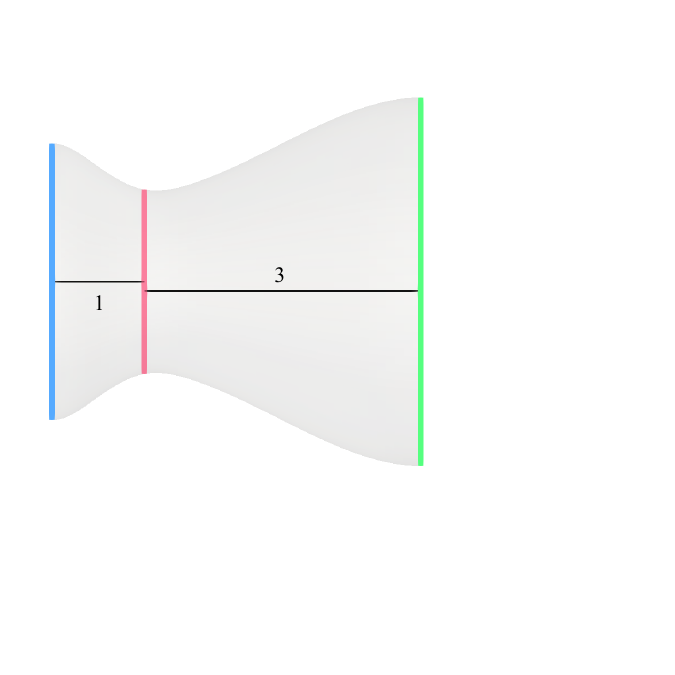}
        \caption{Side view of the nozzle geometry. The contraction occurs at \(x = 1\), followed by an expansion at \(x = 4\).}
        \label{fig:solving-domain-nozzle2}
    \end{subfigure}
    \caption{Nozzle geometry and boundary conditions.}
    \label{fig:nozzle-geometry}
\end{figure}

We conducted simulations of internal flow through a nozzle, as shown in \figref{fig:solving-domain-nozzle1} and \figref{fig:solving-domain-nozzle2}. The inlet cross-section at \(x = 0\) is circular with a radius of 1.5. The nozzle contracts to a radius of 1 at \(x = 1\) before expanding to a radius of 2 at \(x = 4\). The inlet boundary condition imposes a parabolic velocity profile of \(1 - \frac{r^2}{r_{\text{inlet}}^2}\), where \(r\) represents the distance from a point on the inlet plane to the center of the inlet circle, and \(r_{\text{inlet}}\) is the radius of the circle at the inlet plane. The outlet boundary condition sets the pressure to zero, while the remaining boundary surfaces enforce a no-slip Dirichlet condition using the SBM.

\begin{figure}
    \centering
    \begin{subfigure}{0.31\textwidth}
        \centering
        \includegraphics[width=\linewidth,trim=600 130 425 180,clip]{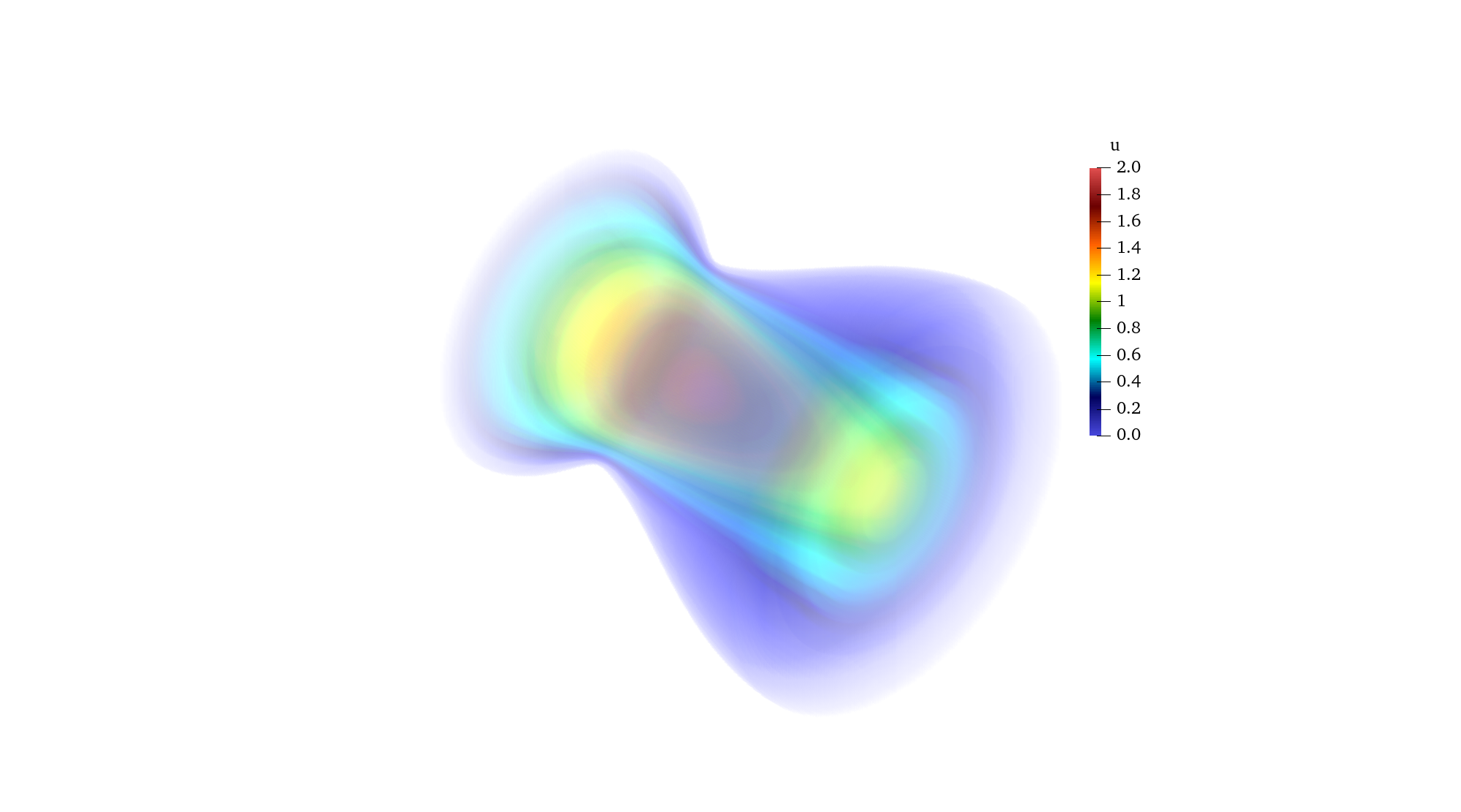}
        \caption{Volume rendering of \(x\)-velocity.}
        \label{fig:volume-render-view-nozzle}
    \end{subfigure}
    \hspace{5pt}
    \begin{subfigure}{0.31\textwidth}
        \centering
        \includegraphics[width=\linewidth,trim=600 130 425 180,clip]{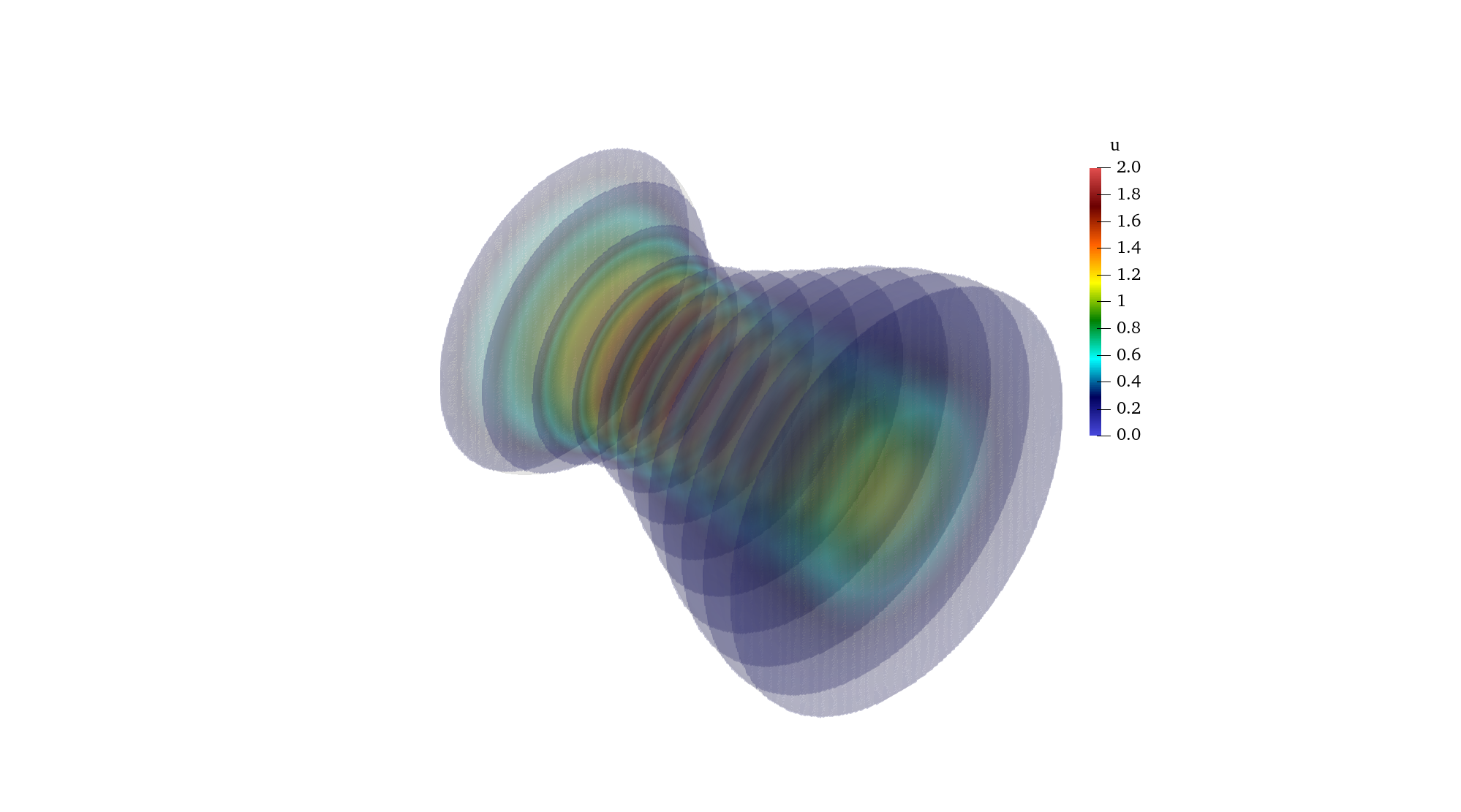}
        \caption{Cross-sectional slices of \(x\)-velocity, taken at 12 equally spaced locations from \(x = 0\) to \(x = 4\).}
        \label{fig:slices-view-nozzle}
    \end{subfigure}
    \hspace{5pt}
    \begin{subfigure}{0.31\textwidth}
        \centering
        \includegraphics[width=\linewidth,trim=600 130 425 180,clip]{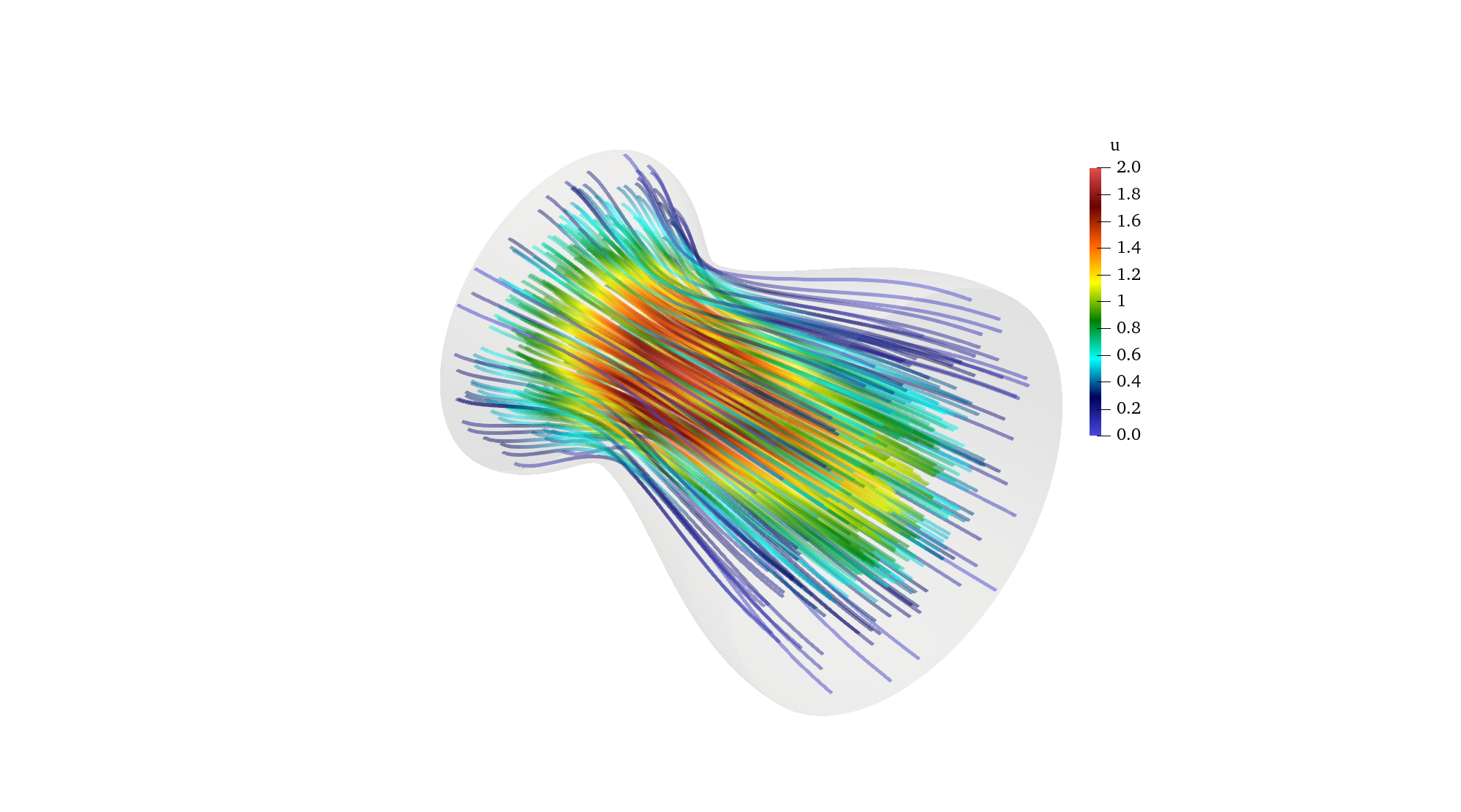}
        \caption{Streamlines showing flow patterns within the nozzle.}
        \label{fig:streamline-view-nozzle}
    \end{subfigure}
    \caption{Visualizations of the flow field through the nozzle.}
    \label{fig:nozzle-visualizations}
\end{figure}

The nozzle geometry was constructed using an incomplete octree framework. Simulations were performed at a Reynolds number of 10 with a non-dimensional time step of 1. Visualizations of the \(x\)-direction velocity (\(\lambda = 0.5\)) at a refinement level of 8 are presented as volume rendering, cross-sectional slices, and streamlines in \figref{fig:volume-render-view-nozzle}, \figref{fig:slices-view-nozzle}, and \figref{fig:streamline-view-nozzle}, respectively.

Simulations were conducted on uniform octree meshes at refinement levels of 4, 5, 6, 7, and 8, resulting in a mesh size of \(4/2^{\text{lvl}}\), where \(\text{lvl}\) denotes the refinement level. The effect of different \(\lambda\) values on the flow was analyzed. To verify mass conservation, we evaluated several cross-sectional slices (\figref{fig:slices-view-nozzle}). The \(x\)-direction velocity was integrated over the area of each slice. 

\begin{figure}[htbp]
    \centering
    \begin{subfigure}{0.49\textwidth}
        \centering
        \includegraphics[width=\linewidth,trim=0 0 0 0,clip]{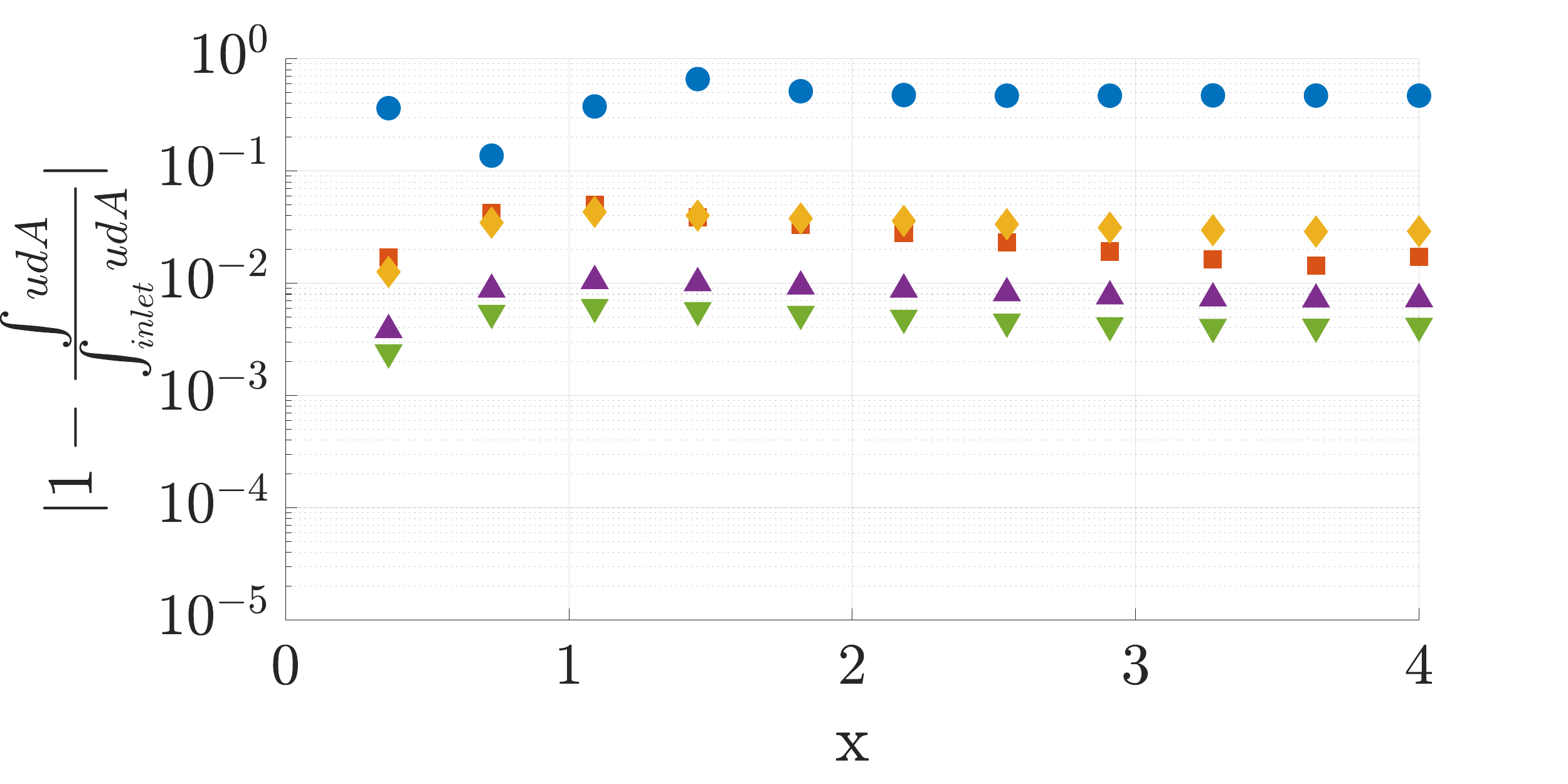}
        \caption{$\lambda = 0$.}
    \end{subfigure}
    \begin{subfigure}{0.49\textwidth}
        \centering
        \includegraphics[width=\linewidth,trim=0 0 0 0,clip]{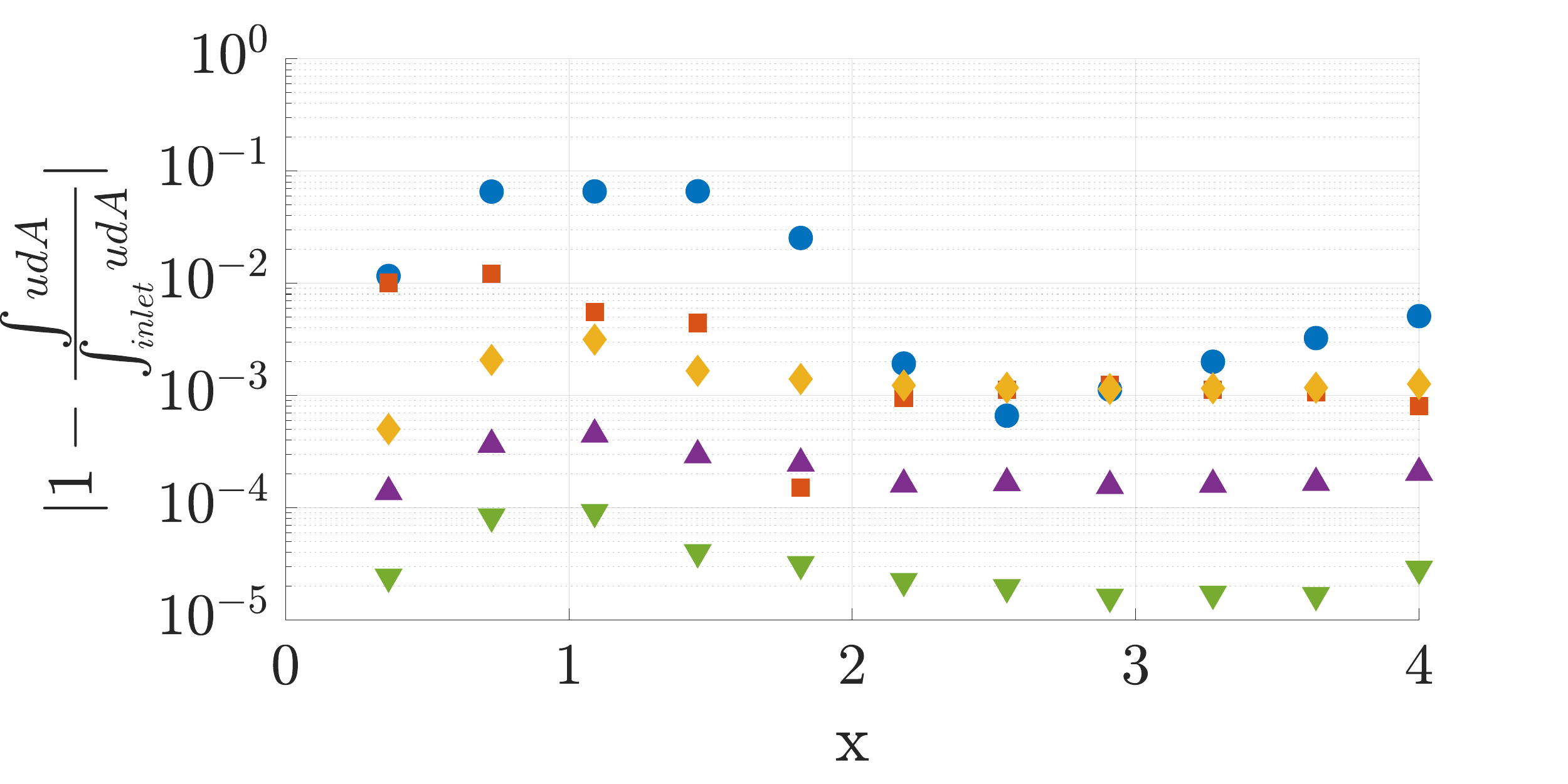}
        \caption{$\lambda = 0.5$.}
    \end{subfigure}
        \begin{subfigure}{0.49\textwidth}
        \centering
        \includegraphics[width=\linewidth,trim=0 0 0 0,clip]{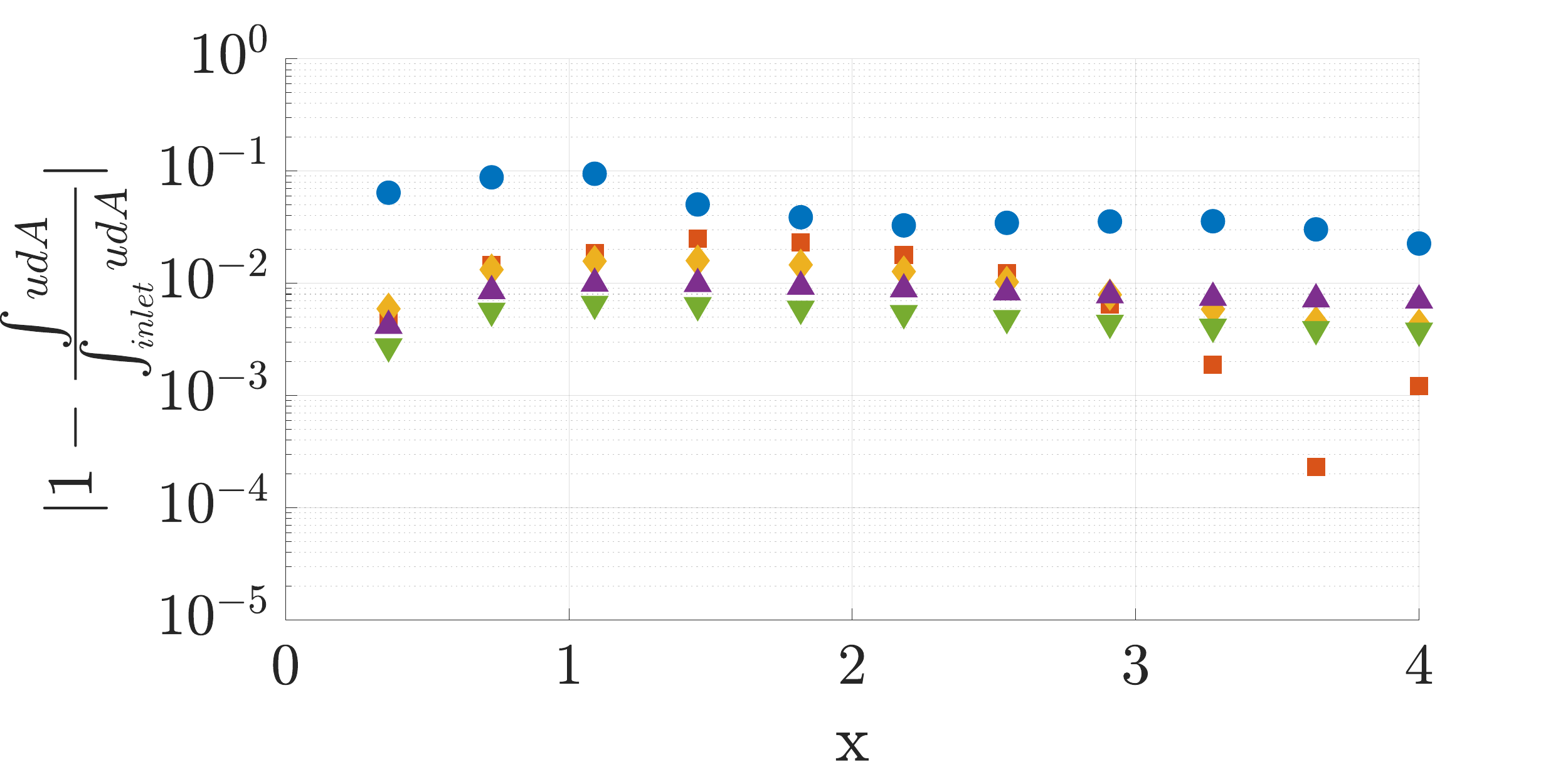}
        \caption{$\lambda = 1$.}
    \end{subfigure}
    \centering
    \includegraphics[width=0.25\linewidth,trim=0 0 0 0,clip]{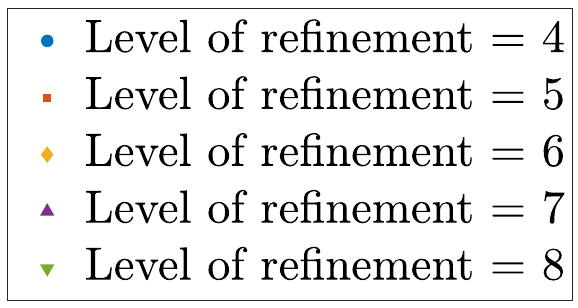}
    \caption{Flux ratio error (\(1 - \frac{\int u dA}{\int_{\text{inlet}} u dA}\)) for different refinement levels and diverse $\lambda$s in internal flow through a nozzle.}
    \label{fig:flux_plots}
\end{figure}

\figref{fig:flux_plots} shows the error \(1 - \frac{\int u dA}{\int_{\text{inlet}} u dA}\), which represents the difference between the ideal and calculated flux ratios. The calculated flux ratio is the flux at a given cross-section divided by the inlet flux. A value closer to zero indicates better mass conservation. It is evident that \(\lambda = 0.5\) achieves superior mass conservation across all refinement levels compared to non-optimal \(\lambda\) values. Notably, \(\lambda = 0\) leads to an increase in flux, while \(\lambda = 1\) results in a flux decrease through the nozzle.

\begin{figure}
    \centering
    \begin{subfigure}{0.45\textwidth}
        \centering
        \includegraphics[width=\linewidth]{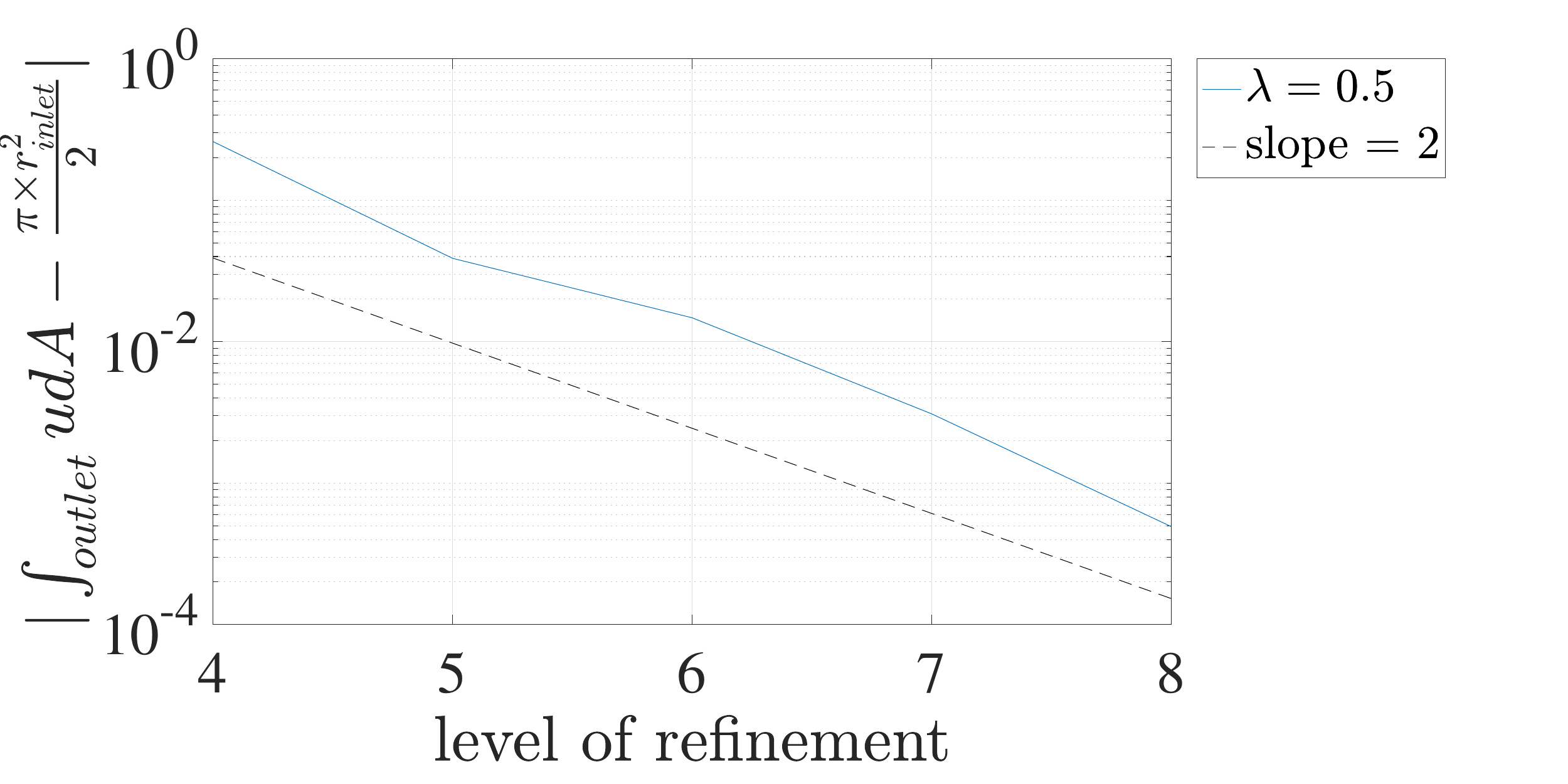}
        \caption{Convergence of outlet flux with mesh refinement. \(\lambda = 0.5\) achieves nearly second-order accuracy.}
        \label{fig:nozzle_convergence}
    \end{subfigure}
    \hspace{5pt}
    \begin{subfigure}{0.45\textwidth}
        \centering
        \includegraphics[width=\linewidth]{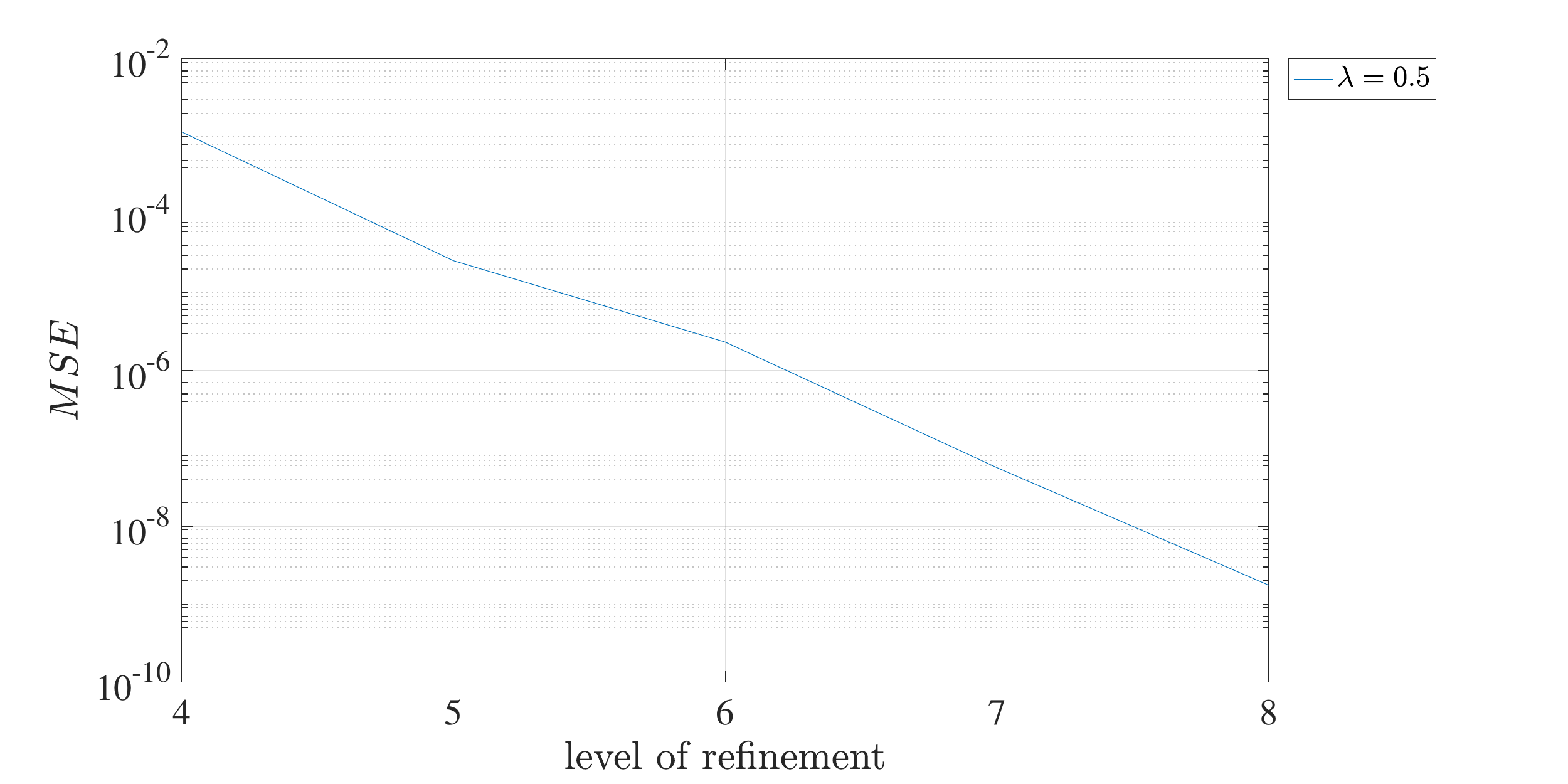}
        \caption{MSE for different \(\lambda\) values across mesh refinement levels. \(\lambda = 0.5\) demonstrates the best convergence.}
        \label{fig:nozzle_MSE}
    \end{subfigure}
    \caption{Flux convergence and MSE analysis for various \(\lambda\) values for internal flow through a nozzle case.}
    \label{fig:nozzle-mass-conservation}
\end{figure}

\figref{fig:nozzle_convergence} illustrates the convergence of the calculated flux at the outlet compared to the exact flux (\(\frac{\pi r_{\text{inlet}}^2}{2}\)) from the parabolic velocity profile at the inlet as the mesh is refined. The results demonstrate that \(\lambda = 0.5\) achieves nearly second-order accuracy. \figref{fig:nozzle_MSE} presents the Mean Squared Error (MSE), calculated as
\[
\text{MSE} = \frac{\sum_{i = 0}^n \left( 1 - \frac{\int_{i} u \, dA}{\int_{\text{inlet}} u \, dA} \right)^2}{n},
\]

where \(n\) represents the number of cross-sectional slices (12 in this study). Here, \(\int_{i} u \, dA\) denotes the flux integrated over the area of the \(i\)-th cross-sectional plane, and \(\int_{\text{inlet}} u \, dA\) corresponds to the flux at the inlet. The MSE quantifies the average error across all cross-sections. The results clearly demonstrate that \(\lambda = 0.5\) achieves the best convergence, with superior mass conservation compared to other values of \(\lambda\).

Using $\lambda = 0.5$ and level of refinement equal to 8, the difference between the outlet flux and the inlet flux is approximately $1 \times 10^{-4}$, which is sufficiently close to zero, demonstrating mass conservation within the surrogate domain. This result is further validated in \secref{sec:SBM_NS}.

\subsubsection{Internal flow through a gyroid}\label{subsubsec:Gyroid}

\begin{figure}[b!]
    \centering
    \begin{subfigure}{0.9\textwidth}
        \centering
        \includegraphics[width=0.5\linewidth,trim=500 100 520 80,clip]{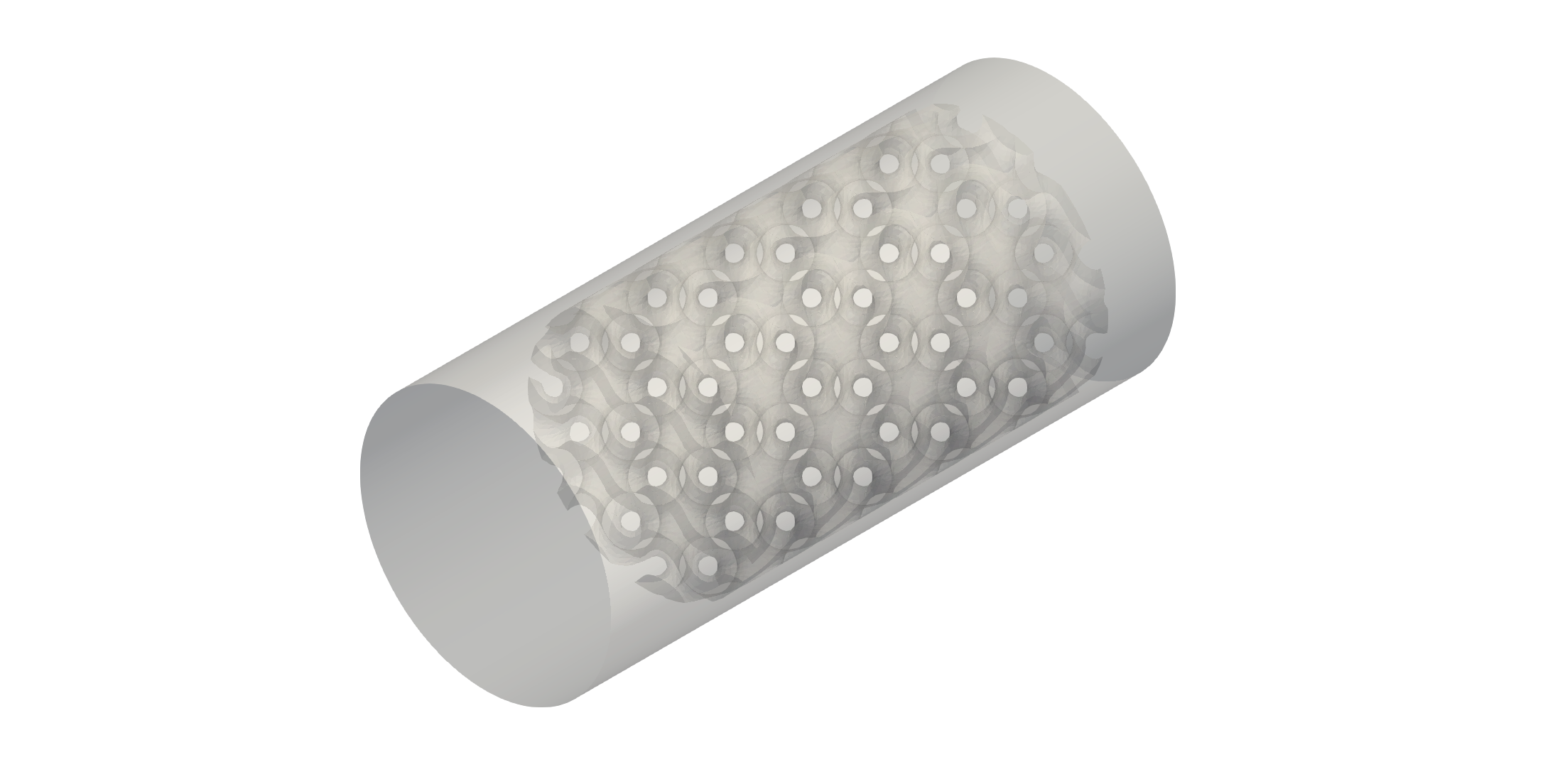}
        \caption{The domain we solve, where a gyroid STL is immersed into the fluid domain.}
        \label{fig:solving-domain}
    \end{subfigure}
    \begin{subfigure}{0.6\textwidth}
        \centering
        \includegraphics[width=0.99\linewidth,trim=0 0 0 0,clip]{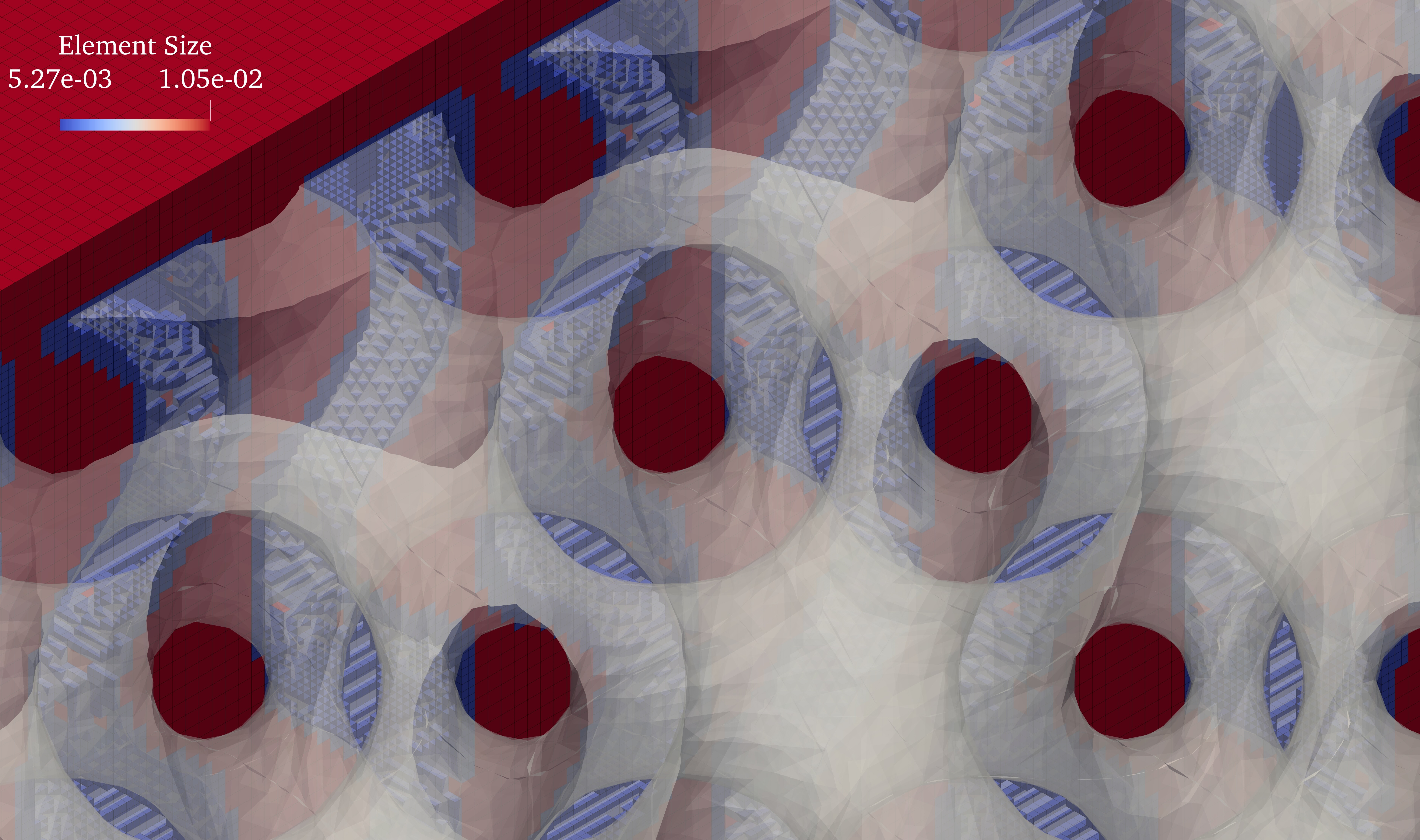}
        \caption{Carved out fluid domain for flow simulation, with finer blue mesh near the gyroid for accurately applying Dirichlet boundary conditions using SBM.}
        \label{fig:carve-out}
    \end{subfigure}
    \hspace{5pt}
    \begin{subfigure}{0.3\linewidth}
        \centering
        \includegraphics[width=0.99\linewidth,trim=0 0 0 0,clip]{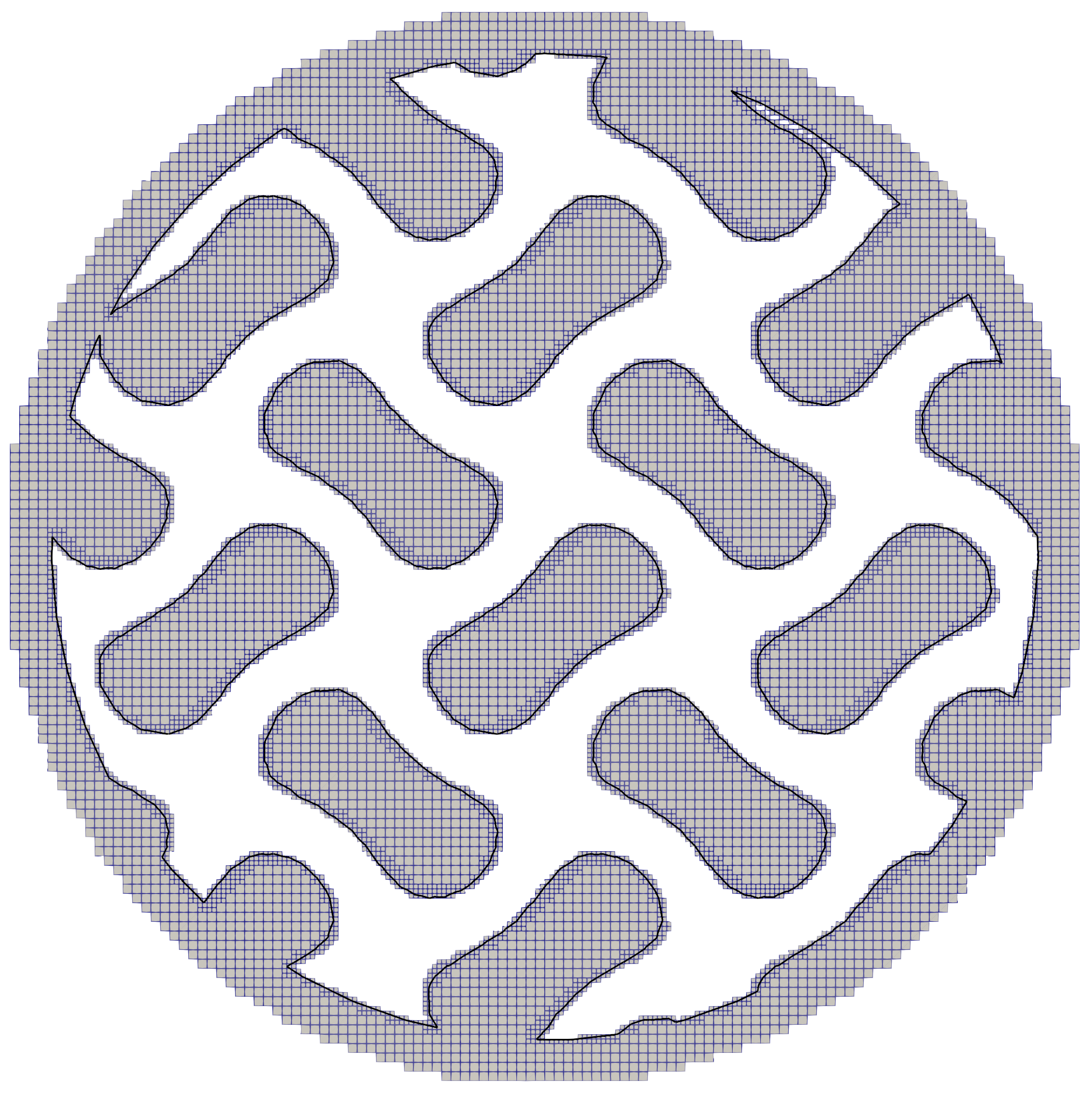}
        \caption{Slice view showing mesh refinement near the gyroid boundary.}
        \label{fig:slice-view}
    \end{subfigure}
    \caption{Flow passing through Gryroid in a pipe: (a) Immersed gyroid STL in fluid domain, (b) Octree-refined mesh of carved-out fluid region, and (c) Cross-sectional view highlighting near-boundary mesh refinement.}
    \label{fig:simulation-domain}
\end{figure}

\begin{figure}[t!]
    \centering
    \begin{subfigure}{0.49\textwidth}
        \centering
        \includegraphics[width=0.99\linewidth,trim=500 180 300 100,clip]{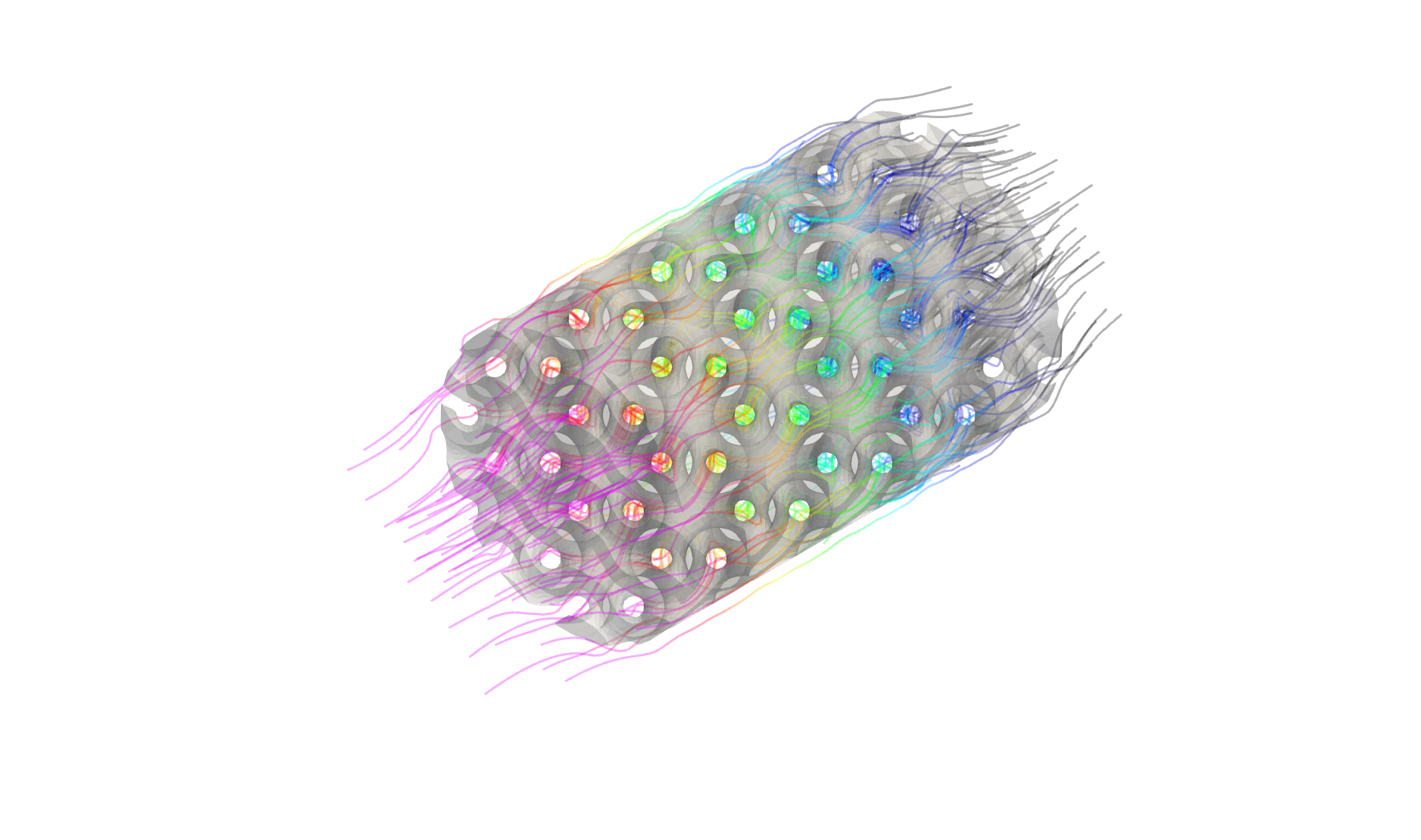}
        \caption{Isometric view with geometry}
        \label{fig:iso-geo}
    \end{subfigure}
    \begin{subfigure}{0.49\textwidth}
        \centering
        \includegraphics[width=0.99\linewidth,trim=500 180 300 100,clip]{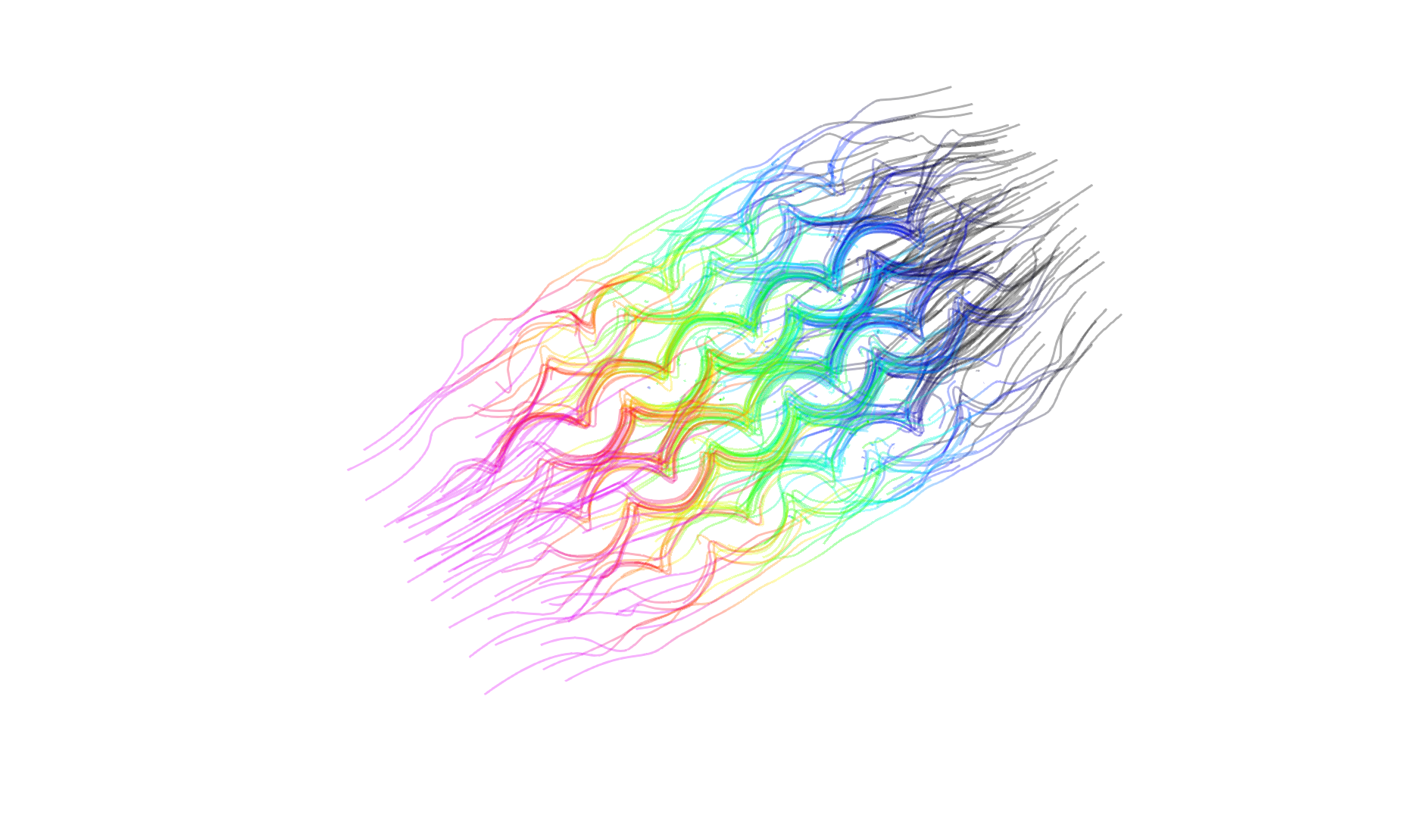}
        \caption{Isometric view without geometry}
        \label{fig:iso-no-geo}
    \end{subfigure}
    \\
    \begin{subfigure}{0.5\textwidth}
        \centering
           \raisebox{0cm}{\includegraphics[width=0.99\linewidth]{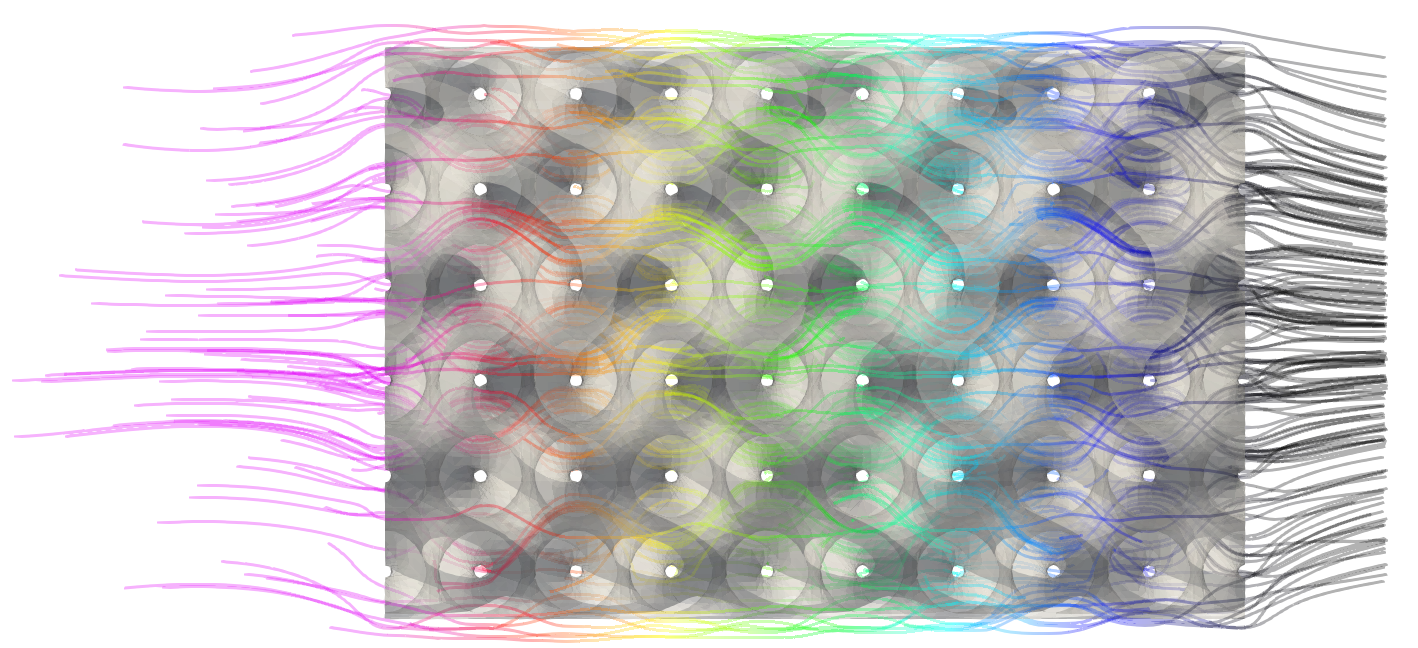}}
        \caption{Top view}
        \label{fig:top-view}
    \end{subfigure}
    \hspace{1pt}
    \begin{subfigure}{0.25\textwidth}
        \centering
        \includegraphics[width=0.99\linewidth,trim=0 0 0 0,clip]{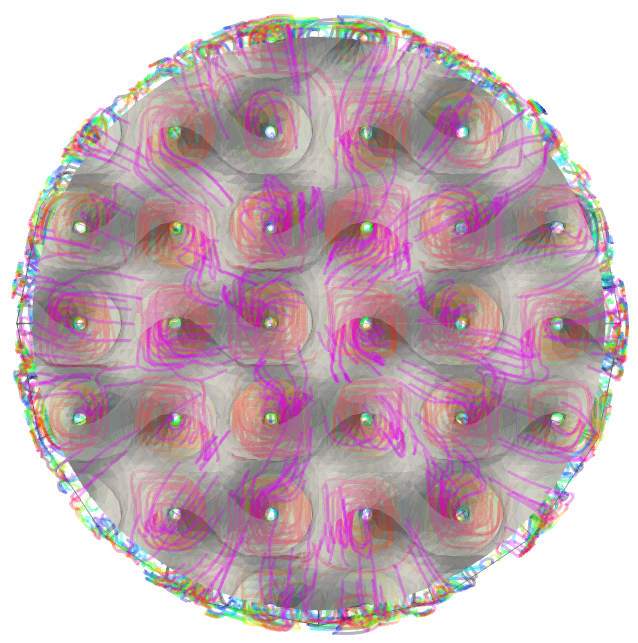}
        \caption{Inlet view}
        \label{fig:inlet-view}
    \end{subfigure}
    \\
    \begin{subfigure}{0.7\textwidth}
        \centering
        \includegraphics[width=0.99\linewidth,trim=0 0 0 0,clip]{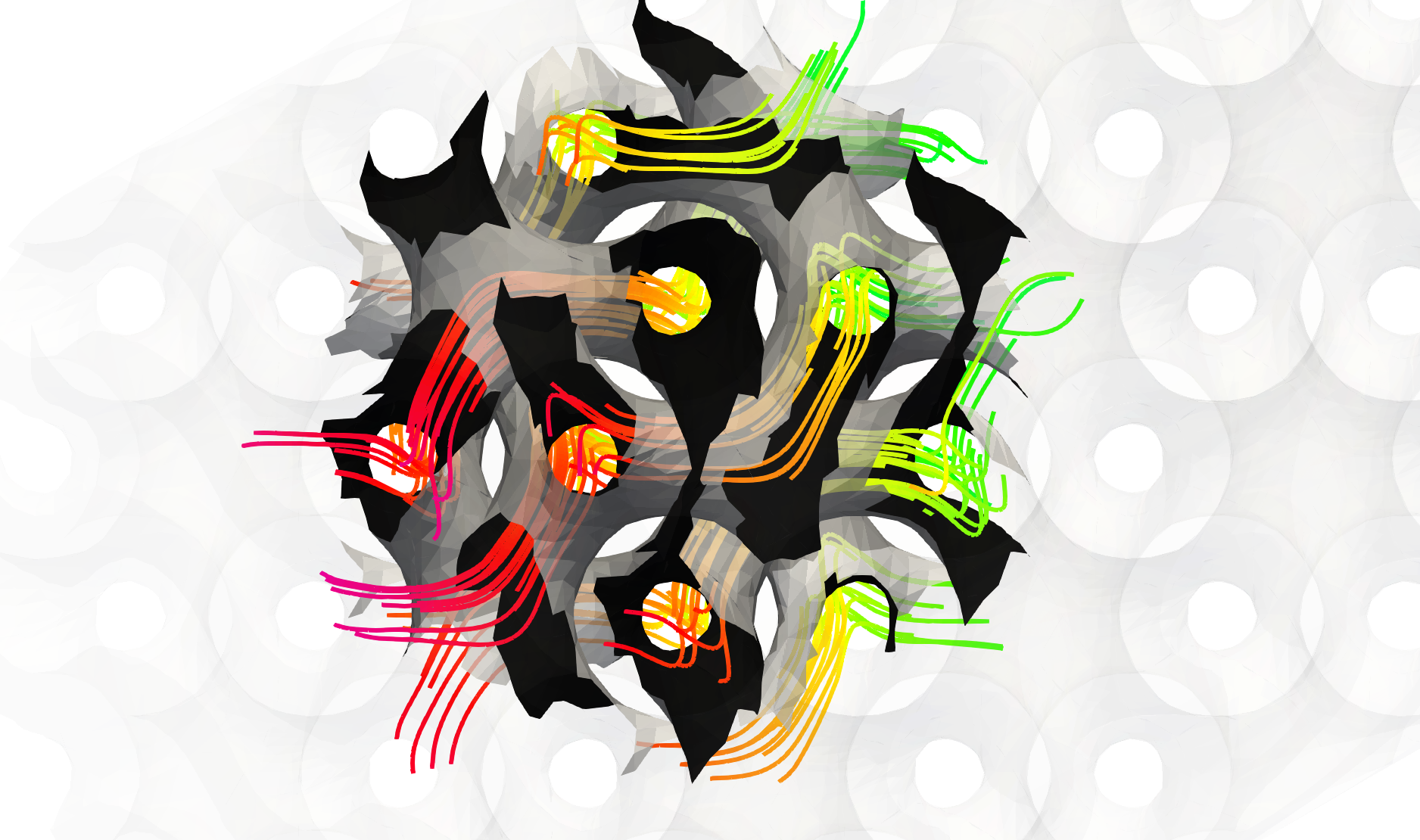}
        \caption{Inside view}
        \label{fig:inside-view}
    \end{subfigure}
    \begin{subfigure}{0.5\textwidth}
        \centering
        \includegraphics[width=0.99\linewidth,trim=0 0 0 0,clip]{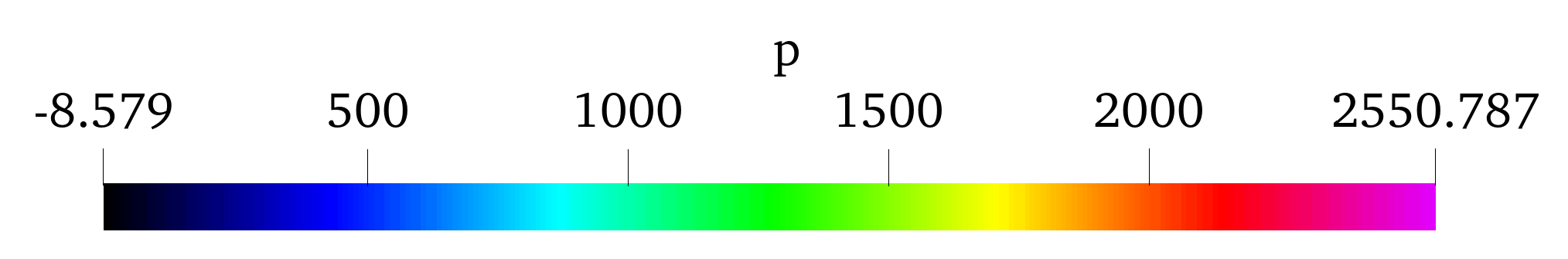}
    \end{subfigure}
    \caption{Multi-perspective streamline visualization of flow passing through a Gyroid structure: (a) Isometric view with geometry, (b) Isometric view without geometry, (c) Top view, (d) Inlet view, and (e) Inside view. Streamlines are color-coded to represent pressure distribution.}
    \label{fig:gyroid-visualization}
\end{figure}

As a final illustrative example of the approach, we conducted flow simulations through a complex geometry. A gyroid -- a bi-continuous, constant curvature surface~\cite{schoen1970infinite} -- was placed within a pipe flow configuration. The fluid domain, illustrated in \figref{fig:solving-domain}, involves applying strong no-slip boundary conditions along the cylinder's lateral surface. Velocity boundary conditions are specified at the inlet as (1, 0, 0), and zero pressure at the outlet. We use SBM to apply Dirichlet conditions on the gyroid's surface. The mesh is constructed using our octree-based meshing, showcased in \figref{fig:carve-out} and \figref{fig:slice-view}. The Reynolds number for the simulation is 10, and the non-dimensional time step is 0.25. We perform simulation till the flow reaches a steady state. The streamline visualizations, which provide us insights into the flow within the gyroid, are plotted in \figref{fig:gyroid-visualization}.

\clearpage


Given the gyroid's intricate structure, the streamlines can be observed swirling and detouring to pass through the gyroid in \figref{fig:iso-geo}, \figref{fig:iso-no-geo}, \figref{fig:top-view}, and \figref{fig:inlet-view}. For a more detailed examination of the flow through the gyroid's internal pipe, \figref{fig:inside-view} offers a closer look at the flow dynamics within this complex structure. 

\subsection{Scaling for parallel computing}\label{subsec:scale}
We conclude with scaling results of our framework on the TACC~\Frontera~system. For the scaling test, we consider the problem described in \secref{subsec:2D_cylinder}. The simulations are conducted at a Reynolds number of 100. Near the circular disk, the maximum refinement level is set to 14 (resulting in a mesh size of $\frac{30}{2^{14}}$), leading to a total of 725053 mesh nodes. Since the problem is time-dependent, we simulate from the initial condition with a non-dimensional time step of 0.01, running the simulations until the non-dimensional total time reaches 0.1. A strong scaling study is performed using 56 $\times$ n processors, which n ranges from 1 to 7.

\begin{figure}[t!]
\centering
\begin{subfigure}{.45\linewidth}
    \centering
    \begin{tikzpicture}
    \begin{loglogaxis}[
        width=0.99\linewidth,
        height=0.75\linewidth,
        xtick={56,112,168,224,280,336, 392},
        xticklabel style={rotate=45,font=\footnotesize},
        xticklabels={$56$,$112$, $168$, $224$,$280$, $336$, $392$},
        scaled y ticks=true,
        xlabel={\footnotesize Number of processors},
        ylabel={\footnotesize Time (s)},
        legend style={at={(0.37,0.2)}, anchor=north, nodes={scale=0.65, transform shape}},
        legend columns=3,
        xmin=50,
        xmax=450,
        ymin=10,
        ymax=1000,
        ]
        \addplot table [x={proc},y={T_0p5_total},col sep=comma] {Scaling_FPC.txt};
        \addplot +[mark=none, black, dashed] [domain=50:500]{10000/x};
    \end{loglogaxis}
    \end{tikzpicture}
    \caption{Total solve time with different number of processors.}
    \label{fig:ScalingFPC_Total}
\end{subfigure}
\begin{subfigure}{.5\linewidth}
    \centering
    \begin{tikzpicture}
    \begin{axis}[
        width=0.99\linewidth,
        height=0.65\linewidth,
        ybar stacked,
        ymin=0,
        ymax=100,
        xticklabel style={rotate=0,font=\footnotesize},
        xticklabels = {$56$,$112$,$168$,$224$,$280$, $336$, $392$}, %
        xtick=data, 
        enlarge x limits={abs=0.5},
        yticklabel style={rotate=0,font=\footnotesize},
        ylabel={\footnotesize Percentage of time $\rightarrow$},
        xlabel={\footnotesize Number of processors $\rightarrow$},
        legend style={at={(0.5,1.35)},anchor= north,legend columns=3},
        ]
        \addplot table[x expr=\coordindex,y expr=\thisrow{T_0p5_OctreeCreate}*100/\thisrow{T_0p5_total},col sep=comma] {Scaling_FPC.txt};
        \addplot table[x expr=\coordindex,y expr=\thisrow{T_0p5_MatAssembly}*100/\thisrow{T_0p5_total},col sep=comma] {Scaling_FPC.txt};
        \addplot table[x expr=\coordindex,y expr=\thisrow{T_0p5_VecAssembly}*100/\thisrow{T_0p5_total},col sep=comma] {Scaling_FPC.txt};
        \addplot table[x expr=\coordindex,y expr=(\thisrow{T_0p5_total}-\thisrow{T_0p5_VecAssembly} - \thisrow{T_0p5_MatAssembly} - \thisrow{T_0p5_OctreeCreate})*100/
        \thisrow{T_0p5_total},col sep=comma] {Scaling_FPC.txt};
        \legend{\tiny Octree Construction, \tiny Matrix Assembly, \tiny Vector Assembly, \tiny Solve}
    \end{axis}
    \end{tikzpicture}
    \caption{Percentage of time for individual steps.}
    \label{fig:ScalingFPC_percentage}
\end{subfigure}
\caption{Scaling performance and time distribution across various stages of the SBM computation for 2D flow past a cylinder, as evaluated on TACC's~\Frontera~supercomputer.}
\label{fig:ScalingBunny}
\end{figure}
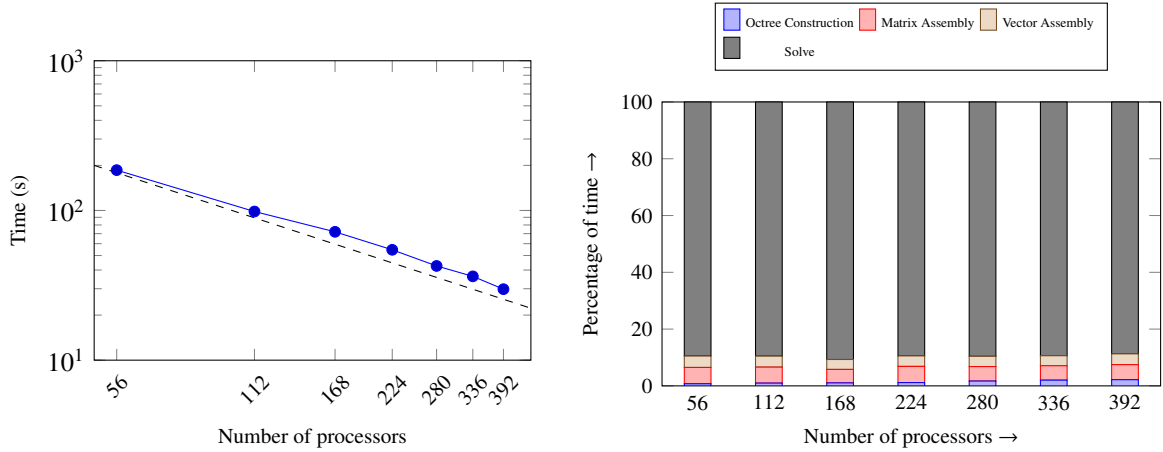

We utilize the BiCGStab (Bi-Conjugate Gradient Stabilized) algorithm, a robust Krylov subspace method, coupled with the Additive Schwarz Method (ASM) preconditioner. \figref{fig:ScalingFPC_Total} demonstrates the scalability of our approach, showing how the total solution time changes with an increasing number of processors. The results reveal a scaling behavior that is reasonably close to ideal, as indicated by the dashed line in \figref{fig:ScalingFPC_Total}. \figref{fig:ScalingFPC_percentage} provides a breakdown of the time allocation across various phases of the simulation process. Notably, the Navier-Stokes solver accounts for the majority of the computational time. This is expected, given that we simulate over 10 time steps and solve the computationally expensive nonlinear Navier-Stokes equations at each step.

\pagebreak
\section{Conclusions and Future Work}
\label{sec:conclusion}

In this work, we advance the field of computational fluid dynamics by applying the shifted boundary method in conjunction with a non-boundary-fitted octree mesh, leveraging an optimal surrogate boundary. This approach builds on previous successes in using SBM with an optimal surrogate boundary for solving Poisson and Linear Elasticity equations~\citep{yang2024optimal}, and extends its application to the Navier-Stokes equations -- a novel venture that integrates the use of octree meshes with an optimal surrogate boundary for the first time. We utilize this sophisticated SBM framework to tackle benchmark fluid dynamics problems.

A selection of fluid dynamics case studies spanning two-dimensional and three-dimensional challenges governed by the Navier-Stokes equations underpins our rigorous evaluation of Octree-SBM's performance. These include flow simulations around circular objects, lid-driven cavity flows with a circular obstacle, various aerodynamic problems, and flow passing through a complex gyroid topology. Our case studies show that the solutions produced by our Octree-SBM approach with optimal surrogate boundary closely align with established reference solutions. This alignment is notable in critical aerodynamic measures like drag coefficient and Strouhal number and effectively captures the pressure distribution near the geometric boundary. Our findings demonstrate that employing an optimal surrogate boundary ensures superior outcomes compared to alternative surrogate boundary choices. This improvement is evident in the reduction of boundary error, as defined in our paper, and in the enhanced accuracy of pressure profile representation near complex geometry boundaries.

Looking ahead, potential research directions could encompass several areas including: (a) Integrating the flow simulation framework with the convection-diffusion equation for thermal incompressible flow analyses. (b) Merging the flow simulation framework with shell formulations or other structural solvers to facilitate fluid-structure interaction studies. (c) Expanding the FEM basis functions within our framework to encompass higher-order basis functions.

\section*{Acknowledgements}
This work was partly supported by the National Science Foundation under the grants LEAP-HI 2053760 (BG, AK, CHY), DMREF 2323715/2323716 (BG, AK, CHY), CNS 1954556 (BG) and DMS 2207164 and DMS 2409919 (GS). BG, AK, and CHY are supported in part by AI Research Institutes program supported by NSF and USDA-NIFA under AI Institute for Resilient Agriculture, grant 2021-67021-35329.

\clearpage
\bibliographystyle{elsarticle-num-names.bst}
\bibliography{Bibs}

\clearpage
\appendix

\section{Boundary-fitted meshes for complex geometries} \label{appendix:LDC-BFM-SBM}
\begin{figure}[t!]
    \centering
    \begin{subfigure}{0.85\textwidth}
        \centering
        \includegraphics[width=0.99\linewidth,trim=0 0 0 0,clip]{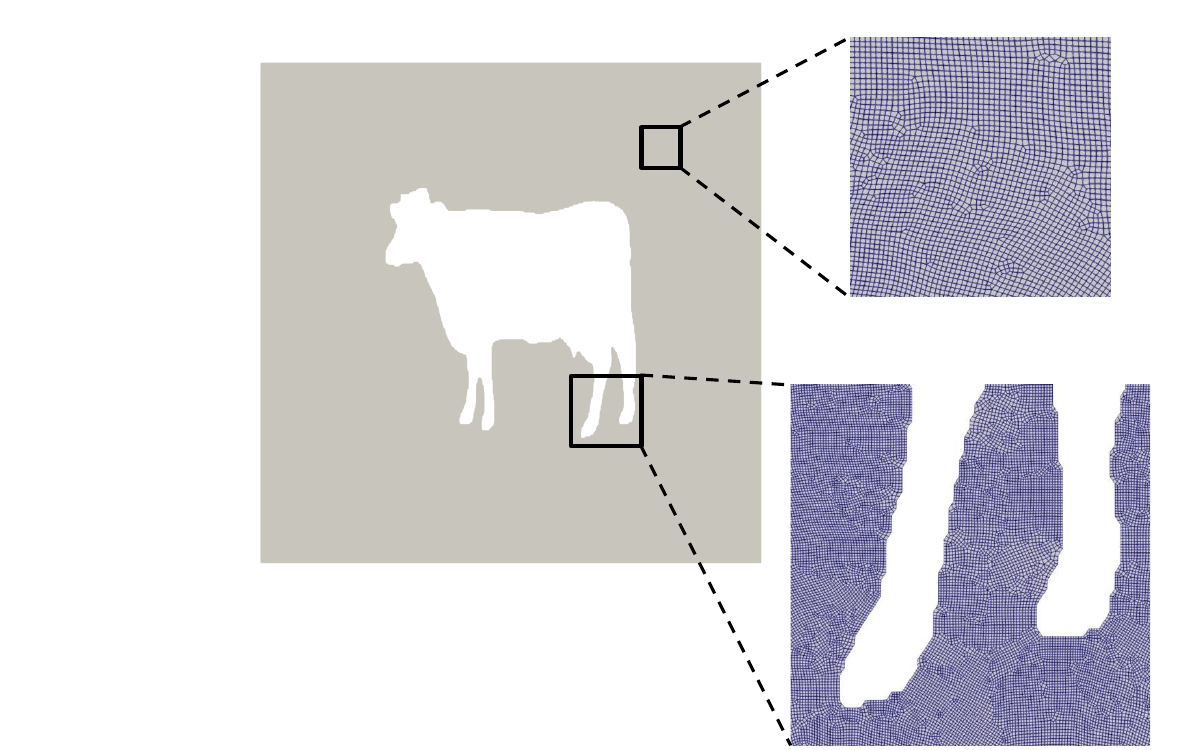}
        \caption{Unstructured boundary-fitted meshes.}
        \label{fig:Mesh-BFM}
    \end{subfigure}
        \hspace{1pt}
    \begin{subfigure}{0.85\textwidth}
        \centering
        \includegraphics[width=0.99\linewidth,trim=0 0 0 0,clip]{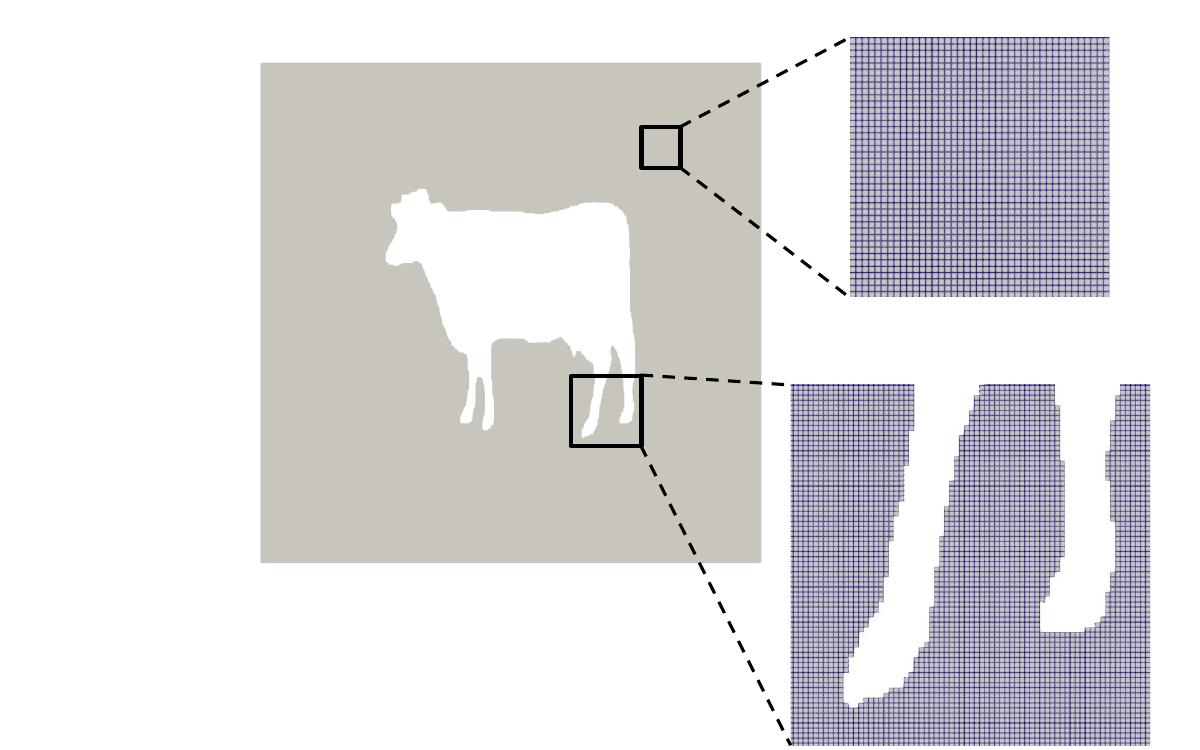}
        \caption{Structured octree meshes.}
        \label{fig:Mesh-SBM}
    \end{subfigure}
    \caption{Comparison between unstructured boundary-fitted meshes and structured octree meshes.}
    \label{fig:mesh-BFM-SBM}
\end{figure}

\begin{figure}[t!]
    \centering
    \begin{subfigure}{0.49\textwidth}
        \centering
        \includegraphics[width=0.99\linewidth,trim=500 0 600 250,clip]{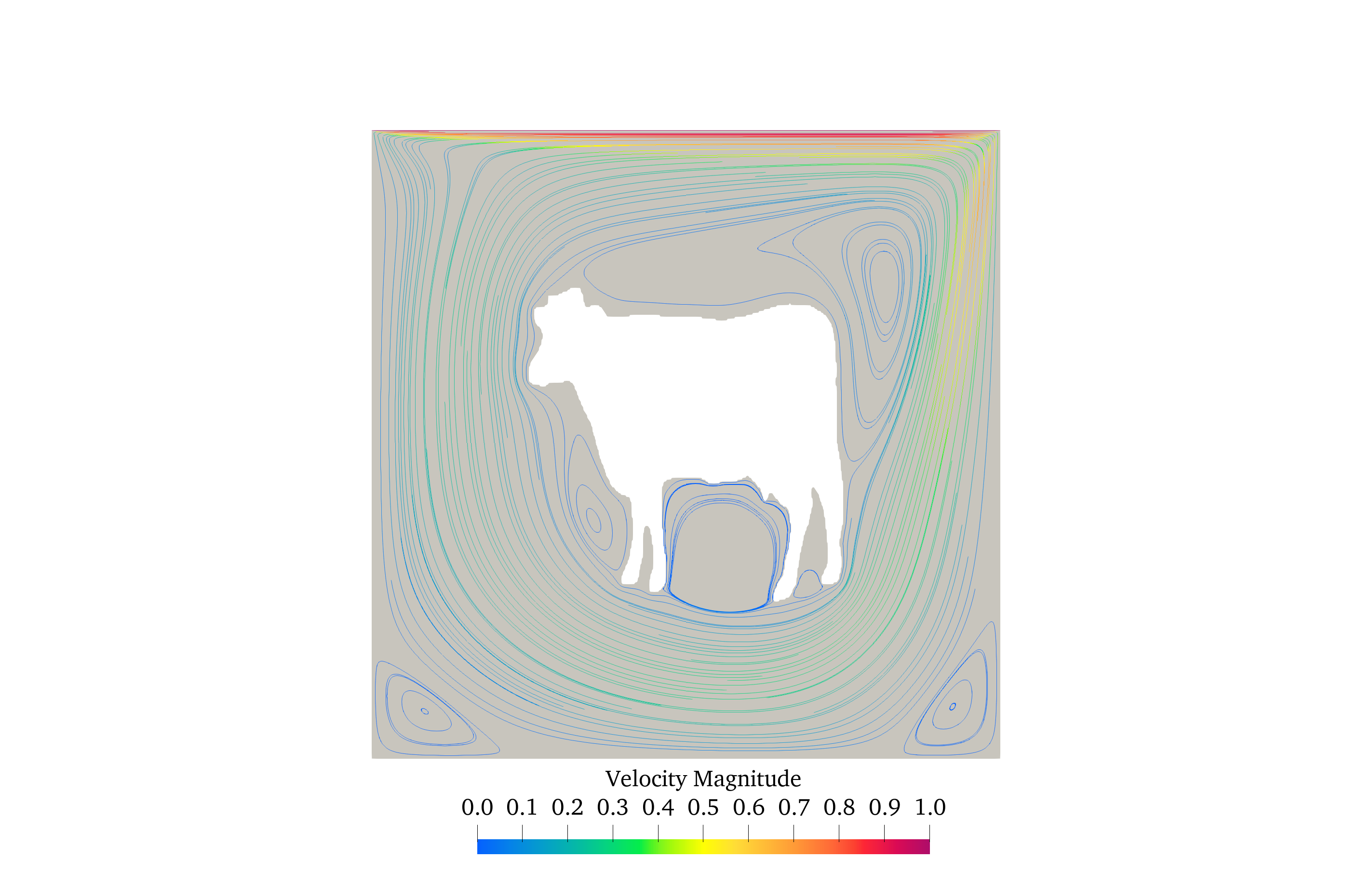}
        \caption{Streamlines of BFM.}
        \label{fig:NS-BFM}
    \end{subfigure}
        \hspace{1pt}
    \begin{subfigure}{0.49\textwidth}
        \centering
        \includegraphics[width=0.99\linewidth,trim=500 0 600 250,clip]{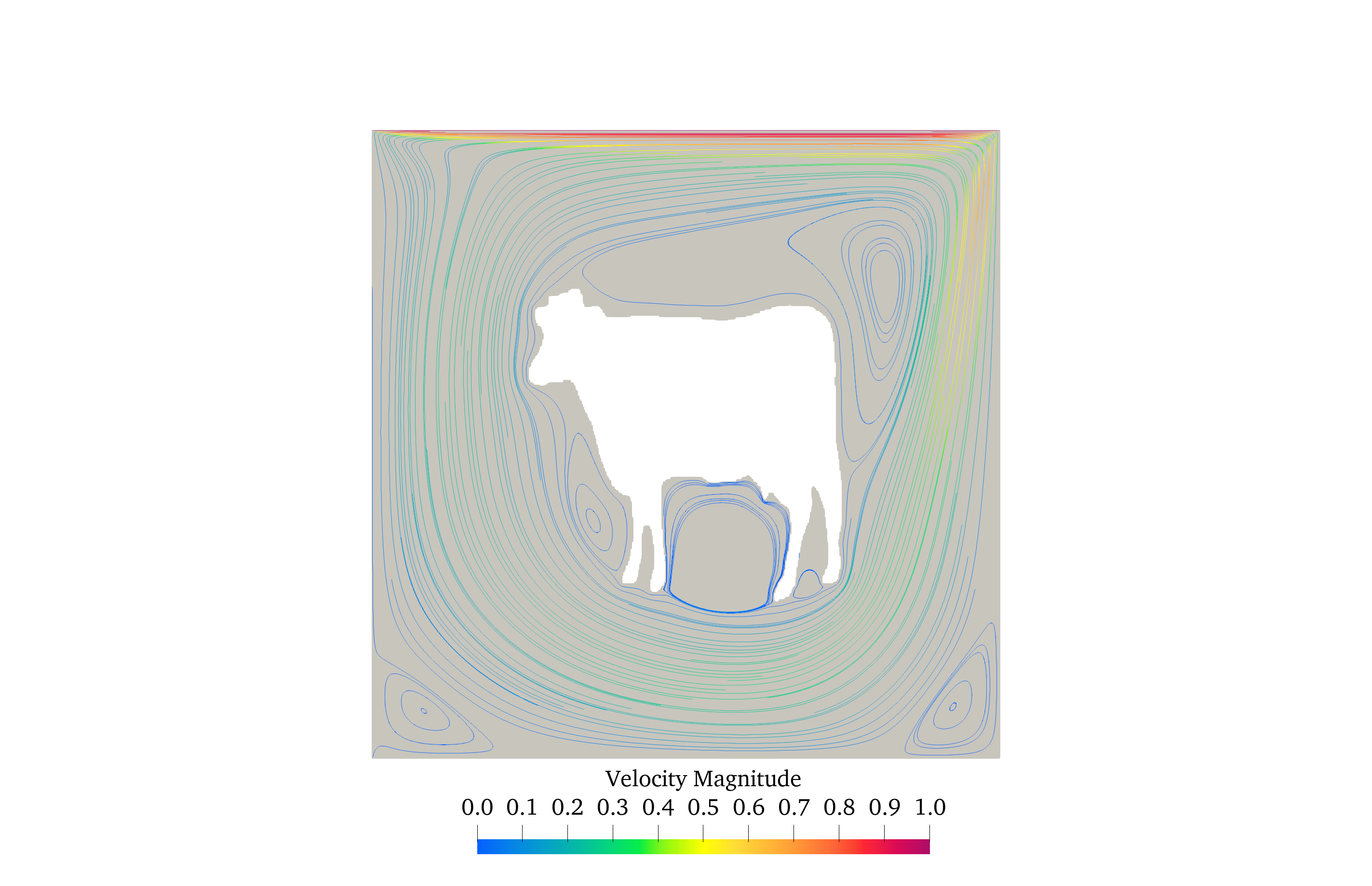}
        \caption{Streamlines of SBM.}
        \label{fig:NS-SBM}
    \end{subfigure}
    \caption{Comparison of streamlines for the lid-driven cavity flow with a cattle-shaped obstacle at $Re = 1000$, colored by velocity magnitude, between the simulation results of BFM and SBM.}
    \label{fig:NS-BFM-SBM}
\end{figure}

In \secref{subsub:ldc}, we conduct lid-driven cavity flow simulations over complex geometries. We select the cattle-shaped case to show and compare unstructured boundary-fitted meshes with structured, non-boundary-fitted octree meshes, as shown in \figref{fig:mesh-BFM-SBM}. Additionally, we present streamlines for both BFM and SBM for the lid-driven cavity flow with a cattle-shaped obstacle at $Re = 1000$ in \figref{fig:NS-BFM-SBM}.

\section{Solver parameters for simulations} \label{appendix:solver}

Scaling performance and time distribution across various stages of the SBM computation for 2D flow past a cylinder, as evaluated on TACC's~\Frontera~supercomputer, is shown in \secref{subsec:scale}.

In this work, the incompressible Navier-Stokes equations are solved using the \petsc~nonlinear solver framework.

We utilize SNES (Scalable Nonlinear Equations Solvers) to handle the Navier-Stokes system, applying a BiCGStab Krylov method (\texttt{ksp\_type = bcgs}) combined with Additive Schwarz preconditioning (\texttt{pc\_type = asm}) and LU factorization for subdomains (\texttt{sub\_pc\_type = lu}). The nonlinear solver is controlled with absolute and relative tolerances (\texttt{snes\_atol} and \texttt{snes\_rtol}) alongside limits on the maximum iterations (\texttt{snes\_max\_it}) and function evaluations (\texttt{snes\_max\_funcs}).

For the linear solve, diagonal scaling (\texttt{ksp\_diagonal\_scale}) is applied, with adjustments for potential scaling issues (\texttt{ksp\_diagonal\_scale\_fix}). The convergence of the linear solver is guided by the specified absolute and relative tolerances (\texttt{ksp\_atol} and \texttt{ksp\_rtol}). Additionally, the solver configuration includes options for monitoring convergence progress using SNES and KSP diagnostic tools (\texttt{snes\_monitor}, \texttt{ksp\_monitor\_short}).

The solver settings used in this study are carefully configured to ensure robust convergence and efficient computation. The key parameters are summarized in the code snippet below:

\begin{lstlisting}
solver_options_ns = {
  snes_atol = 1e-8
  snes_rtol = 1e-8
  snes_stol = 1e-8
  snes_max_it = 10
  snes_max_funcs = 1000
  snes_max_linear_solve_fail = 4

  ksp_diagonal_scale = True
  ksp_diagonal_scale_fix = True
  ksp_max_it = 1000
  ksp_atol = 1e-10
  ksp_rtol = 1e-10
  ksp_type = "bcgs"
  pc_type = "asm"
  sub_pc_type = "lu"

  snes_monitor = ""
  snes_converged_reason = ""
  ksp_monitor_short = ""
  ksp_converged_reason = ""
};
\end{lstlisting}

\end{document}